\documentclass[12pt]{article}

\usepackage{amssymb}
\usepackage{amsmath}
\usepackage{amscd}
\usepackage{latexsym}
\usepackage{graphicx}

\numberwithin{equation}{section}
\usepackage{cite}

\newcommand{\be}{\begin{equation}}
\newcommand{\ee}{\end{equation}}
\newcommand{\Dlt}{\Delta}
\newcommand{\dlt}{\delta}
\newcommand{\prt}{\partial}
\newcommand{\br}{{\bf r}}
\newcommand{\bk}{{\bf k}}
\newcommand{\bx}{{\bf x}}
\newcommand{\bfe}{{\bf e}}
\newcommand{\ba}{{\bf a}}
\newcommand{\bp}{{\bf p}}
\newcommand{\bd}{{\bf d}}
\newcommand{\bn}{{\bf n}}
\newcommand{\bP}{{\bf P}}

\newcommand{\bv}{{\bf v}}
\newcommand{\bt}{\beta}
\newcommand{\vp}{\varphi}
\newcommand{\ep}{\varepsilon}
\newcommand{\al}{\alpha}
\newcommand{\ra}{\rightarrow}
\newcommand{\sgm}{\sigma}

\newcommand{\gm}{\gamma}
\newcommand{\om}{\omega}
\newcommand{\Om}{\Omega}
\newcommand{\Gm}{\Gamma}
\newcommand{\dgr}{\dagger}
\newcommand{\lbd}{\lambda}
\newcommand{\Lbd}{\Lambda}
\newcommand{\bI}{{\bf I}}
\newcommand{\bS}{{\bf S}}
\newcommand{\bB}{{\bf B}}
\newcommand{\bF}{{\bf F}}
\newcommand{\bJ}{{\bf J}}
\newcommand{\bL}{{\bf L}}
\newcommand{\rgl}{\rangle}
\newcommand{\lgl}{\langle}

\begin{document}

{\bf Review}

\vskip 0.5cm

\begin{center}

{\Large{\bf Dipolar and spinor bosonic systems} \\ [5mm]

V.I. Yukalov} \\ [3mm]

{\it Bogolubov Laboratory of Theoretical Physics, \\
Joint Institute for Nuclear Research, Dubna 141980, Russia \\ [2mm]
                                         and \\ [2mm]
Instituto de Fisica de S\~ao Calros, Universidade de S\~ao Paulo, \\
CP 369,  S\~ao Carlos 13560-970, S\~ao Paulo, Brazil }

\end{center}

\vskip 5mm

\begin{abstract}

The main properties and methods of describing dipolar and spinor atomic systems,
composed of bosonic atoms or molecules, are reviewed. The general approach for the
correct treatment of Bose-condensed atomic systems with nonlocal interaction potentials
is explained. The approach is applied to Bose-condensed systems with dipolar interaction
potentials. The properties of systems with spinor interaction potentials are described.
Trapped atoms and atoms in optical lattices are considered. Effective spin Hamiltonians
for atoms in optical lattices are derived. The possibility of spintronics with cold
atom is emphasized. The present review differs from the previous review articles by 
concentrating on a thorough presentation of basic theoretical points, helping the 
reader to better follow mathematical details and to make clearer physical
conclusions.

\end{abstract}

\vskip 3mm
{\parindent =0pt


}

\vskip 1cm

{\parindent =0pt
{\bf Contents}

\vskip 2mm

{\bf 1}. {\bf Introduction}

\vskip 4mm
{\bf 2}. {\bf Nonlocal interaction potentials}

}

\vskip 2mm
   2.1. General approach

\vskip 2mm
   2.2. Grand Hamiltonian

\vskip 2mm
   2.3. Condensate function

\vskip 2mm
   2.4. Mean-field approximation

\vskip 2mm
   2.5. Uniform system

\vskip 2mm
   2.6. Thermodynamic characteristics

\vskip 2mm
   2.7. Local-density approximation

\vskip 2mm
   2.8. Grand potential

\vskip 2mm
   2.9. Bogolubov approximation

\vskip 2mm
   2.10. Stability conditions

\vskip 2mm
   2.11. Superfluid density

\vskip 2mm
   2.12. Optical lattices

\vskip 2mm
   2.13. Condensate in lattice

\vskip 2mm
   2.14. Lattice Hamiltonian

\vskip 2mm
   2.15. Rotating systems

\vskip 4mm
{\parindent =0pt
{\bf 3}. {\bf Dipolar interaction potentials} 
}

\vskip 2mm
   3.1. Scattering lengths

\vskip 2mm
   3.2. Interaction potentials

\vskip 2mm
   3.3. Regularized potential

\vskip 2mm
   3.4. Excitation spectrum

\vskip 2mm
   3.5. Bogolubov approximation

\vskip 2mm
   3.6. Roton instability

\vskip 2mm
   3.7. Dipolar instability

\vskip 2mm
   3.8. Trapped atoms

\vskip 2mm
   3.9. Geometric stabilization

\vskip 2mm
   3.10. Thomas-Fermi approximation

\vskip 2mm
   3.11. Quantum ferrofluid

\vskip 2mm
   3.12. Magnetic induction

\vskip 2mm
   3.13. Optical lattices

\vskip 2mm
   3.14. Deep lattice

\vskip 2mm
   3.15. Spin dynamics

\vskip 4mm
{\parindent =0pt
{\bf 4}. {\bf Spinor interaction potentials}
}

\vskip 2mm
   4.1. Hyperfine structure

\vskip 2mm
   4.2. Multicomponent versus fragmented

\vskip 2mm
   4.3. Spinor Hamiltonian

\vskip 2mm
   4.4. Grand Hamiltonian

\vskip 2mm
   4.5. Atoms with $F = 1$

\vskip 2mm
   4.6. Equilibrium properties

\vskip 2mm
   4.7. Stratification of components

\vskip 2mm
   4.8. Nonequilibrium properties

\vskip 2mm
   4.9. Optical lattices

\vskip 2mm
   4.10. Insulating lattice

\vskip 2mm
   4.11. Spin dynamics

\vskip 2mm
   4.12. Stochastic quantization

\vskip 2mm
   4.13. Scale separation

\vskip 2mm
   4.14. Spin waves

\vskip 2mm
   4.15. Influence of tunneling

\vskip 4mm
{\parindent =0pt
{\bf 5}. {\bf Conclusion}

}

\newpage

\section{Introduction}

Cold atomic and molecular bosonic systems have recently been the objects of intensive
research, both experimental and theoretical.  First, the attention has been concentrated
on dilute systems, composed of particles characterized by local interaction potentials,
independent of spins, whose properties are well described by the $s$-wave scattering
length. There are several books \cite{Lieb_1,Letokhov_2,Pethick_3,Pitaevskii_4}
and review articles \cite{Courteille_5,Andersen_6,Yukalov_7,Bongs_8,Yukalov_9,
Posazhennikova_10,Morsch_11,Yukalov_12,Moseley_13,Bloch_14,Proukakis_15,
Yurovsky_16,Yukalov_17,Yukalov_18,Yukalov_201} covering different aspects of
such bosonic systems with spin-independent local interaction potentials.

In the present review, bosonic atoms or molecules are treated interacting through nonlocal
interaction potentials, such as dipolar potential, and interacting through spinor forces that
are local but depending on the effective spins related to hyperfine states. Although there
exist reviews devoted to dipolar \cite{Griesmaier_19,Baranov_20,Pupillo_21,Lahaye_22,
Baranov_23,Gadway_24} and spinor \cite{Kurn_25,Kurn_26} systems
(see also \cite{Pethick_3,Ueda_27})  the present paper differs from them in the following
aspects. First, our aim here is not a brief enumeration of particular cases, numerical
calculations, and different experiments, but the attention here is concentrated on the
principal theoretical  points allowing for the correct treatment of dipolar and spinor systems.
Second, more attention is payed to the derivation of effective spin Hamiltonians for atoms
and molecules in optical lattices. Third, the possibility of spintronics with cold atoms and
molecules, interacting through dipolar and spinor forces, is discussed.

The exposition of the material is organized so that the main mathematical points be clear
to the reader. More technical details can be found in the tutorials \cite{Yukalov_28,Yukalov_29}.
Throughout the paper, the system of units is employed where the Boltzmann and Planck
constants are set to one, $k_B = 1$ and $\hbar = 1$.

\section{Nonlocal interaction potentials}

\subsection{General approach}

The system of particles interacting through a nonlocal potential $\Phi({\bf r})$ is described
by the energy Hamiltonian
$$
\label{2.1}
 \hat H = \int \hat\psi^\dgr(\br) \left ( -\; \frac{\nabla^2}{2m} + U
\right ) \hat\psi(\br) \; d\br \; +
$$
\be
+ \;
\frac{1}{2}
\int \hat\psi^\dgr(\br) \hat\psi^\dgr(\br') \Phi(\br-\br')
\hat\psi(\br')\hat\psi(\br)\; d\br d\br' \; ,
\ee
in which $U = U({\bf r},t)$ is an external potential, generally, depending
on spatial, ${\bf r}$, and temporal, $t$, variables. The field operators
$\hat{\psi}({\bf r})=\hat{\psi}({\bf r},t)$ contain time that, for short, is
not written explicitly. Bose-Einstein statistics is assumed.

When Bose-Einstein condensate appears in the system, the global gauge symmetry
becomes broken, being the necessary and sufficient condition for Bose-Einstein
condensation \cite{Lieb_1,Yukalov_12,Yukalov_18}. The gauge symmetry breaking is
the most conveniently realized by the Bogolubov \cite{Bogolubov_30,Bogolubov_31}
shift
\be
\label{2.2}
 \hat\psi(\br) = \eta(\br) + \psi_1(\br) \;  ,
\ee
where $\eta({\bf r})$ is the condensate function and $\hat{\psi}_1({\bf r})$ is the operator
of uncondensed particles. These variables are orthogonal to each other,
\be
\label{2.3}
 \int \eta^*(\br) \psi_1(\br) \; d\br = 0 \;  .
\ee
Note that shift (\ref{2.2}) is an exact canonical transformation, but not an approximation,
as one sometimes incorrectly writes.

The condensate function plays the role of the system order parameter defined as the
statistical average
\be
\label{2.4}
 \eta(\br) = \lgl  \hat\psi(\br) \rgl \; ,
\ee
which implies that
\be
\label{2.5}
 \lgl  \psi_1(\br) \rgl = 0  \;  .
\ee

The condensate function is normalized to the number of condensed particles,
\be
\label{2.6}
 N_0 = \int | \eta(\br) |^2 \; d\br \;  ,
\ee
while the number of uncondensed particles is the average
\be
\label{2.7}
N_1 = \lgl \hat N_1 \rgl
\ee
of the operator
\be
\label{2.8}
  \hat N_1 \equiv \int \psi_1^\dgr(\br) \psi_1(\br) \; d\br \;  .
\ee

More generally, condition (\ref{2.5}) is represented as the average
\be
\label{2.9}
 \lgl \hat\Lbd \rgl = 0
\ee
of the operator
\be
\label{2.10}
 \hat\Lbd = \int \left [ \lbd(\br) \psi_1^\dgr(\br) +
\lbd^*(\br) \psi_1(\br) \right ] \; d\br \;  .
\ee

The effective action functional, taking into account conditions (\ref{2.6}), (\ref{2.7}),
and (\ref{2.9}), is
\be
\label{2.11}
 A[\eta, \; \psi_1 ] = \int \left ( \hat L [\hat\psi] + \mu_0 N_0 + \mu_1 \hat N_1 +
\hat\Lbd \right ) \; dt \;  ,
\ee
with the  Lagrangian
\be
\label{2.12}
 \hat L [\hat\psi] = \int  \hat \psi^\dgr(\br) \left ( i\; \frac{\prt}{\prt t} \right )
\hat\psi(\br) \; d\br \; - \; \hat H  \; .
\ee
The action functional can also be written as
\be
\label{2.13}
 A[\eta, \; \psi_1 ] = \int L [\eta, \; \psi_1 ]\; dt \;  ,
\ee
through the generalized Lagrangian
\be
\label{2.14}
 L [\eta, \; \psi_1 ] = \int \hat \psi^\dgr(\br) \left ( i\; \frac{\prt}{\prt t} \right )
\hat\psi(\br) \; d\br \; - \;  H  \;  ,
\ee
with the grand Hamiltonian
\be
\label{2.15}
 H = \hat H - \mu_0 N_0 - \mu_1 \hat N_1  -\hat\Lbd \; .
\ee
The quantities $\mu_0$, $\mu_1$, and $\lambda({\bf r})$ play the role of Lagrange
multipliers.

The evolution equations for the variables $\eta({\bf r},t)$ and $\psi_1({\bf r},t)$
are given by the extremization of the action functional:
\be
\label{2.16}
\left \lgl \frac{\dlt A[\eta,\;\psi_1]}{\dlt\eta^*(\br,t) } \right \rgl = 0
\ee
and
\be
\label{2.17}
 \frac{\dlt A[\eta,\;\psi_1]}{\dlt\psi_1^\dgr(\br,t) } = 0 \; .
\ee
The extremization equations can be rewritten in the form
\be
\label{2.18}
 i\; \frac{\prt}{\prt t}\; \eta(\br,t) =
\left \lgl \frac{\dlt H}{\dlt\eta^*(\br,t) } \right \rgl
\ee
and
\be
\label{2.19}
 i\; \frac{\prt}{\prt t}\; \psi_1(\br,t) = \frac{\dlt H}{\dlt\psi_1^\dgr(\br,t) } \;   .
\ee
Note that the variational derivatives are related to commutators by the equation
\cite{Yukalov_18,Yukalov_32}
$$
 \frac{\dlt H}{\dlt\psi_1(\br,t) } = [ \psi_1(\br,t) ,\; H ] \;  .
$$
Hence equation (\ref{2.19}) is equivalent to the Heisenberg equation of motion.

The above equations make it straightforward to derive the evolution equations for
various operators. For instance, the evolution equation for the operator density
of uncondensed particles is
$$
i\; \frac{\prt}{\prt t}\; \left [ \psi_1^\dgr(\br,t)\psi_1(\br,t) \right ] =
 \left [ \psi_1^\dgr(\br,t)\psi_1(\br,t) , \; H \right ] =
$$
\be
\label{2.20}
 = \psi_1^\dgr(\br,t) \; \frac{\dlt H}{\dlt\psi_1^\dgr (\br,t) } \; - \;
\frac{\dlt H}{\dlt\psi_1(\br,t) }\; \psi_1(\br,t) \; .
\ee

\subsection{Grand Hamiltonian}

With the Bogolubov shift (\ref{2.2}), the grand Hamiltonian(\ref{2.15}) becomes the sum
of five terms distinguished by the power of the operators $\psi_1$,
\be
\label{2.21}
 H = \sum_{n=0}^4 H^{(n)} \;  .
\ee
The zero-order term contains no operators $\psi_1$,
\be
\label{2.22}
 H^{(0)} = \int \eta^*(\br) \left ( - \; \frac{\nabla^2}{2m} + U - \mu_0 \right )
\eta(\br) \; d\br + \frac{1}{2} \int \Phi(\br-\br') |\eta(\br)|^2  |\eta(\br')|^2 \;
d\br d\br' \; .
\ee
To satisfy condition (\ref{2.5}), there should be no first-order terms with respect to
$\psi_1$
in the Hamiltonian \cite{Yukalov_33}, so that
\be
\label{2.23}
 H^{(1)} = 0 \;  ,
\ee
which is achieved by fixing the Lagrange multiplier
$$
 \lbd(\br) = \left ( - \; \frac{\nabla^2}{2m} + U \right ) \eta(\br) +
\int \Phi(\br-\br') | \eta(\br')|^2 \eta(\br) \; d\br' \;  .
$$
The second-order term is
$$
H^{(2)} = \int \psi_1^\dgr(\br) \left ( - \; \frac{\nabla^2}{2m} + U - \mu_1 \right )
\psi_1(\br) \; d\br \; +
$$
$$
+\; \; \int \Phi(\br-\br') \left [ | \eta(\br) |^2 \psi_1^\dgr(\br')\psi_1(\br') +
\eta^*(\br)\eta(\br')\psi_1^\dgr(\br')\psi_1(\br) + \right.
$$
\be
\label{2.24}
 + \left.
\frac{1}{2} \; \eta^*(\br)\eta^*(\br')\psi_1(\br')\psi_1(\br) +
\frac{1}{2} \; \eta(\br)\eta(\br')\psi_1^\dgr(\br')\psi_1^\dgr(\br) \right ] \; d\br d\br' \;.
\ee
The third-order term reads as
\be
\label{2.25}
 H^{(3)} = \int \Phi(\br-\br')  \left [
\eta^*(\br) \psi_1^\dgr(\br')\psi_1(\br')\psi_1(\br) +
\psi_1^\dgr(\br)\psi_1^\dgr(\br')\psi_1(\br')\eta(\br) \right ] \; d\br d\br' \; .
\ee
And the fourth-order term becomes
\be
\label{2.26}
 H^{(4)} = \frac{1}{2} \int \psi_1^\dgr(\br)\psi_1^\dgr(\br') \Phi(\br-\br')
\psi_1(\br')\psi_1(\br)\; d\br d\br' \;  .
\ee

The densities of condensed and uncondensed particles are
\be
\label{2.27}
 \rho_0(\br) \equiv | \eta(\br)|^2 \; , \qquad
\rho_1(\br) \equiv \lgl  \psi_1^\dgr(\br)\psi_1(\br) \rgl \; .
\ee
The numbers of condensed and uncondensed particles are given by the integrals
\be
\label{2.28}
 N_0 = \int\rho_0(\br) \; d\br \; , \qquad  N_1 = \int\rho_1(\br) \; d\br \; .
\ee
Respectively, the total particle density is
\be
\label{2.29}
 \rho(\br) = \rho_0(\br) +\rho_1(\br) \;  ,
\ee
defining the total number of particles
\be
\label{2.30}
 N = \int \rho(\br) \; d\br =  N_0 + N_1 \;  .
\ee

In equilibrium, the condensate function does not depend on time and is real,
$$
 \eta(\br,t) = \eta(\br) = \eta^*(\br) \;  .
$$
And, since
$$
 \frac{\prt}{\prt t} \; \lgl \psi_1^\dgr(\br,t) \psi_1(\br,t) \rgl = 0 \;  ,
$$
from equation (\ref{2.20}) it follows that the correlation functions
$$
\lgl \psi_1^\dgr(\br) \psi_1(\br') \rgl = \lgl \psi_1^\dgr(\br') \psi_1(\br) \rgl \; ,
\qquad
\lgl \psi_1(\br) \psi_1(\br') \rgl = \lgl \psi_1^\dgr(\br') \psi_1^\dgr(\br) \rgl \; ,
$$
$$
\lgl \psi_1^\dgr(\br) \psi_1(\br)\psi_1(\br') \rgl =
\lgl \psi_1^\dgr(\br') \psi_1^\dgr(\br)\psi_1(\br) \rgl
$$
are also real, provided all operators $\psi_1$ are taken at the same time $t$.

\subsection{Condensate function}

The evolution equation for the condensate function is given by equation (\ref{2.18}).
This equation, with the notation for the single-particle density matrix
\be 
\label{2.31}
\rho_1(\br,\br') \equiv \lgl \psi_1^\dgr(\br')\psi_1(\br) \rgl \;   ,
\ee
the anomalous average
\be
\label{2.32}
\sgm_1(\br,\br') \equiv \lgl \psi_1(\br')\psi_1(\br) \rgl \;     ,
\ee
and for the anomalous average
\be
\label{2.33}
 \xi(\br,\br') \equiv \lgl \psi_1^\dgr(\br')\psi_1(\br') \psi_1(\br) \rgl \;  ,
\ee
takes the form
$$
i\; \frac{\prt}{\prt t} \; \eta(\br) = 
\left ( - \; \frac{\nabla^2}{2m} + U - \mu_0 \right ) \eta(\br) +
$$
\be
\label{2.34}
 +
\int \Phi(\br-\br') \left [ \rho(\br')\eta(\br) + \rho(\br,\br')\eta(\br')
+ \sgm_1(\br,\br')\eta^*(\br') + \xi(\br,\br') \right ] \; d\br' \; .
\ee

In equilibrium,
$$
\frac{\prt}{\prt t} \; \eta(\br) = 0 \;   ,
$$
which yields the equation
$$
\mu_0 \eta(\br) =  \left ( - \; \frac{\nabla^2}{2m} + U  \right ) \eta(\br) +
$$
\be
\label{2.35}
 +
\int \Phi(\br-\br') \left [ \rho(\br')\eta(\br) + \rho_1(\br,\br')\eta(\br')
+ \sgm_1(\br,\br')\eta^*(\br') + \xi(\br,\br') \right ] \; d\br' \; .
\ee

For a uniform system, without external potentials, the condensate function turns
into a constant
$$
 \eta(\br) = \eta = \sqrt{\rho_0} \qquad ( U = 0 ) \;  .
$$
The particle density is also a constant $\rho({\bf r}) = \rho$, and equation (\ref{2.35})
reduces to the equation
\be
\label{2.36}
 \mu_0 \eta = \rho \Phi_0\eta + \int \Phi(\br) [ \rho_1(\br,0)\eta +
\sgm_1(\br,0)\eta^* + \xi(\br,0) ] \; d\br \;  ,
\ee
in which
$$
 \Phi_0 \equiv \int \Phi(\br) \; d\br \;  ,
$$
assuming that the interaction potential is integrable.

In the Fock space $\mathcal{F}(\psi_1)$, generated by the field operator $\psi_1$,
the vacuum state is defined by the requirement
\be
\label{2.37}
  \psi_1(\br,t) | \eta \rgl = 0 \; ,
\ee
from which it follows that the state $|\eta \rangle$ is the coherent state with respect
to the filed operator $\hat{\psi}$, since
\be
\label{2.38}
 \hat \psi(\br,t) | \eta \rgl = \eta(\br,t) | \eta \rgl \; .
\ee
Thus, the condensate function has the meaning of the {\it coherent field}
\cite{Yukalov_28,Yukalov_29}.

The vacuum state $|\eta \rangle$ of the Fock space $\mathcal{F}(\psi_1)$ is the eigenstate
of the grand Hamiltonian (\ref{2.21}),
\be
\label{2.39}
 H | \eta \rgl =  H^{(0)} | \eta \rgl \;  ,
\ee
with $H^{(0)}$ being the eigenvalue.

It is important to stress that the Fock space $\mathcal{F}(\psi_1)$, where the gauge
symmetry is broken, is principally different from the Fock space $\mathcal{F}(\psi)$,
without gauge symmetry breaking, generated by the field operator $\psi$ for which
the Bogolubov shift is not applicable. In the Fock space $\mathcal{F}(\psi)$, without
the symmetry breaking, the Hamiltonian $H[\psi]$ has the same form (\ref{2.1}), but
its eigenstates, given by the eigenproblem
$$
 H[\psi] | n \rgl = E_n | n \rgl \;  ,
$$
are different from the state $|\eta \rangle$. The latter is not an eigenstate of $H[\psi]$.
 Actually, the Fock spaces $\mathcal{F}(\psi_1)$ and $\mathcal{F}(\psi)$ are orthogonal
in thermodynamic limit \cite{Yukalov_29,Yukalov_34}. So, the states $|n \rangle$,
pertaining to $\mathcal{F}(\psi)$, and $|\eta \rangle$, pertaining to the orthogonal Fock
space $\mathcal{F}(\psi_1)$, are also orthogonal.

Averaging equation (\ref{2.18}) over the vacuum coherent state $|\eta \rangle$ yields
the equation for the vacuum coherent field
\be
\label{2.40}
 i\; \frac{\prt}{\prt t} \; \eta(\br,t) = 
\left ( - \; \frac{\nabla^2}{2m} + U - \mu_0 \right ) \eta(\br,t) +
\int \Phi(\br-\br') |\eta(\br',t)|^2 \eta(\br,t) \; d\br' \;  .
\ee
As is evident, the equation for the vacuum coherent field (\ref{2.40}) differs form the
equation (\ref{2.34}) for the condensate function that is a coherent field, but, generally,
not the vacuum coherent field. Equation (\ref{2.40}) is a particular case of (\ref{2.34}),
where there are no uncondensed particles, so that $\rho_1$, $\sigma_1$, as well as
$\xi$, are zero.

One often confuses equation  (\ref{2.40}) for the vacuum coherent field with mean-filed
approximation. This is not correct. The mean-field approximation for a Bose-condensed
system is the Hartree-Fock-Bogolubov approximation, where $\xi$ is zero, but  $\rho_1$
and $\sigma_1$ are not \cite{Yukalov_28,Yukalov_29}.

By its mathematical structure, the vacuum coherent field equation (\ref{2.40}) is the
nonlinear Schr\"{o}dinger (NLS) equation. First, equation (\ref{2.40}) was advanced
by Bogolubov \cite{Bogolubov_35} in his well known Lectures on Quantum Statistics
published in 1949 and then republished numerous times
(e.g. \cite{Bogolubov_30,Bogolubov_31}). Solutions to this equation have been studied
in many papers starting from \cite{Gross_36,Gross_37,Gross_38,Gross_39,Wu_40,
Pitaevskii_41,Gross_42}.

\subsection{Mean-field approximation}

The mean-field approximation for a system with Bose-Einstein condensate is the
Hartree-Fock-Bogolubov (HFB) approximation, whose details are thoroughly expounded
in \cite{Yukalov_28,Yukalov_29}. In this approximation, Hamiltonian (\ref{2.21})
becomes
$$
H_{HFB} = E_{HFB} + 
\int \psi_1^\dgr(\br) \left ( - \; \frac{\nabla^2}{2m} + U - \mu_1 
\right ) \psi_1(\br) \; d\br \; + 
$$
$$
+ \; \int \Phi(\br-\br') \left [ \rho(\br') \psi_1^\dgr(\br) \psi_1(\br) +
 \rho(\br',\br) \psi_1^\dgr(\br') \psi_1(\br) + \right.
$$
\be
\label{2.41}
  + \left.
\frac{1}{2} \; \sgm(\br,\br') \psi_1^\dgr(\br') \psi_1^\dgr(\br) +
\frac{1}{2} \; \sgm^*(\br,\br') \psi_1(\br') \psi_1(\br) \right ] \; d\br d\br' \; ,
\ee
where the first term is
\be
\label{2.42}
 E_{HFB} = H^{(0)} - \; \frac{1}{2} \int \Phi(\br-\br') \left [ \rho_1(\br) \rho_1(\br')
+ | \rho_1(\br,\br')|^2 +   | \sgm_1(\br,\br')|^2 \right ] \; d\br d\br' \; ,
\ee
 with the notations
\be
\label{2.43}
\rho(\br,\br') \equiv \eta(\br) \eta^*(\br') + \rho_1(\br,\br')
\ee
and
\be
\label{2.44}
 \sgm(\br,\br') \equiv \eta(\br) \eta(\br') + \sgm_1(\br,\br')  .
\ee

The condensate-function equation (\ref{2.34}) in the Hartree-Fock-Bogolubov
approximation reads as
$$
i\; \frac{\prt}{\prt t} \; \eta(\br) = 
\left ( - \; \frac{\nabla^2}{2m} + U - \mu_0 \right ) \eta(\br) \; +
$$
\be
\label{2.45}
  + \;
\int \Phi(\br-\br') \left [ \rho(\br')\eta(\br) + \rho_1(\br,\br')\eta(\br') +
 \sgm_1(\br,\br')\eta^*(\br') \right ] \; d\br' \;    .
\ee
Here, generally, the quantities $\rho_1$ and $\sigma_1$ are not zero.

In equilibrium, this equation reduces to the equation
$$
\mu_0 \eta(\br) = \left ( - \; \frac{\nabla^2}{2m} + U  \right ) \eta(\br) \; +
$$
\be
\label{2.46}
   + \;
 \int \Phi(\br-\br') \left [ \rho(\br')\eta(\br) + \rho_1(\br,\br')\eta(\br') +
 \sgm_1(\br,\br')\eta^*(\br') \right ] \; d\br' \; .
\ee

For a uniform system, when $U = 0$, the latter equation becomes
\be
\label{2.47}
  \mu_0 = \rho \Phi_0 + 
\int \Phi(\br) [ \rho_1(\br,0) + \sgm_1(\br,0) ] \; d\br \; .
\ee

The Hamiltonian (\ref{2.41}) can be diagonalized by means of the Bogolubov
canonical transformations
\be
\label{2.48}
\psi_1(\br) = \sum_k [ b_k u_k(\br) + b_k^\dgr v_k(\br) ]
\ee
and
\be
\label{2.49}
 b_k = \int \left [ u_k^*(\br) \psi_1(\br) - v_k^*(\br) \psi_1^\dgr(\br) 
\right ] \; d\br \;  ,
\ee
where $k$ is a multi-index. Since the field operators $\psi_1$, as well as $b_k$,
obey the Bose-Einstein statistics, we have the equations
\be
\label{2.50}
\sum_k \left [ u_k(\br) u_k^*(\br') - v_k^*(\br) v_k(\br') \right ] = 
\dlt(\br-\br') \; ,   \qquad
\sum_k \left [ u_k(\br) v_k^*(\br') - v_k^*(\br) u_k(\br') \right ] = 0 \; ,
\ee
and
\be
\label{2.51}
 \int \left [ u_k^*(\br) u_p(\br) - v_k^*(\br) v_p(\br) \right ]\; d\br  = 
\dlt_{kp} \; ,   \qquad
\int \left [ u_k(\br) v_p(\br) - v_k(\br) u_p(\br) \right ] \; d\br = 0 \;  .
\ee

For convenience, we introduce the notations
\be
\label{2.52}
\om(\br,\br') = \left [ -\; \frac{\nabla^2}{2m} + U(\br) - \mu_1 +
\int \Phi(\br-\br') \rho(\br') \; d\br' \right ] \dlt(\br-\br') +
\Phi(\br-\br') \rho(\br,\br')
\ee
and
\be
\label{2.53}
 \Dlt(\br-\br') \equiv \Phi(\br-\br') \sgm(\br,\br') \;  .
\ee

Hamiltonian (\ref{2.41}) is diagonalized under the Bogolubov equations
$$
\int \left [ \om(\br,\br') u_k(\br') + \Dlt(\br,\br') v_k(\br')\right ]\; d\br'  = 
\ep_k u_k(\br) \; , 
$$
\be
\label{2.54}
\int \left [ \om^*(\br,\br') v_k(\br') + \Dlt^*(\br,\br') u_k(\br') \right ] \; d\br' = 
- \ep_k v_k(\br)  \;  .
\ee
As a result, Hamiltonian (\ref{2.41}) reduces to the Bogolubov form
\be
\label{2.55}
 H_B = E_B + \sum_k \ep_k b_k^\dgr b_k \;  ,
\ee
in which
\be
\label{2.56}
 E_B = E_{HFB} - \sum_k \ep_k \int | v_k(\br)|^2 \; d\br \;  .
\ee

The normal and anomalous density matrices become
\be
\label{2.57}
\rho_1(\br,\br') = \sum_k \left [ \pi_k u_k(\br) u_k^*(\br') + 
(1+\pi_k) v_k^*(\br) v_k(\br')  \right ] 
\ee
and, respectively,
\be
\label{2.58}
\sgm_1(\br,\br') = \sum_k \left [ \pi_k u_k(\br) v_k^*(\br') + 
(1+\pi_k) v_k^*(\br) u_k(\br')  \right ]  \;  ,
\ee
where
\be
\label{2.59}
\pi_k \equiv \lgl b_k^\dgr b_k \rgl = \left ( e^{\bt\ep_k} -1 \right )^{-1} \; .
\ee
The density of uncondensed particles, being the diagonal element of (\ref{2.57}), is
\be
\label{2.60}
 \rho_1(\br) = \sum_k \left [ \pi_k | u_k(\br) |^2 + (1+\pi_k) | v_k(\br)|^2
\right ] \;  .
\ee

\subsection{Uniform System}

In the absence of external potentials, the system is uniform, and the field operator
can be expanded over plane waves,
\be
\label{2.61}
\psi_1(\br) = \frac{1}{\sqrt{V}} \; \sum_{k\neq 0} a_k e^{i\bk\cdot\br} \; ,
\ee
with $V$ being the system volume (or quantization volume). Assuming that the
interaction potential can be Fourier transformed, one has
\be
\label{2.62}
 \Phi(\br) = \frac{1}{V} \; \sum_k \Phi_k e^{i\bk\cdot\br} \; ,  \qquad
  \Phi_k = \int \Phi(\br) e^{-i\bk\cdot\br}\; d\br \; .
\ee
The long-wave limit of the Fourier transform, defined as
$$
 \lim_{k\ra 0} \Phi_k = \int \lim_{k\ra 0} \Phi(\br) e^{-i\bk\cdot\br}\; d\br \;  ,
$$
gives the limiting value
\be
\label{2.63}
\Phi_0 =  \lim_{k\ra 0} \Phi_k = \int \Phi(\br) \; d\br \; .
\ee
Conditions, imposed on the interaction potential, allowing for the existence of
its Fourier transform are discussed in \cite{Yukalov_43}.

The single-particle density matrix becomes
\be
\label{2.64}
 \rho_1(\br,\br') = \frac{1}{V} \sum_{k\neq 0} n_k  e^{i\bk\cdot(\br-\br')}\; ,
\ee
where
\be
\label{2.65}
n_k  = \lgl a_k^\dgr a_k \rgl
\ee
is the momentum distribution of particles. The anomalous matrix reads as
\be
\label{2.66}
 \sgm_1(\br,\br') = \frac{1}{V} \sum_{k\neq 0} \sgm_k  e^{i\bk\cdot(\br-\br')}\;  ,
\ee
with the anomalous average
\be
\label{2.67}
 \sgm_k  = \lgl a_k a_{-k} \rgl \; .
\ee
Recall that the absolute value of the diagonal element $|\sigma_1({\bf r})|$
describes the density of pair-correlated particles. Hence $|\sigma_k|$ has the
meaning of the momentum distribution of pair-correlated particles \cite{Yukalov_29}.

For a uniform system, the diagonal elements of the single-particle density matrix,
\be
\label{2.68}
 \rho_1 \equiv \rho_1(\br,\br) = \frac{1}{V} \sum_{k\neq 0} n_k \;   ,
\ee
and of the anomalous matrix,
\be
\label{2.69}
 \sgm_1 \equiv \sgm_1(\br,\br) = \frac{1}{V} \sum_{k\neq 0} \sgm_k \; ,
\ee
do not depend on spatial variables. The total density is a constant
\be
\label{2.70}
 \rho = \rho_0 + \rho_1 \qquad ( \rho_0 \equiv | \eta|^2 ) \;  .
\ee

Passing to Fourier transforms in the Bogolubov equations (\ref{2.54}), we get
\be
\label{2.71}
\om_k = \frac{k^2}{2m} + \rho\Phi_0 + \rho_0 \Phi_k + 
\frac{1}{V} \sum_{p\neq 0} n_p \Phi_{k+p} - \mu_1 
\ee
and
\be
\label{2.72}
\Dlt_k = \rho_0 \Phi_k + \frac{1}{V} \sum_{p\neq 0} \sgm_p \Phi_{k+p} \;  .
\ee
This yields the spectrum of collective excitations
\be
\label{2.73}
 \ep_k = \sqrt{\om_k^2 - \Dlt_k^2} \;  .
\ee

The momentum distribution (\ref{2.65}) becomes
\be
\label{2.74}
n_k = \frac{\om_k}{2\ep_k} \; \coth\left ( \frac{\ep_k}{2T}\right ) - \; \frac{1}{2} \; ,
\ee
and the anomalous average (\ref{2.67}) is
\be
\label{2.75}
 \sgm_k = -\;\frac{\Dlt_k}{2\ep_k} \; \coth\left ( \frac{\ep_k}{2T}\right )  \;  .
\ee
Here $T$ is temperature. Notice that $n_k$ and $\sigma_k$ are connected by the
relation
$$
 \sgm_k = -\;\frac{\Dlt_k}{2\om_k} \; (1 + 2n_k) \;  .
$$

The condensate chemical potential (\ref{2.47}) becomes
\be
\label{2.76}
 \mu_0 = \rho \Phi_0 + \frac{1}{V} \sum_{k\neq 0} ( n_k + \sgm_k) \Phi_k \;  .
\ee

The condition for Bose condensate existence \cite{Yukalov_17,Yukalov_18,Yukalov_29}
tells us that in thermodynamic limit
$$
 \lim_{k\ra 0} \; \frac{1}{n_k} ~ \propto ~ \frac{1}{N} \ra 0 \qquad 
( n_k \geq 0 \; , \;\; N\ra \infty) \;  .
$$
This implies the condition
\be
\label{2.77}
  \lim_{k\ra 0}\ep_k = 0 \qquad ( \ep_k \geq 0 ) \;   .
\ee
In particular, from (\ref{2.74}) it is seen that
$$
 n_k \simeq \frac{T\om_0}{\ep_k^2} \qquad ( \ep_k \ra 0 ) \;  ,
$$
where
$$
 \om_0 = \rho_0 \Phi_0 + \frac{1}{V} \sum_{p\neq 0} \sgm_p \Phi_p \;  .
$$
Therefore
$$
 \lim_{k\ra 0} \ep_k ~ \propto ~ \frac{1}{\sqrt{N} } \ra 0 \qquad (  N\ra \infty) \;   .
$$
A gapless spectrum of collective excitations in a system with global gauge
symmetry breaking, agreeing with condition (\ref{2.77}), is also required by the
Hugenholtz-Pines \cite{Hugenholtz_44} and Bogolubov \cite{Bogolubov_31}
theorems. Equating $\varepsilon_k$ to zero at $k = 0$ results in the expression
\be
\label{2.78}
\mu_1 = \rho \Phi_0 + \frac{1}{V} \sum_{k\neq 0} ( n_k - \sgm_k) \Phi_k \;   .
\ee
Then equation (\ref{2.71}) reduces to
\be
\label{2.79}
 \om_k = \frac{k^2}{2m} + \rho_0 \Phi_k + \frac{1}{V} \sum_{p\neq 0}
(n_p \Phi_{k+p} - n_p \Phi_p + \sgm_p \Phi_p ) \;  .
\ee
The spectrum of collective excitations (\ref{2.73}) is defined by the equation
$$
\ep_k^2 = \left [ \frac{k^2}{2m} + \frac{1}{V} \sum_{p\neq 0} ( n_p - \sgm_p)
(\Phi_{k+p} - \Phi_p ) \right ] \times
$$
\be
\label{2.80}
 \times
\left \{ \frac{k^2}{2m} + 2\rho_0 \Phi_k + \frac{1}{V} \sum_{p\neq 0} 
\left [ (n_p + \sgm_p)\Phi_{k+p} - (n_p - \sgm_p)\Phi_p \right ] \right \} \;  .
\ee

It is important to stress that the anomalous average $\sigma_k$ is either zero or
not together with the condensate density. Thus if $\sigma_k \ra 0$, then, according
to (\ref{2.75}), we have $\Delta_k \ra 0$. Then (\ref{2.72}) gives $\rho_0 \ra 0$. That
is, the assumption that the anomalous average is zero is equivalent to stating that
there is no condensate in the system,
\be
\label{2.81}
  \rho_0 = 0 \qquad ( \sgm_k = 0 ) \; .
\ee

To simplify the above formulas, it is possible to make the approximation that can be
called the central-peak approximation \cite{Yukalov_43}. Taking into account that
$n_p$ and $\sigma_p$ are maximal at $p = 0$, we may write
\be
\label{2.82}
  \sum_p n_p \Phi_{k+p} ~ \cong ~ \Phi_k \sum_p n_p \; , \qquad
 \sum_p \sgm_p \Phi_{k+p} ~ \cong ~ \Phi_k \sum_p \sgm_p \; .
\ee
Of course, this approximation is sensible provided that it does not result in unphysical
divergences that cannot be regularized.

Then the chemical potentials (\ref{2.76}) and (\ref{2.78}) reduce to the form
\be
\label{2.83}
 \mu_0 = (\rho + \rho_1 + \sgm_1 ) \Phi_0 \; , \qquad 
 \mu_1 = (\rho + \rho_1 - \sgm_1 ) \Phi_0 \;  .
\ee
Equations (\ref{2.79}) and (\ref{2.72}) become
\be
\label{2.84}
 \om_k = \frac{k^2}{2m} + \rho \Phi_k - (\rho_1 - \sgm_1) \Phi_0 \; , \qquad
\Dlt_k =  (\rho_0 + \sgm_1) \Phi_k \;  .
\ee
And for the spectrum of collective excitations in (\ref{2.80}), we obtain
\be
\label{2.85}
\ep_k^2 = \left [ \frac{k^2}{2m} + (\rho_1 - \sgm_1) (\Phi_k - \Phi_0 ) \right ] 
\left \{ \frac{k^2}{2m} + (\rho + \rho_0 + \sgm_1)\Phi_k - 
(\rho_1 - \sgm_1) \Phi_0 \right\} \;   .
\ee

\subsection{Thermodynamic characteristics}

The grand thermodynamic potential
\be
\label{2.86}
\Om = - T \ln {\rm Tr}\; e^{-\bt H} \qquad ( \bt T = 1 )
\ee
defines the pressure
\be
\label{2.87}
 p = - \frac{\Om}{V} = \frac{T}{V} \; \ln {\rm Tr}\; e^{-\bt H} \;  .
\ee
The latter equation is termed the equation of state.

In the HFB approximation, one has
\be
\label{2.88}
 \Om = E_B + T \sum_k \ln \left ( 1 - e^{-\bt\ep_k} \right ) \;  ,
\ee
where
$$
E_B = -\; \frac{1}{2} \; N\rho \Phi_0 - \rho_0 \sum_p ( n_p + \sgm_p) \Phi_p \; -
$$
\be
\label{2.89}
 - \; 
\frac{1}{2V} \sum_{kp} ( n_k n_p + \sgm_k \sgm_p ) \Phi_{k+p} + 
\frac{1}{2} \sum_k ( \ep_k - \om_k) \;  .
\ee
Employing the central-peak approximation (\ref{2.82}) gives
\be
\label{2.90}
 E_B = - \; \frac{1}{2} \; V\Phi_0 \left [ \rho^2 + 2\rho_0 (\rho_1 + \sgm_1) +
\rho_1^2 + \sgm_1^2 \right ] + \frac{1}{2} \sum_k (\ep_k - \om_k) \;  .
\ee

The average of the grand Hamiltonian
\be
\label{2.91}
\lgl H \rgl = \lgl \hat H \rgl - \mu_0 N_0 - \mu_1 N_1
\ee
can also be written as
\be
\label{2.92}
 \lgl H \rgl = \lgl \hat H \rgl - \mu N \;  ,
\ee
which defines the system chemical potential
\be
\label{2.93}
 \mu = \mu_0 n_0 + \mu_1 n_1 \;  ,
\ee
in which the condensate fraction and the fraction of uncondensed particles are
\be
\label{2.94}
 n_0 \equiv \frac{N_0}{N} \; , \qquad n_1 \equiv \frac{N_1}{N} \;  .
\ee

In the HFB approximation, the chemical potential reads as
\be
\label{2.95}
 \mu = \rho \Phi_0 + \frac{1}{V} 
\sum_{k\neq 0} [ n_k + ( n_0 - n_1 ) \sgm_k ] \Phi_k \;  .
\ee
And the central-peak approximation (\ref{2.82}) gives
$$
 \mu = \rho \Phi_0 [ 1 + n_1 ( 1 - \sgm ) + n_0\sgm ] \;  ,
$$
where $\sigma \equiv \sigma_1/ \rho$.

\subsection{Local-density approximation}

When the external potential varies slowly in space (for details see reviews
\cite{Yukalov_18,Yukalov_29}), one can extend the use of the formulas for the
uniform matter by assuming that the dependence on the spatial variable comes
from the slowly varying local density. Then the local density of uncondensed
particles reads as
\be
\label{2.96}
\rho_1(\br) = \frac{1}{V} \sum_{k\neq 0} n_k(\br)
\ee
and the local anomalous average is
\be
\label{2.97}
\sgm_1(\br) = \frac{1}{V} \sum_{k\neq 0} \sgm_k(\br) \; .
\ee
In the HFB approximation, one has
\be
\label{2.98}
n_k(\br) = \frac{\om_k(\br)}{2\ep_k(\br)} \; 
\coth \left [ \frac{\ep_k(\br)}{2T} \right ] - \; \frac{1}{2}
\ee
and
\be
\label{2.99}
 \sgm_k(\br) = -\; \frac{\Dlt_k(\br)}{2\ep_k(\br)} \; 
\coth \left [ \frac{\ep_k(\br)}{2T} \right ] \;  ,
\ee
with the notations
\be
\label{2.100}
\om_k(\br) = \frac{k^2}{2m} + U(\br) + \rho(\br) \Phi_0 + \rho_0(\br) \Phi_k
+ \frac{1}{V} \sum_{p\neq 0} n_p(\br) \Phi_{k+p} \; - \; \mu_1(\br)
\ee
and
\be
\label{2.101}
 \Dlt_k(\br) = \rho_0(\br) \Phi_k + 
\frac{1}{V} \sum_{p\neq 0} \sgm_p(\br) \Phi_{k+p} \;  .
\ee
The local spectrum of collective excitations is
\be
\label{2.102}
 \ep_k(\br) = \sqrt{\om_k^2(\br) - \Dlt_k^2(\br) } \;  .
\ee

The condition of condensate existence, or the requirement of the gapless spectrum,
now takes the local form
\be
\label{2.103}
 \lim_{k\ra 0} \ep_k(\br) = 0 \; , \qquad \ep_k(\br) \geq 0 \;  .
\ee
This gives
\be
\label{2.104}
 \mu_1(\br) = U(\br) + \rho(\br) \Phi_0 + 
\frac{1}{V} \sum_{k\neq 0} [ n_k(\br) - \sgm_k(\br) ] \Phi_k \;  ,
\ee
which reduces equation (\ref{2.100}) to
\be
\label{2.105}
  \om_k(\br) = \frac{k^2}{2m} + \rho_0(\br) \Phi_k + 
\frac{1}{V} \sum_{p\neq 0} [ n_p(\br) ( \Phi_{k+p} - \Phi_p ) + \sgm_p(\br) \Phi_p ] \; .
\ee
Then spectrum (\ref{2.102}) is defined by the equation
$$
\ep_k^2(\br) = \left \{ \frac{k^2}{2m} + 
\frac{1}{V} \sum_{p\neq 0} [ n_p(\br) - \sgm_p(\br)] ( \Phi_{k+p} - \Phi_p ) \right \}
\times
$$
\be
\label{2.106}
 \times
\left \{ \frac{k^2}{2m} + 2\rho_0(\br) \Phi_k + 
\frac{1}{V} \sum_{p\neq 0} [ n_p(\br) + \sgm_p(\br)] \Phi_{k+p} -
[ n_k(\br) - \sgm_p(\br) ] \Phi_p \right \} \; .
\ee

In the central-peak approximation (\ref{2.82}), the chemical potential (\ref{2.104}) is
\be
\label{2.107}
 \mu_1(\br) = U(\br) + [ \rho(\br) + \rho_1(\br) - \sgm_1(\br) ] \Phi_0 \;  .
\ee
Expression (\ref{2.105}) transforms into
\be
\label{2.108}
 \om_k(\br) = \frac{k^2}{2m} + \rho(\br)\Phi_k - [\rho_1(\br) - \sgm_1(\br) ] \Phi_0 \;  .
\ee
And equation (\ref{2.101}) becomes
\be
\label{2.109}
 \Dlt_k(\br) = [ \rho_0(\br) + \sgm_1(\br) ] \Phi_k \;  .
\ee
Then the local spectrum (\ref{2.106}) is defined by the equation
$$
\ep_k^2(\br) = \left \{ \frac{k^2}{2m} +  
[\rho_1(\br) - \sgm_1(\br) ](\Phi_k - \Phi_0) \right \} \times
$$
\be
\label{2.110}
 \times 
\left \{ \frac{k^2}{2m} +  [\rho(\br) +\rho_0(\br) + \sgm_1(\br) ]\Phi_k - 
[\rho_1(\br) - \sgm_1(\br) ]\Phi_0 \right \} \; .
\ee

Summation over momentum, as usual, can be replaced by integrals. So, the density
of uncondensed particles (\ref{2.96}) can be written as
\be
\label{2.111}
\rho_1(\br) = \int n_k(\br) \; \frac{d\bk}{(2\pi)^3}
\ee
and the anomalous average (\ref{2.97}) as
\be
\label{2.112}
 \sgm_1(\br) = \int \sgm_k(\br) \; \frac{d\bk}{(2\pi)^3} \; .
\ee

The Fourier transform of the interaction potential is invariant with respect
to the momentum inversion, $\Phi_{-k} = \Phi_k$. But the interactions can be
anisotropic. So the long-wave behavior of $\Phi_k$, generally, looks like
\be
\label{2.113}
 \Phi_k \simeq \Phi_0 + \frac{1}{2} \; A_k k^2 \qquad ( k \ra 0 ) \;  ,
\ee
where $A_k$ is caused by the anisotropy of the interactions and is not expandable
in powers of $k$. Then the long-wave limit of spectrum, given by (\ref{2.110}), reads 
as
$$
\ep_k^2(\br) ~ \simeq ~ c_k^2(\br) k^2 \; +
$$
\be
\label{2.114}
 + \;
\left \{ 1 + [ \rho_1(\br) -\sgm_1(\br) ] m A_k \right \}
\left \{ 1 + [ \rho(\br) + \rho_0(\br) + \sgm_1(\br) ] m A_k \right \} 
\left ( \frac{k^2}{2m} \right )^2 \; ,
\ee
with the sound velocity defined by the equation
\be
\label{2.115}
 c_k^2(\br) = \frac{\Phi_0}{m} \; 
[ \rho_0(\br) + \sgm_1(\br) ] \left \{ 1 + [\rho_1(\br) - 
\sgm_1(\br) ] A_k \right \} \;  .
\ee

\subsection{Grand potential}

In the local-density approximation, the grand thermodynamic potential
\be
\label{2.116}
\Om = - \int p(\br) \; d\br 
\ee
is the integral over the local pressure
\be
\label{2.117}
 p(\br) = - E_B(\br) - T \int \ln [ 1 - \exp\{-\bt \ep_k(\br) \} ] \;
\frac{d\bk}{(2\pi)^3} \;  ,
\ee
where
$$
 E_B(\br) = - \; \frac{1}{2} \; \Phi_0 \left \{ \rho^2(\br) + 
2\rho_0(\br) [ \rho_1(\br) + \sgm_1(\br) ] + \rho^2_1(\br) + \sgm^2_1(\br)
\right \} \; + 
$$
\be
\label{2.118}
  + \; \frac{1}{2} \int [ \ep_k(\br) - \om_k(\br) ] \; 
\frac{d\bk}{(2\pi)^3} \;  .
\ee
With the notation
\be
\label{2.119}
 E_B = \int E_B(\br) \; d\br \;  ,
\ee
potential (\ref{2.116}) takes the form
\be
\label{2.120}
\Om = E_B + T \int \ln [ 1 - \exp\{-\bt \ep_k(\br) \} ] \;
\frac{d\bk}{(2\pi)^3}\; d\br  \;  .
\ee

The system internal energy is
\be
\label{2.121}
  E = \lgl \hat H \rgl = \lgl H \rgl + \mu N \; .
\ee
The average of the grand Hamiltonian reads as
\be
\label{2.122}
 \lgl H \rgl = E_B + \int \ep_k(\br) \pi_k(\br) \; 
\frac{d\bk}{(2\pi)^3} \; d\br \; ,
\ee
where
\be
\label{2.123}
 \pi_k(\br) = [ \exp\{\bt \ep_k(\br) \} - 1 ]^{-1} \; .
\ee

The system chemical potential becomes
\be
\label{2.124}
 \mu = \mu_0 n_0 + \frac{1}{N} \int \mu_1(\br) \rho_1(\br) \; d\br \;  .
\ee
The fractions of condensed and uncondensed particles are
\be
\label{2.125}
n_0 = \frac{1}{N} \int \rho_0(\br) \; d\br \; , \qquad
n_1 = \frac{1}{N} \int \rho_1(\br) \; d\br \;   ,
\ee
respectively.

\subsection{Bogolubov approximation}

At very low temperature and asymptotically weak interactions, almost all particles
are Bose-condensed. In that case, the contribution of uncondensed particles
can be neglected, which means that $\rho_1$ and $\sigma_1$ are negligibly small,
as compared to the density of condensed particles $\rho_0 \sim \rho$. Neglecting
in spectrum (\ref{2.106}) the terms containing $n_p$ and $\sigma_p$ yields the
Bogolubov form of the spectrum
\be
\label{2.126}
\ep_k(\br) = \left [ \rho(\br) \; \frac{\Phi_0}{m} \; k^2 +  \left (
\frac{k^2}{2m} \right )^2 \right ]^{1/2} \;   .
\ee
The spectrum long-wave limit is described by the equation
\be
\label{2.127}
 \ep^2_k(\br) \simeq c^2(\br) k^2 + [ 1 + 2\rho(\br) m A_k ] 
\left ( \frac{k^2}{2m} \right )^2 \; ,
\ee
with the sound velocity
\be
\label{2.128}
 c(\br) = \sqrt{ \rho(\br) \; \frac{\Phi_0}{m} } \;  .
\ee

If the interaction potential is anisotropic, then, in the Bogolubov approximation,
spectrum (\ref{2.126}) is also anisotropic. However the sound velocity (\ref{2.128})
is isotropic.

Neglecting $\rho_1$ and $\sigma_1$ in the condensate-function equation (\ref{2.45})
results in the vacuum coherent-field equation (\ref{2.40}). Hence, the latter can be
interpreted as the Bogolubov approximation of the condensate-function equation.

In the Thomas-Fermi approximation, where the kinetic energy is neglected, the
condensate-function equation (\ref{2.40}) becomes
$$
 \left [ U(\br) + \int \Phi(\br-\br') \rho_0(\br')\; d\br' \right ] \eta(\br) =
\mu_0 \eta(\br) \;  .
$$
For nonzero $\eta$, this reduces to the equality
\be
\label{2.129}
 \int \Phi(\br-\br') \rho_0(\br')\; d\br'  = \mu_0 - U(\br) \;  .
\ee
The latter, together with the normalization
$$
\int  \rho_0(\br)\; d\br  = N_0 \; , \qquad  \rho_0(\br) \equiv | \eta(\br) |^2 \;  ,
$$
defines the condensate density and the chemical potential.

If the Thomas-Fermi approximation is not involved, then the chemical potential can
be represented as
\be
\label{2.130}
 \mu_0 = \frac{1}{N} \int \eta^*(\br) \left [ - \; \frac{\nabla^2}{2m} + 
U(\br) \right ] \eta(\br) \; d\br + 
\int \Phi(\br-\br') \rho_0(\br) \rho_0(\br')\; d\br d\br' \;  .
\ee

\subsection{Stability conditions}

As follows from equation (\ref{2.128}), the sound velocity is defined, provided
the interaction potential is effectively repulsive, such that
\be
\label{2.131}
 \Phi_0 \equiv \int \Phi(\br) \; d\br > 0 \;  .
\ee
This requirement is valid for uniform systems and large traps with a slowly varying
trapping potential, when the local-density approximation is applicable. Generally, finite
traps can confine atoms with attractive interactions, if the number of atoms is limited
by a critical number \cite{Pethick_3,Pitaevskii_4,Courteille_5,Yukalov_29}.

The interaction potential is to be integrable, so that
\be
\label{2.132}
 \left  | \int \Phi(\br) \; d\br \right | < \infty \;  .
\ee
If the interaction potential of bare particles is not integrable, it is necessary to take
account of particle correlations and use an integrable pseudopotential \cite{Yukalov_45}.

The general condition of thermodynamic stability requires that the fluctuations of
extensive observable quantities be {\it thermodynamically normal}
\cite{Yukalov_7,Yukalov_17,Yukalov_18,Yukalov_28,Yukalov_29}. This implies that,
if $\hat{A}$ is the operator of an extensive observable, then its variance has to satisfy
the condition
\be
\label{2.133}
  0 \leq \frac{{\rm var}(\hat A)}{N} < \infty \; ,
\ee
in which
$$
 {\rm var}(\hat A) \equiv \lgl \hat A^2 \rgl - \lgl \hat A \rgl^2 \;  .
$$
For instance, particle fluctuations are to be thermodynamically normal,
\be
\label{2.134}
  0 \leq \frac{{\rm var}(\hat N)}{N} < \infty \; ,
\ee
where
$$
\hat N =  \int \hat\psi^\dgr(\br) \hat\psi(\br) \; d\br \;  .
$$

The variance of the number-of-particle operator, can be represented in the form
\be
\label{2.135}
 {\rm var}(\hat N) = N + \int \rho(\br) \rho(\br') 
[ g(\br,\br') -1 ]  \; d\br d\br' \; ,
\ee
with the pair correlation function
\be
\label{2.136}
 g(\br,\br') \equiv 
\frac{\lgl \hat\psi^\dgr(\br)\hat\psi^\dgr(\br')\hat\psi(\br')\hat\psi(\br)\rgl}
{\rho(\br)\rho(\br')} \; ,
\ee
in which
$$
 \rho(\br) = \rho_0(\br) + \rho_1(\br) \; .
$$

In the HFB approximation, the Hamiltonian terms of fourth order with respect to the
field operators $\psi_1$, are reduced to the terms of second order with respect to
these operators. In that sense, the HFB approximation can be called an approximation
of second order. Therefore the expressions of higher orders may be not well defined
in this approximation and can be omitted. Thus, omitting the terms of fourth order,
with respect to $\psi_1$, in the pair correlation function, we obtain
\be
\label{2.137}
 \frac{{\rm var}(\hat N)}{N}  = 1 + \frac{2}{N} \int \rho(\br) \lim_{k\ra 0}
[ n_k(\br) + \sgm_k(\br) ] \; d\br \;  .
\ee
The details can be found in \cite{Yukalov_17,Yukalov_18,Yukalov_29}.

According to (\ref{2.102}),
\be
\label{2.138}
 \om_k^2(\br) = \Dlt_k^2(\br) + \ep_k^2(\br) \;  .
\ee
We also know that in the long-wave limit $k \ra 0$ the spectrum is gapless, with
$\varepsilon_k \ra 0$. Then for small $\varepsilon_k$, such that
\be
\label{2.139}
 \ep_k(\br) \ll | \Dlt_k(\br) | \; ,
\ee
equation (\ref{2.98}) gives
\be
\label{2.140}
 n_k(\br) \simeq \frac{T\Dlt_k(\br)}{\ep_k^2(\br)} \; + \; \frac{\Dlt_k(\br)}{12T} \;
+ \; \frac{T}{2\Dlt_k(\br)} \; - \; \frac{1}{2} \; ,
\ee
while equation (\ref{2.99}) yields
\be
\label{2.141}
\sgm_k(\br) \simeq -\;\frac{T\Dlt_k(\br)}{\ep_k^2(\br)} \; - \; 
\frac{\Dlt_k(\br)}{12T} \; .
\ee
From here, we find
\be
\label{2.142}
 \lim_{k\ra 0}\; [ n_k(\br) + \sgm_k(\br) ] = 
\frac{1}{2} \left [ \frac{T}{\Dlt(\br)} \; - \; 1 \right ] \; ,
\ee
where
\be
\label{2.143}
 \Dlt(\br) \equiv \lim_{k\ra 0} \Dlt_k(\br) = \rho_0(\br) \Phi_0 + 
\int \sgm_p(\br) \Phi_p \; \frac{d\bp}{(2\pi)^3} \;  .
\ee
In the central-peak approximation (\ref{2.82}), this simplifies to
\be
\label{2.144}
\Dlt(\br) =  [ \rho_0(\br) + \sgm_1(\br) ] \Phi_0 \; .
\ee

In this way, we obtain
\be
\label{2.145}
\frac{ {\rm var}(\hat N)}{N} = \frac{T}{N} 
\int \frac{\rho(\br)}{\Dlt(\br)} \; d\br \; .
\ee
To be non-negative and finite, this expression requires that $\Phi_0$ be
nonzero and positive. It is also important to notice that the anomalous average
cannot be omitted, otherwise the integral (\ref{2.137}) becomes divergent.

\subsection{Superfluid density}

Superfluidity is defined as dissipationless flow of a fluid. The fluid motion with
velocity ${\bf v}$ can be initiated by the velocity boost
\be
\label{2.146}
 B_v =  e^{im\bv\cdot\br} \;  .
\ee
The field operator for a moving system is
$$
 \hat\psi_v(\br) = B_v \hat\psi(\br) \; .
$$
The momentum density operator for a moving system becomes
$$
\hat\bP_v(\br) = \hat\psi^\dgr_v(\br) \hat\bp_v(\br) \hat\psi_v(\br) = 
 \hat\bP(\br) + m \bv \hat\psi^\dgr(\br) \hat\psi(\br) \; ,
$$
where
$$
\hat\bP(\br) =  \hat\psi^\dgr (\br) \hat\bp(\br) \hat\psi(\br)\; , \qquad
\hat\bp(\br) \equiv - i \nabla \; .
$$
The momentum operator of the whole moving system is
$$
 \hat\bP_v = \int \hat\psi^\dgr_v(\br) \hat\bp(\br) \hat\psi_v(\br)\; d\br =
 \hat\bP + m \bv \hat N \; ,
$$
with
$$
 \hat\bP = \int \hat\bP(\br)\; d \br \; .
$$
The system kinetic energy reads as
$$
 \hat K_v = 
\int \hat\psi^\dgr_v(\br)\; \frac{\hat\bp^2}{2m}\; \hat\psi_v(\br)\; d\br =
\int \hat\psi^\dgr(\br)\; \frac{(\hat\bp+m\bv)^2}{2m}\; \hat\psi(\br)\; d\br \; .
$$
The energy Hamiltonian becomes
$$
 \hat H_v = \hat H[\hat\psi_v] = \hat H + \int \hat\psi^\dgr(\br) 
\left ( \bv \cdot \hat\bp  + \frac{mv^2}{2} \right ) \hat\psi(\br) \; d\br \; .
$$
Statistical averages of operators $\hat{A}_v$ are given by the formula
$$
\lgl \hat A_v \rgl_v \equiv {\rm Tr}\hat\rho_v \hat A_v \; , \qquad
\hat\rho_v \equiv \frac{\exp(-\bt\hat H_v)}{{\rm Tr}\exp(-\bt\hat H_v)} \;  .
$$

The superfluid fraction is defined as
\be
\label{2.147}
 n_s \equiv \lim_{v\ra 0} \; \frac{\frac{\prt}{\prt\bv}\cdot\lgl\hat\bP_v\rgl_v}
{\lgl \frac{\prt}{\prt\bv}\cdot\hat\bP_v\rgl_v} \; .
\ee
This leads to the expression
\be
\label{2.148}
  n_s = 1 - \; \frac{2Q}{Td} \;  ,
\ee
where $d$ is the space dimensionality and the dissipated heat is
$$
 Q  \equiv \frac{{\rm var}(\hat\bP)}{2mN} \; .
$$
In equilibrium, when $v = 0$, one has
$$
 {\rm var}(\hat\bP) = \lgl \hat\bP^2 \rgl \qquad ( \lgl \hat\bP \rgl = 0 ) \; .
$$

Similarly it is straightforward to characterize local superfluidity by the local quantity
\be
\label{2.149}
n_s(\br) = \lim_{v\ra 0} \;
\frac{\frac{\prt}{\prt\bv}\cdot\lgl\hat\bP_v(\br)\rgl_v}
{\lgl \frac{\prt}{\prt\bv}\cdot\hat\bP_v(\br)\rgl_v} \;
\ee
defining the local superfluid density
$$
 \rho_s(\br) = n_s(\br) \rho(\br) \;  .
$$
Then the latter takes the form
\be
\label{2.150}
 \rho_s(\br) = \rho(\br) - \; \frac{2Q(\br)}{Td}  \;  ,
\ee
with the dissipated heat density
$$
 Q(\br) = \frac{1}{2m} \; {\rm cov}(\hat\bP(\br),\hat\bP ) \;  ,
$$
in which
$$
 {\rm cov}(\hat A,\hat B) \equiv \frac{1}{2}\lgl \hat A\hat B + \hat B\hat A\rgl
- \lgl \hat A \rgl \lgl \hat B \rgl \; .
$$

In three-dimensional space, employing the HFB and local-density approximations,
we have
$$
 Q(\br) = \int \frac{k^2}{2m} \; \left [ n_k(\br) + n_k^2(\br) - \sgm_k^2(\br)
\right ] \; \frac{\bk}{(2\pi)^3} \; .
$$
Using the equality
$$
 n_k(\br) + n_k^2(\br) - \sgm_k^2(\br) = \frac{1}{4\sinh^2[\bt\ep_k(\br)/2]} \;  ,
$$
this can be rewritten as
$$
Q(\br) = \frac{1}{8m} 
\int \frac{k^2}{\sinh^2[\bt\ep_k(\br)/2]} \; \frac{\bk}{(2\pi)^3} \; .
$$

The local quantities are connected to the global ones by the relations
$$
 n_s = \frac{1}{N} \int \rho_s(\br) \; d\br \; , \qquad
Q =  \frac{1}{N} \int Q(\br) \; d\br \; .
$$

It is worth stressing that the expressions for the superfluid fraction (\ref{2.147})
and superfluid density (\ref{2.150}) are valid for any direction of the boost velocity
and for any system, isotropic or anisotropic. In all the cases, these quantities,
as is evident, are scalars.

\subsection{Optical lattices}

Optical lattices are formed by interfering laser beams creating standing waves
\cite{Morsch_11,Moseley_13,Bloch_14,Yukalov_17} or by magnetic fields
\cite{Fernholz_46}. A lattice, with the lattice vector
$$
 \ba = \left\{ a_\al = \frac{\lbd_\al}{2} \; : ~ \al = 1,2,\ldots,d \right\} \;  ,
$$
where $d$ is space dimensionality, can be created by the laser wave vector
$$
 \bk_0 = \left\{ k_0^\al = \frac{2\pi}{\lbd_\al} = \frac{\pi}{a_\al} \right\} \;  .
$$
The typical lattice potential has the form
\be
\label{2.151}
 V_L(\br) = \sum_{\al=1}^d V_\al \sin^2(k_0^\al r_\al) \;  .
\ee
The effective lattice depth is characterized by the parameter
\be
\label{2.152}
V_0 \equiv \frac{1}{d} \sum_{\al=1}^d V_\al \;   .
\ee
Depending on the balance between the lattice depth and the recoil energy
\be
\label{2.153}
  E_R \equiv \frac{k_0^2}{2m} \qquad 
\left ( k_0^2 \equiv |\bk_0 |^2 \right )  \; ,
\ee
atoms can be either localized in the vicinity of a lattice site or delocalized,
moving through the whole system. The single-atom Hamiltonian is
\be
\label{2.154}
\hat H(\br) = -\; \frac{\nabla^2}{2m} + V_L(\br) + U(\br) \;   ,
\ee
where $U({\bf r})$ is an additional, say trapping, potential superimposed on
the lattice potential.

The field operators can be expanded over Wannier functions that can be
chosen to be real and well localized \cite{Marzari_47},
\be
\label{2.155}
 \hat\psi(\br) = \sum_{nj} \hat c_{nj} w_n(\br-\ba_j) \;  ,
\ee
where $n$ is the band index and $j$ enumerates lattice sites. Using this expansion
in the system Hamiltonian, we meet the following matrix elements: the local field at
the $j$-th lattice site
\be
\label{2.156}
 h_j^{mn} \equiv \int w_m^*(\br-\ba_j) \hat H(\br) w_n(\br-\ba_j) \; d\br \; ,
\ee
the tunneling matrix
\be
\label{2.157}
J_{ij}^{mn} \equiv - \int w_m^*(\br-\ba_i) \hat H(\br) w_n(\br-\ba_j) \; d\br \; ,
\ee
with $i \neq j$, and the interaction matrix
\be
\label{2.158}
 U_{j_1j_2j_3j_4}^{n_1n_2n_3n_4} \equiv 
\int w_{n_1}^*(\br-\ba_{j_1}) w_{n_2}^*(\br'-\ba_{j_2}) \Phi(\br-\br') 
w_{n_3}(\br'-\ba_{j_3}) w_{n_4}(\br-\ba_{j_4}) \; d\br d\br' \; .
\ee
Then the energy Hamiltonian takes the form
$$
\hat H = - \sum_{i\neq j} \sum_{mn}  J_{ij}^{mn} \hat c^\dgr_{mi} \hat c_{nj} +
\sum_{j} \sum_{mn} h_j^{mn} \hat c^\dgr_{mj} \hat c_{nj} \; +
$$
\be
\label{2.159}
   + \;
\frac{1}{2} \sum_{\{j\} } \sum_{\{n\} } U_{j_1j_2j_3j_4}^{n_1n_2n_3n_4} 
\hat c^\dgr_{n_1j_1} \hat c^\dgr_{n_2j_2} \hat c_{n_3j_3} \hat c_{n_4j_4}  \; .
\ee

One usually considers the single-band approximation, keeping in mind the
lowest band and omitting the band index, which is equivalent to the use of
the expansion
$$
 \hat\psi(\br) = \sum_j \hat c_j w(\br-\ba_j) \;  ,
$$
instead of expansion (\ref{2.155}). The single-site interaction energy is denoted
as $U_j \equiv U_{jjjj}$. And the two-site interaction potential, including both the
direct as well as exchange interactions, is written as
$$
 U_{ij} \equiv U_{ijji} + U_{ijij} \;  .
$$
Then Hamiltonian (\ref{2.159}) simplifies to the extended Hubbard model
\be
\label{2.160}
 \hat H = - \sum_{i\neq j} J_{ij} \hat c^\dgr_i \hat c_j + \sum_j h_j \hat c^\dgr_j \hat c_j
+ \frac{1}{2} \sum_j U_j \hat c^\dgr_j \hat c^\dgr_j \hat c_j \hat c_j +
  \frac{1}{2} \sum_{i\neq j} U_{ij} \hat c^\dgr_i \hat c^\dgr_j \hat c_j \hat c_i \; .
\ee

One often considers the tunneling only between the nearest neighbors. In the case
of an ideal lattice, the values $h_j$ and $U_j$ do not depend on the lattice index.

Note that, instead of expanding the field operators over Wannier functions, one could
expand them over Bloch functions, thus deriving a different representation for the
system Hamiltonian \cite{Yukalov_17,Yukalov_48}. But the Hubbard form is more
convenient for treating localized particles.

\subsection{Condensate in lattice}

Let the lattice, loaded with $N$ atoms, have $N_L$ lattice sites, so that the filling
factor be
\be
\label{2.161}
 \nu \equiv \frac{N}{N_L} \;  .
\ee
Accomplishing the Bogolubov shift $\hat{\psi} = \eta + \psi_1$ and comparing it
with the expansion over Wannier functions (\ref{2.155}), we get the Wannier
expansion for the condensate function
\be
\label{2.162}
\eta(\br) = \sqrt{\frac{N_0}{N_L}} \; \sum_j w_0(\br-\ba_j)
\ee
and for the field operator of uncondensed atoms
\be
\label{2.163}
\psi_1(\br) = \sum_{nj} c_{nj} w_n(\br-\ba_j) \;  .
\ee
The Wannier function $w_0$ is assumed to pertain to the lowest band. In this way,
\be
\label{2.164}
 \hat c_{nj} = \sqrt{\frac{N_0}{N_L}} \; \dlt_{n0} + c_{nj} \;  .
\ee

The condition of orthogonality of $\eta$ and $\psi_1$ yields the equation
\be
\label{2.165}
 \sum_j c_{0j} = 0 \;  ,
\ee
which gives
\be
\label{2.166}
  \sum_j \hat c_{0j} = \sqrt{N_0 N_L} \; .
\ee

And condition (\ref{2.5}) implies
\be
\label{2.167}
  \lgl c_{nj} \rgl = 0 \; ,
\ee
which results in the order parameter
\be
\label{2.168}
  \lgl \hat c_{nj} \rgl = \sqrt{\frac{N_0}{N_L}} \; \dlt_{n0} \; .
\ee

In the single-band approximation, instead of (\ref{2.164}), we have
\be
\label{2.169}
 \hat c_j = \sqrt{\nu n_0} + c_j \; , \qquad 
\left ( n_0 \equiv \frac{N_0}{N} \right ) \; .
\ee
The site operators have to satisfy the conditions
$$
\frac{1}{N_L} \sum_j \hat c_j = \sqrt{\nu n_0} \; , \qquad \sum_j c_j = 0 \; ,
$$
\be
\label{2.170}
  \lgl \hat c_j \rgl = \sqrt{\nu n_0} \; , \qquad \lgl c_j \rgl = 0 \; .
\ee

\subsection{Lattice Hamiltonian}

Substituting into the grand Hamiltonian the expansion of field operators (\ref{2.155}),
we shall limit ourselves by the single-band approximation. The symmetry $U_{ij} = U_{ji}$
and $J_{ij} = J_{ji}$ will be used, which is evident for real Wannier functions.

Using the Wannier representation for the number-of-particle operators gives
\be
\label{2.171}
 N_0 =  \nu n_0 N_L \; , \qquad \hat N_1 = \sum_j c^\dgr_j c_j \; .
\ee
Operator (\ref{2.10}), removing the linear in $\psi_1$ terms, reads as
\be
\label{2.172}
 \hat\Lbd = \sum_j \left ( \lbd_j c_j^\dgr + \lbd^*_j c_j \right ) \;  ,
\ee
with
$$
\lbd_j \equiv \int \lbd(\br) w^*(\br-\ba_j) \; d\br \;   .
$$

The grand Hamiltonian (\ref{2.15}) is the sum
\be
\label{2.173}
 H = H^{(0)} +  H^{(2)} + H^{(3)} +  H^{(4)} \; ,
\ee
similar to (\ref{2.21}), where the terms are distinguished by the order with respect to
the operators $c_j$. The term, containing no such operators, is
\be
\label{2.174}
 H^{(0)} = \nu n_0 \left ( \sum_j h_j - \sum_{i\neq j} J_{ij} \right ) - \mu_0 N_0 +
\frac{(\nu n_0)^2}{2} \left ( \sum_j U_j + \sum_{i\neq j} U_{ij} \right ) \; .
\ee
The first-order term is canceled by setting
$$
 \lbd_j =  \sqrt{\nu n_0} \; \left ( \sum_j h_j - \sum_{i(\neq j)} J_{ij} \right )
+  (\nu n_0)^{3/2}  \left ( \sum_j U_j + \sum_{i(\neq j)} U_{ij} \right )\; ,
$$
which gives $H^{(1)} = 0$. The second-order term is
$$
H^{(2)} = - \sum_{i\neq j} J_{ij} c_i^\dgr c_j + 
\sum_j (h_j - \mu_1) c_j^\dgr c_j + \frac{\nu n_0}{2} 
\sum_j U_j \left ( 4 c_j^\dgr c_j + c_j^\dgr c_j^\dgr + c_j c_j\right ) \; +
$$
\be
\label{2.175}
 + \; \frac{\nu n_0}{2} 
\sum_{i\neq j} U_{ij} \left ( 2 c_i^\dgr c_j + 2c_j^\dgr c_j^\dgr + c_i^\dgr c_j^\dgr  
+ c_j c_i \right ) \; .
\ee
The third-order term reads as
\be
\label{2.176}
 H^{(3)} = \sqrt{\nu n_0} 
\sum_j U_j \left ( c_j^\dgr c_j^\dgr c_j + c_j^\dgr c_j c_j\right ) 
+ \sqrt{\nu n_0} 
\sum_{i\neq j} U_{ij} \left ( c_i^\dgr c_j^\dgr c_j + c_j^\dgr c_j c_i\right ) \;  .
\ee
And the fourth-order term becomes
\be
\label{2.177}
 H^{(4)} = \frac{1}{2} \sum_j U_j c_j^\dgr c_j^\dgr c_j c_j +
\frac{1}{2} \sum_{i\neq j} U_{ij} c_i^\dgr c_j^\dgr c_j c_i  \; .
\ee

For an ideal lattice, it is convenient to resort to the Fourier transform
$$
 c_j = \frac{1}{\sqrt{N_L}} \sum_k a_k e^{i\bk\cdot\ba_j} \; , \qquad 
 a_k = \frac{1}{\sqrt{N_L}} \sum_j c_j e^{-i\bk\cdot\ba_j} \; .
$$
In view of conditions (\ref{2.170}), one has
\be
\label{2.178}
 a_0 = \frac{1}{\sqrt{N_L}} \sum_j c_j = 0 \; .
\ee

The evolution equations for the operators $c_j$ are
$$
 i\; \frac{\prt c_j}{\prt t} = [ c_j , \; H ] = \frac{\dlt H}{\dlt c_j^\dgr} \; , 
\qquad
i\; \frac{\prt c_j^\dgr}{\prt t} = [ c_j^\dgr , \; H ] = -\; \frac{\dlt H}{\dlt c_j} \; , 
$$
$$
i\; \frac{\prt}{\prt t}\; \left ( c_i^\dgr c_j\right ) = 
c_i^\dgr \; \frac{\dlt H}{\dlt c_j^\dgr} \; - \; \frac{\dlt H}{\dlt c_j}\; c_j \; .
$$
From the latter equation, in equilibrium, when
$$
 \frac{\prt}{\prt t} \; \lgl c_i^\dgr c_j \rgl = 0 \;  ,
$$
we have the relations
$$
 \lgl c_i^\dgr c_j \rgl = \lgl c_j^\dgr c_i \rgl = \lgl c_i^\dgr c_j \rgl^* \; ,
$$
$$
\lgl c_i c_j \rgl = \lgl c_j^\dgr c_i^\dgr \rgl = \lgl c_i c_j \rgl^* \; ,
$$
$$
\lgl c_i^\dgr c_i c_j \rgl = \lgl c_j^\dgr c_i^\dgr c_i \rgl = 
\lgl c_i^\dgr c_i c_j \rgl^* \; ,
$$
where the operators $c_j$ are taken at the same moment of time. These relations
mean that in equilibrium the corresponding correlation functions are real.

In equilibrium, we also have
\be
\label{2.179}
 \left \lgl \frac{\dlt H}{\dlt n_0} \right \rgl = 0 \;  .
\ee
This leads to the expression for the condensate chemical potential
$$
\mu_0 = \frac{1}{N_L} \left ( \sum_j h_j - \sum_{i\neq j} J_{ij} \right ) + 
\frac{\nu n_0}{N_L} \left ( \sum_j U_j + \sum_{i\neq j} U_{ij} \right ) +
$$
$$
+ \frac{1}{N_L} \sum_j U_j \lgl 2 c_j^\dgr c_j + c_j^\dgr c_j^\dgr \rgl +
  \frac{1}{N_L} \sum_{i\neq j} U_{ij} \lgl c_i^\dgr c_j + c_j^\dgr c_j + 
  c_i^\dgr c_j^\dgr \rgl + 
$$
\be
\label{2.180}
+ \frac{1}{\sqrt{\nu n_0} \;N_L} 
\sum_j U_j \lgl c_j^\dgr c_j^\dgr c_j \rgl +
\frac{1}{\sqrt{\nu n_0} \;N_L} 
\sum_{i\neq j} U_{ij} \lgl c_i^\dgr c_j^\dgr c_j \rgl \; .
\ee

\subsection{Rotating systems}

The study of superfluid systems often requires to consider system rotation.
Then one can treat the system either in the laboratory frame at rest or in the
rotating frame, where the system is immovable. It is usually convenient to treat
the system in the rotating frame involving the transformation from the laboratory
frame to the noninertial rotating frame \cite{Aharonov_49,Takagi_50,Klink_51}.

Let the spatial variable in the laboratory frame at rest be $\br=\{x,y,z\}=\{r_\al\}$.
The system is assumed to rotate counter-clockwise around the axis ${\bf e}_z$.
In the rotating frame, where the system is immovable, the spatial variable is denoted
by ${\bf x} = \{x_1, x_2, x_3\} = \{x_\alpha\}$, with $\alpha = 1, 2, 3$.  And let the
rotation angle between the axes $x$ and $x_1$ be denoted by $\varphi = \varphi(t)$.
The transformation from the frame at rest to the rotating frame is realized by the
transformation matrix
\begin{eqnarray}
\label{2.181}
R(\vp) = \left [ \begin{array}{ccc}
\cos\vp & - \sin\vp & 0 \\
\sin\vp &   \cos\vp & 0 \\
0       &     0     & 1 \end{array} 
\right ]  \; ,
\end{eqnarray}
so that
\begin{eqnarray}
\nonumber
\left [ \begin{array}{c}
x_1 \\
x_2 \\
x_3 \end{array} \right ] = R(\vp) 
\left [ \begin{array}{c}
x \\
y \\
z \end{array} \right ] \; ,
\end{eqnarray}
or explicitly
$$
 x_1 = x\cos\vp - y \sin\vp \; , \qquad 
 x_2 = x\sin\vp + y \cos\vp \; , \qquad     x_3 = z \; .
$$
The inverse rotation matrix is
\begin{eqnarray}
\label{2.182}
R^{-1}(\vp) = R(-\vp) = \left [ \begin{array}{ccc}
\cos\vp  &  \sin\vp & 0 \\
-\sin\vp &  \cos\vp & 0 \\
0       &     0     & 1 \end{array} 
\right ]  \; ,
\end{eqnarray}
giving
\begin{eqnarray}
\nonumber
\left [ \begin{array}{c}
x \\
y \\
z \end{array} \right ] = R^{-1}(\vp) 
\left [ \begin{array}{c}
x_1 \\
x_2 \\
x_3 \end{array} \right ] \; ,
\end{eqnarray}
or explicitly
$$
x = x_1\cos\vp + x_2 \sin\vp \; , \qquad 
y = -x_1\sin\vp + x_2 \cos\vp \; , \qquad     z = x_3 \;   .
$$

Let the orthonormal right-handed triad in the laboratory frame be denoted
as ${\bf e}_\alpha$ and the orthonormal right-handed triad rotating with angular
velocity $\Omega = \Omega(t)$ be ${\bf e}_\alpha(t)$, with the initial condition
${\bf e}_\alpha(0) = {\bf e}_\alpha$. The angular velocity is connected with the
rotation angle by the integral
\be
\label{2.183}
 \vp(t) = \int_0^t \Om(t') \; dt' \;  .
\ee
The rotating triad  satisfies the equation
\be
\label{2.184}
 \frac{d}{dt}\; \bfe_\al(t) = {\bf\Om} \times \bfe_\al(t) \;  ,
\ee
with ${\bf \Om}=\Om\bfe_z$. Any spatial vector ${\bf r}$ can be represented either in 
the laboratory frame or in the rotating frame, respectively, as
$$
 \br = \sum_\al r_\al e_\al = \sum_\al x_\al e_\al(t) \;  .
$$
The variables in the rotating frame, spatial $x_\alpha$ and temporal $\tau$, are
connected with the variables of the laboratory frame $r_\alpha$ and $t$ by
the equations
\be
\label{2.185}
 x_\al =\br \cdot \bfe_\al(t) \; , \qquad \tau = t \;  .
\ee

The interaction energy does not depend on whether the system rotates or not.
Therefore it is illustrative to consider, first, the single-particle picture in terms of
the Schr\"{o}dinger equation. The wave function $\Psi$ in the rotating frame is
connected with the wave function $\psi$ in the laboratory frame by the relation
\be
\label{2.186}
 \Psi(x,\tau) = \psi\left (\sum_\al x_\al \bfe_\al(\tau),\tau\right ) \;  .
\ee
In the laboratory frame, the single-particle Schr\"{o}dinger equation
\be
\label{2.187}
i\; \frac{\prt}{\prt t} \; \psi(\br,t) = \hat H(\br,t) \psi(\br,t)
\ee
contains the Hamiltonian
\be
\label{2.188}
\hat H(\br,t) = \frac{\hat\bp^2}{2m} + U(\br,t) \qquad 
\left ( \hat\bp \equiv - i \; \frac{\prt}{\prt\br} \right ) \;   .
\ee
Differentiating the wave function $\Psi$, we employ the relations between the
spatial derivatives
$$
\frac{\prt}{\prt\br}  = \sum_\al \bfe_\al(\tau)\; \frac{\prt}{\prt x_\al} \; ,
\qquad
\sum_\al \frac{\prt^2}{\prt r_\al^2} = \sum_\al \frac{\prt^2}{\prt x_\al^2} \; ,
$$
and the temporal derivatives
$$
 \frac{\prt}{\prt t} =  \frac{\prt\tau}{\prt t}  \frac{\prt}{\prt\tau} + 
\frac{\prt{\bf x}}{\prt t} \cdot \frac{\prt}{\prt{\bf x}} \; , 
\qquad
\frac{\prt\tau}{\prt t} = 1 \; , 
\qquad 
\frac{\prt{\bf x}}{\prt t} \cdot \frac{\prt}{\prt{\bf x}} = 
- i {\bf\Om}\cdot \hat{\bf L}(\bx) \; ,
$$ 
with the angular momentum operator in the rotating frame
$$
 \hat{\bf L}(x) \equiv \bx \times \hat\bp' \qquad 
\left ( \hat\bp' \equiv - i \; \frac{\prt}{\prt\bx} \right ) \; .
$$
Therefore we obtain
$$
 \frac{\prt}{\prt t} =  \frac{\prt}{\prt\tau} - i{\bf\Om} \cdot \hat{\bf L}(x) \;  .
$$
Then the Schr\"{o}dinger equation in the rotating frame reads as
\be
\label{2.189}
 i\; \frac{\prt}{\prt\tau}\; \Psi(\bx,\tau) = \hat H_{eff}(\bx,\tau) \Psi(\bx,\tau) \; ,
\ee
with the effective Hamiltonian
\be
\label{2.190}
 \hat H_{eff}(\bx,\tau) = \hat H(\bx,\tau) - {\bf\Om} \cdot \hat{\bf L}(x) \;  ,
\ee
in which
\be
\label{2.191}
\hat H(\bx,\tau) = -\; \frac{1}{2m} \sum_\al \frac{\prt^2}{\prt x_\al^2} + U(\bx,\tau)
\ee
and ${\bf x} = \sum_\alpha x_\alpha {\bf e}_\alpha(t)$. Thus the effective Hamiltonian
acquires a Coriolis term.

The above consideration can be summarized by involving the unitary transformation
\be
\label{2.192}
 \hat U = \hat T \exp\left\{ - i \int_0^\tau {\bf\Om}(t') \cdot 
\hat{\bf L}(x) \; dt' \right \} \; ,
\ee
where $\hat{T}$ denotes time ordering, which transforms the wave function in the
laboratory frame into the wave function in the rotating frame
\be
\label{2.193}
  \Psi(\bx,\tau) = \hat U^{+}\psi(\br,\tau) \; .
\ee
Then, substituting $\psi = \hat{U} \Psi$ into equation (\ref{2.187}) gives
equation (\ref{2.189}), with the effective Hamiltonian
\be
\label{2.194}
 \hat H_{eff} = \hat U^{+} \hat H \hat U -
i \hat U^{+} \; \frac{\prt\hat U}{\prt t} \; .
\ee
Since
\be
\label{2.195}
\hat U^{+} \hat H(\br,t) \hat U = \hat H(\bx,\tau)
\ee
and
\be
\label{2.196}
 -i \hat U^{+} \frac{\prt\hat U}{\prt t} = - {\bf\Om} \cdot \hat{\bf L}(x) \; ,
\ee
the effective Hamiltonian takes the same form as in (\ref{2.190}).

These considerations are straightforwardly generalized to a many-particle system
with the Hamiltonian in the laboratory frame
\be
\label{2.197}
 \hat H = \int \hat\psi^\dgr(\br) \hat H(\br,t) \hat\psi(\br) \; d\br + 
\frac{1}{2} \int \hat\psi^\dgr(\br) \hat\psi^\dgr(\br') 
\Phi(\br-\br') \hat\psi(\br') \hat\psi(\br)\; d\br d\br' \; .
\ee
In the rotating frame, we obtain
\be
\label{2.198}
 \hat H_\Om = \hat H - {\bf\Om} \cdot \hat{\bf L} \;  ,
\ee
with the angular momentum operator
\be
\label{2.199}
 \hat{\bf L} = \int \hat\psi^\dgr(\br) \hat{\bf L}(\br) \hat\psi(\br) \; d\br \;   ,
\ee
where $\hat{L}({\bf r}) ={\bf r} \times \hat{\bf p}$.

Note that in the integrals entering (\ref{2.198}) and (\ref{2.199}), the variable
of integration ${\bf x}$ is replaced by ${\bf r}$.

When the rotation axis is ${\bf e}_z$, then the angular momentum $\hat{\bf L}({\bf r})$
reduces to
\be
\label{2.200}
 \hat L_z(\br) = - i \; \frac{\prt}{\prt\vp} \qquad
({\bf\Om} = \Om\bfe_z ) \;  ,
\ee
where $\varphi$ is the angle in the cylindrical system of coordinates.

\section{Dipolar interaction potentials}

\subsection{Scattering lengths}

An important class of nonlocal interactions includes dipolar interactions.
Dipolar interactions arise between atoms or molecules possessing electric, $d_0$,
or magnetic, $\mu_0$, dipoles. Similarly to local atomic interactions that are
described by a scattering length $a_s$, the strength of dipolar interactions can be
characterized by an effective dipolar scattering length, or just dipolar length
\begin{eqnarray}
\nonumber
a_D \equiv \left \{ \begin{array}{ll}
md_0^2/\hbar^2 , ~ & ~ {\rm electric \; dipoles} \\
                   &                            \\
m\mu_0^2/\hbar^2 , ~ & ~ {\rm magnetic \; dipoles}
\end{array} \right. \; .
\end{eqnarray}
The value of electric dipoles is measured in Debye units $D$,
$$
 1D = 10^{-18} \; \frac{{\rm erg}}{{\rm G}} \qquad 
\left ( 1{\rm G}^2 = 1 \frac{{\rm erg}}{{\rm cm^3}} \right ) \; ,
$$
while that of magnetic dipoles, in Bohr magnetons $\mu_B$,
$$
 \mu_B = 0.927401 \times 10^{-20} {\rm erg}/{\rm G} \; .
$$
So that the relation holds:
$$
1D = 107.8282 \mu_B \;   .
$$

The scattering length of local interactions, generally, depends on the value of the dipoles
of scattered particles, $a_s = a_s(d_0)$.  Scattering lengths are often expressed in units
of the Bohr radius $a_B = 0.529177 \times 10^{-8}$ cm.

For example, an atom of $^{87}$Rb, with mass $m = 1.443 \times 10^{-22}$ g, has the
magnetic moment $\mu_0 = 0.5 \mu_B$ and the dipolar length
$$
a_D = 2.79 \times 10^{-9} {\rm cm} = 0.527 a_B  \; .
$$
An atom of $^{52}$Cr, with $m = 0.863 \times 10^{-22}$ g, possesses the magnetic moment
$\mu_0 = 6 \mu_B$ and the dipolar length
$$
a_D = 2.4 \times 10^{-7} {\rm cm} = 45.4 a_B \; .
$$
An atom of $^{168}$Er, with $m = 2.777 \times 10^{-22}$ g, has the magnetic  moment
$\mu_0 = 7 \mu_B$ and the dipolar length
$$
a_D = 1.052 \times 10^{-6} {\rm cm} = 199 a_B \; .
$$
An atom of $^{164}$Dy, with $m = 2.698 \times 10^{-22}$ g, possesses the magnetic moment
$\mu_0 = 10 \mu_B$ and the dipolar length
$$
a_D = 2.087 \times 10^{-6} {\rm cm} = 395 a_B \; .
$$
Polar molecules and Rydberg atoms can have electric dipole moments
$d_0 \sim (0.1 - 10)$ D and the dipolar lengths $a_D \sim (10^{-4} - 10^{-2})$ cm.
More details can be found in reviews
\cite{Griesmaier_19,Baranov_20,Pupillo_21,Lahaye_22,Baranov_23,Gadway_24}.

There also exist hyperfine dipolar interactions between electrons of an atom and protons
of a nucleus, which is modeled by a local form. The effective scattering length of such
local hyperfine interactions is
$$
 a_H = \frac{m\mu_n\mu_e}{\hbar^2} \;  ,
$$
where $\mu_n = 1.411 \times 10^{-23}$ erg/G is the proton magnetic moment and
$\mu_e = 0.928 \times 10^{-20}$ erg/G is the electron magnetic moment. Thus for
$^{87}$Rb, one has $a_H = 1.699 \times 10^{-11}$ cm, which is much smaller than
$a_D = 2.79 \times 10^{-9}$ cm.

\subsection{Interaction potentials}

Generally, interactions between atoms, or molecules, contain several terms. There exists
the local, or contact, interaction
\be
\label{3.1}
 \Phi_{loc}(\br) = 4\pi \hbar^2 \; \frac{a_s}{m} \; \dlt(\br) \;  ,
\ee
described by the scattering length $a_s$ that, strictly speaking, also depends on the dipole
moment. There is the hyperfine contact interaction
\be
\label{3.2}
A(\br) = -\; \frac{8\pi}{3} \; \vec{\mu}_n \cdot \vec{\mu}_e \dlt(\br) \; ,
\ee
in which
$$
\vec{\mu}_n = \hbar \gm_n \bI \; , \qquad \vec{\mu}_e = - \hbar \gm_e \bS \; ,
$$
are the proton and electron magnetic moments, $\gamma_n$ and $\gamma_e$ are
the proton and electron giromagnetic ratios, and ${\bf I}$ and ${\bf S}$ are the proton
and electron spins, respectively. So that the hyperfine interaction can be written as
$$
 A(\br) = \frac{8\pi}{3} \; \hbar^2 \gm_n \gm_e \; \bI \cdot \bS \; \dlt(\br) \;   .
$$
The hyperfine interactions are rather small. Thus
$$
 \frac{8\pi}{3}\; | \vec{\mu}_n | | \vec{\mu}_e | = 
1.097 \times 10^{-42} \; {\rm erg}\; {\rm cm}^3  \;   .
$$
This should be compared with the factor $4 \pi \hbar^2 a_s/m$ in the interaction
potential (\ref{3.1}). For instance, the scattering length of $^{87}$Rb is
$a_s = 103 a_B = 0.545 \times 10^{-6}$ cm, which gives
$$
4\pi \hbar^2 \; \frac{a_s}{m} = 0.883 \times 10^{-37} \; {\rm erg \; cm}^3  \;  .
$$

Similarly, one can consider electric hyperfine interactions \cite{Parker_52}.

In what follows, all contact interactions are assumed to be combined and characterized
by an effective scattering length $a_s$, in general, depending on dipole moments.

One often needs to use Fourier transforms of interaction potentials. Thus for the local
interaction potential, one has
\be
\label{3.3}
 \Phi_k^{loc} \equiv \int \Phi_{loc}(\br) e^{-i\bk\cdot\br} \; d\br \;  ,
\ee
which gives
\be
\label{3.4}
\Phi_k^{loc} = 4\pi \hbar^2 \;  \frac{a_s}{m} \;   .
\ee

Note that the sufficient condition for the existence of the Fourier transform of a potential
$\Phi({\bf r})$, being a function of bounded variation, is its absolute integrability,
\be
\label{3.5}
  \int | \Phi(\br) | \; d\br < \infty \; ,
\ee
which is called the Dirichlet theorem \cite{Champeney_53}.

The dipole-dipole interaction between two dipoles is given by the potential
\be
\label{3.6}
D(\br) = \frac{1}{r^3} \left [ (\bd_1\cdot\bd_2) - 
3 (\bd_1\cdot\bn) (\bd_2\cdot\bn) \right ] \; ,
\ee
in which
$$
 \br \equiv \br_1 - \br_2 \; , \qquad \bn \equiv \frac{\br}{r} \; , \qquad
r \equiv |\br | \;  .
$$
For concreteness, we write here electric dipoles ${\bf d}$, keeping in mind that the
same formulas are valid for magnetic dipoles $\vec{\mu}$. If all dipoles are polarized
in one direction ${\bf e}_d$ and have the same absolute dipole moment, so that
$$
 \bd_i = d_0 \bfe_d \qquad (d_0 \equiv |\bd_i| ) \;  ,
$$
then the interaction potential becomes
\be
\label{3.7}
 D(\br) = \frac{d_0^2}{r^3} \; \left ( 1 - 3\cos^2\vartheta \right ) \qquad
\left ( \cos\vartheta \equiv \frac{\br\cdot\bfe_d}{r} \right ) \;  .
\ee

The Fourier transform of potential (\ref{3.6}),
\be
\label{3.8}
 D_k \equiv \int D(\br) e^{-i\bk\cdot\br}\; d\br \;  ,
\ee
reads as
\be
\label{3.9}
 D_k = \frac{4\pi}{3} \left [ \frac{3 (\bd_1\cdot\bk) (\bd_2\cdot\bk)}{k^2} \; - \;
 (\bd_1\cdot\bd_2) \right ] \;  ,
\ee
with $k \equiv |{\bf k}|$. And the Fourier transform of potential (\ref{3.7}) for polarized
dipoles is
\be
\label{3.10}
 D_k = \frac{4\pi}{3}\; d_0^2  \left ( 3\cos^2\vartheta_k -1 \right ) \qquad 
\left ( \cos\vartheta_k \equiv \frac{\bk\cdot\bfe_d}{k} \right ) \;  .
\ee

However, Fourier transforms (\ref{3.9}) and (\ref{3.10}), as is seen, are not defined
for $k\ra 0$ as well as for $k \ra \infty$. This happens because the dipole-dipole
interaction potential is not absolutely integrable, that is, does not satisfy the
Dirichlet theorem and condition (\ref{3.5}). Really, in three dimensions,
$$
 \int | D(\br) | \; d\br ~ \sim ~
d_0^2\; \lim_{R\ra\infty}\; \lim_{b\ra 0}\; \ln \frac{R}{b} \ra \infty \; .
$$
While in one and two dimensions,
$$
 \int | D(\br) | \; d\br ~ \sim  ~ d_0^2\; \lim_{b\ra 0} b^{d-3} \ra \infty \qquad
(d=1,2) \;  .
$$
This means that the Fourier transform for the dipole-dipole interaction potential is
not well defined.

\subsection{Regularized potential}

In order to have a well defined Fourier transform, the dipole-dipole interaction
potential needs to be regularized. In dimensions one and two, it is sufficient to
regularize the potential only at short distance. But in three dimensions, it is
necessary to regularize it both at short and at long-range distance.

The physical origin of such a regularization is quite clear. The dipole-dipole
interaction potential (\ref{3.6}) is written for point-like particles, while real atoms
and molecules have finite sizes. Hence form (\ref{3.6}) is meaningful only for $r$
larger than a short distance $b$ of the order of the size of the interacting particles.
And at long-range distance, there exists particle screening attenuating the bare
interaction. Therefore the regularized dipolar potential can be written as \cite{Yukalov_43}
\be
\label{3.11}
 D(\br,b,\varkappa) = \Theta(r-b) D(\br) e^{-\varkappa r} \; ,
\ee
where $\Theta(r)$ is the unit-step function and the screening parameter $\varkappa$
defines the screening radius $r_s = 1/\varkappa$.

The dipolar potential (\ref{3.11}) is absolutely integrable, hence enjoys a well defined
Fourier transform
\be
\label{3.12}
 D_k(b,\varkappa) \equiv \int D(\br,b,\varkappa) e^{-i\bk\cdot\br} \; d\br \;  .
\ee
Keeping in mind the case of polarized dipoles in three dimensions, we have
\be
\label{3.13}
 D_k(b,\varkappa) = D_k I_k(b,\varkappa) \;  ,
\ee
where $D_k$ is given by (\ref{3.10}) and
\be
\label{3.14}
 I_k(b,\varkappa) =  9kb \int_1^\infty \left [ \frac{\sin(kbx)}{(kbx)^4} \; - \;
\frac{\cos(kbx)}{(kbx)^3} \; - \; 
\frac{\sin(kbx)}{3(kbx)^2} \right ] e^{-\varkappa bx} \; dx \; .
\ee

We may notice that integral (\ref{3.14}), actually, depends on two parameters,
\be
\label{3.15}
q \equiv kb \; , \qquad \zeta \equiv \varkappa b = \frac{b}{r_s} \;   .
\ee
So that it can be rewritten as
\be
\label{3.16}
I_k(b,\varkappa) = J(q,\zeta)
\ee
through the integral
\be
\label{3.17}
 J(q,\zeta) = 9q \int_1^\infty \left [ \frac{\sin(qx)}{(qx)^4} \; - \;
\frac{\cos(qx)}{(qx)^3} \; - \; 
\frac{\sin(qx)}{3(qx)^2} \right ] e^{-\zeta x} \; dx \;  .
\ee

At small $q$, we have
\be
\label{3.18}
J(q,\zeta) ~ \simeq ~
\frac{(1+\zeta)e^{-\zeta}}{5\zeta^2}\; q^2  \qquad  (q\ra 0) \;  ,
\ee
which yields
\be
\label{3.19}
D_k(b,\varkappa) ~ \simeq ~
D_k\; \frac{(1+\varkappa b)e^{-\varkappa b}}{5\varkappa^2}\; k^2  \qquad  (k\ra 0) \;  .
\ee
Therefore the short-wave limit is well defined:
\be
\label{3.20}
 D_0(b,\varkappa) \equiv \lim_{k\ra 0} D_k(b,\varkappa) = 0 \;  .
\ee

In the opposite limit, we find
\be
\label{3.21}
 J(q,\zeta) \simeq -3e^{-\zeta}\; \frac{\cos q}{q^2} \qquad (q\ra \infty) \;  ,
\ee
which gives
\be
\label{3.22}
D_k(b,\varkappa) \simeq - 3 D_k e^{-\varkappa b} \; \frac{\cos(kb)}{(kb)^2} \qquad 
(k\ra \infty)
\ee
going to zero as $k \ra \infty$.

In the general case, the interaction potential is the sum of the contact interaction
term and the dipolar term,
\be
\label{3.23}
 \Phi(\br) =  4\pi \hbar^2 \; \frac{a_s}{m}\; \dlt(\br) + D(\br,b,\varkappa) \;  .
\ee
Then its Fourier transform writes as
\be
\label{3.24}
 \Phi_k =  4\pi \hbar^2 \; \frac{a_s}{m} + D_k(b,\varkappa) \;  .
\ee
And the long-wave limit of the Fourier transform is
\be
\label{3.25}
 \Phi_0 \equiv \lim_{k\ra 0} \Phi_k =  4\pi \hbar^2 \; \frac{a_s}{m} \;  .
\ee

It is worth stressing that the use of an effective regularized potential is well justified
and allows for the construction of a uniquely defined perturbation theory \cite{Yukalov_45}.

\subsection{Excitation spectrum}

For compactness, in this and following sections, we use the system of units with
the Boltzmann constant $k_B$ and Planck constant $\hbar$ set to unity.

The Fourier transform of the interaction potential (\ref{3.23}) can be written in the form
\be
\label{3.26}
\Phi_k = \Phi_0 + f_k \;   ,
\ee
with the notation
\be
\label{3.27}
 f_k \equiv D_k(b,\varkappa) = D_k I_k(b,\varkappa) = D_k J(q,\zeta) \;  .
\ee

The spectrum of collective excitations can be found by using the formulas of the
previous Sec. 2. Taking into account that $f_k \ra  0$ as $k \ra 0$, from (\ref{2.110})
we find
\be
\label{3.28}
 \ep_k^2 = \left [ \frac{k^2}{2m} + (\rho_1-\sgm_1) f_k \right ]
\left [ \frac{k^2}{2m} + 2(\rho_0 + \sgm_1) \Phi_0 +
(2\rho_0 + \rho_1 +\sgm_1) f_k \right ] \;  .
\ee
In the local-density approximation, the quantities $\rho_0$, $\rho_1$, and $\sigma_1$
are functions of ${\bf r}$. The long-wave limit, in view of the asymptotic form
$$
f_k \simeq \frac{1+\varkappa b}{5\varkappa^2 e^{\varkappa b}} \; D_k k^2 \qquad (k\ra 0) \;   ,
$$
results in the phonon spectrum
\be
\label{3.29}
\ep_k \simeq c_k k \qquad  (k\ra 0) \;   ,
\ee
with the anisotropic sound velocity given by the equation
\be
\label{3.30}
 c_k^2 = (\rho_0 + \sgm_1 ) \frac{\Phi_0}{m} \left [ 1 + (\rho_1-\sgm_1)\;
\frac{2m(1+\varkappa b)}{5\varkappa^2 e^{\varkappa b}} \; D_k \right ] \; .
\ee

\subsection{Bogolubov approximation}

The Bogolubov approximation is applicable for low temperatures and very weak
interactions, when it is admissible to neglect both $\rho_1$ and $\sigma_1$, as
compared to $\rho_0 \sim \rho$. Recall that omitting $\sigma_1$ but keeping
$\rho_1$ would be principally wrong, since these quantities are of the same order
of magnitude \cite{Yukalov_17,Yukalov_18,Yukalov_29,Yukalov_33,Yukalov_54}.

In the Bogolubov approximation, spectrum (\ref{3.28}) reduces to
\be
\label{3.31}
 \ep_k = 
\sqrt{ \frac{\rho}{m}\; ( \Phi_0 + f_k ) k^2 + \left ( \frac{k^2}{2m}\right )^2 } \;  .
\ee
The spectrum is anisotropic because of the presence of the dipolar part $f_k$.
In the long-wave limit, the spectrum is phononic,
\be
\label{3.32}
 \ep_k \simeq c_B k \qquad ( k \ra 0 ) \;  ,
\ee
with the isotropic sound velocity
\be
\label{3.33}
  c_B \equiv \sqrt{ \frac{\rho}{m}\; \Phi_0 } = 
\frac{\sqrt{4\pi\rho a_s}}{m} \; .
\ee
Note that there is no the so-called phonon instability in the long-wave limit,
which would arise if the screening would be neglected.

The balance between the dipolar interactions, with the effective strength
$4 \pi d_0^2 / 3$, and the local interaction, characterized by $\Phi_0$, is
measured by the parameter
\be
\label{3.34}
 \ep_D \equiv \frac{4\pi d_0^2}{3\Phi_0} = \frac{a_D}{3a_s} \qquad 
( a_D \equiv m d_0^2 ) \;  .
\ee
Then, keeping in mind polarized dipoles, for spectrum (\ref{3.31}) we get
\be
\label{3.35}
 \ep_k^2 = 
c_B^2 \left [ 1 + 
\ep_D \left ( 3\cos^2\vartheta_k - 1\right ) J(q,\zeta) \right ] k^2 +
\left ( \frac{k^2}{2m} \right )^2 \; ,
\ee
where $q = k b$ and $\zeta = \varkappa b$.

The spectrum can be represented in a dimensionless form by employing the healing
length $\xi$ that is defined by the equality of the effective sound-wave energy $mc^2$,
where $c$ is sound velocity, with the effective kinetic energy $1 / 2 m \xi^2$, which,
in the Bogolubov approximation, gives
\be
\label{3.36}
 \xi \equiv \frac{1}{\sqrt{2}\; m c_B} = \frac{1}{\sqrt{8\pi\rho a_s} } \;  .
\ee

Let us introduce the dimensionless momentum
\be
\label{3.37}
p \equiv k \xi
\ee
and the dimensionless spectrum
\be
\label{3.38}
  \ep(p) \equiv \frac{\xi}{c_B} \; \ep_k \; .
\ee
Also, for short, let us use the notation
\be
\label{3.39}
J_p \equiv J(q,\zeta) \qquad \left ( q = \frac{b}{\xi} \; p \right )
\ee
defining the dipolar part of the interaction
$$
 f_k = D_k J_p \qquad (p = k\xi ) \;  .
$$
Then the dimensionless form of spectrum in (\ref{3.35}) writes as
\be
\label{3.40}
  \ep(p) = 
p \sqrt{1 + \ep_D ( 3\cos^2\vartheta_p -1) J_p + \frac{1}{2}\; p^2 } \; ,
\ee
where
$$
 \cos\vartheta_p \equiv \frac{\bp\cdot\bfe_d}{p} \qquad (\bp \equiv \xi\bk ) \;  .
$$

It is important to emphasize that in the long-wave limit we have
$$
 \lim_{p\ra 0} J_p = \lim_{q\ra 0} J(q,\zeta) = \lim_{k\ra 0} I_k(b,\varkappa) = 0 \;  ,
$$
provided the screening effect is taken into account, so that $\varkappa \neq 0$.
But if the screening would be neglected, we would have
$$
 J(q,0) =  \lim_{\zeta\ra 0} J(q,\zeta) = \frac{3}{q^3} \; (\sin q - q \cos q) \; ,
$$
which leads to
$$
\lim_{q\ra 0} J(q,0) =  1\;  .
$$
However then the Fourier transform (\ref{3.24}) is not defined for $k \ra 0$.
Moreover, as follows from Sec. 2, thermodynamic quantities, containing $\Phi_0$,
such as the chemical potential, grand potential, and so on, would not be defined
as scalars \cite{Yukalov_43}. Therefore taking account of screening is compulsory
for having well defined thermodynamic characteristics. So, generally, we have to
deal with the integral
$$
J_p = 9p \int_\dlt^\infty \left [ \frac{\sin(pz)}{(pz)^4} \; - \; 
\frac{\cos(pz)}{(pz)^3} \; - \; 
\frac{\sin(pz)}{3(pz)^2} \right ]\; e^{-\varkappa \xi z} \; dz \; ,
$$
in which
$$
\delta \equiv \frac{b}{\xi} \; .
$$

The short-range cutoff $b$ can be treated to be of the order of the scattering
length $a_s$ that depends on the dipolar interactions \cite{Bortolotti_55,Ronen_56}.
If the screening length is close to the healing length, then
$\varkappa \xi = \xi / r_s \approx 1$.

The integral $J_p$, for $\varkappa \xi = 1$ can be written in the form
$$
J_p = - 3\pi\; \frac{1+p^2}{2p^3} + 3i \; \frac{1+p^2}{4p^3} \left \{ {\rm Ei}[(-1-ip)\dlt] -
{\rm Ei}[(-1+ip)\dlt] \right\} +
$$
$$
 + 
\frac{3e^{-\dlt}}{2p^3\dlt^3} \left [ \left ( 2 -\dlt+\dlt^2 \right ) \sin(p\dlt) -
p(2-\dlt)\dlt \cos(p\dlt) \right ] \; ,
$$
where ${\rm Ei}[\cdot]$ is the exponential integral function \cite{Yukalov_43}.
The behavior of the integral $J_p$ as a function of $p$, for $\delta = 0.1$ and
$\varkappa = 1/\xi$, is shown in Fig. 1.

\begin{figure}[ht!]
\vspace{9pt}
\centerline{
\includegraphics[width=10cm]{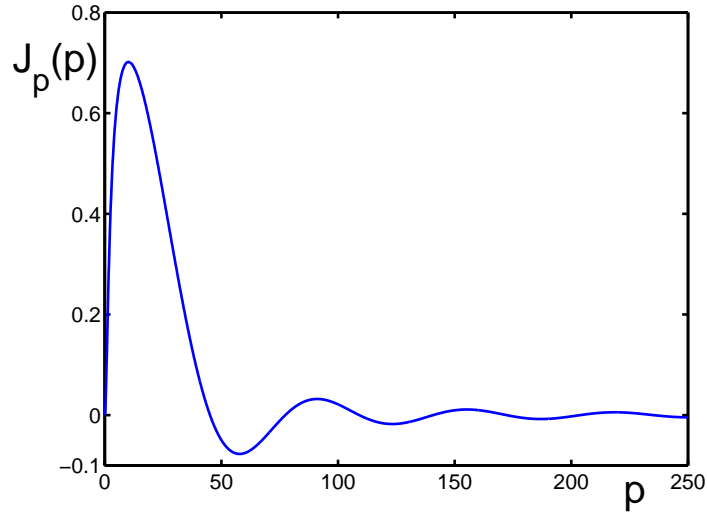} }
\caption{Behaviour of the integral $J_p$ as a function of the dimensionless momentum $p$ for
$\dlt=0.1$.
}
\label{fig:Fig.1}
\end{figure}

\subsection{Roton instability}

Spectrum (\ref{3.40}) is anisotropic, depending on the direction of momentum
with respect to the dipole polarization. When the momentum is parallel to the dipole
polarization, which is denoted as
\be
\label{3.41}
 \ep_{||}(p) = \ep(p) \qquad (\vartheta_p = 0 ) \;  ,
\ee
we have
\be
\label{3.42}
 \ep_{||}(p) = p\sqrt{1+2\ep_D J_p + \frac{1}{2}\; p^2 } \; .
\ee
While when the momentum is orthogonal to the dipole polarization, which is denoted as
\be
\label{3.43}
 \ep_\perp(p) = \ep(p) \qquad \left ( \vartheta_p = \frac{\pi}{2} \right ) \;  ,
\ee
we get
\be
\label{3.44}
 \ep_\perp(p) =  p\sqrt{1 - \ep_D J_p + \frac{1}{2}\; p^2 } \;  .
\ee

In  any case, in the long-wave limit $k \xi \ll 1$, the spectrum is acoustic, being
proportional to $p$. And in the short-wave limit $k \xi \gg 1$, it is of single-particle
type $p^2/2m$. But at an intermediate momentum, spectrum (\ref{3.44}) can have
a roton minimum. Roton instability develops when the roton minimum touches zero
at some critical momentum $p_c$, where $\varepsilon_\perp(p_c) = 0$. This
defines the critical momentum by the equation
\be
\label{3.45}
p^2 - 2\ep_D J_p + 2 = 0 \qquad (p = p_c) \;   .
\ee
A real-valued solution to this equation exists when $\varepsilon_D$ is of order of
unity or larger. If $\varepsilon_D \ll 1$, the roton minimum does not touch zero and
there is no roton instability.

The change of spectrum (\ref{3.44}) under increasing $\varepsilon_D$ is illustrated
in Fig. 2, where there is no roton minimum, Fig. 3, where there is roton minimum,
and Fig. 4, where the spectrum touches zero, thus corresponding to the appearing
roton instability. Varying the value $\delta$ between $0.001$ and $0.1$ does not
influence much the behavior of the spectrum.

\begin{figure}[ht!]
\vspace{9pt}
\centerline{
\includegraphics[width=10cm]{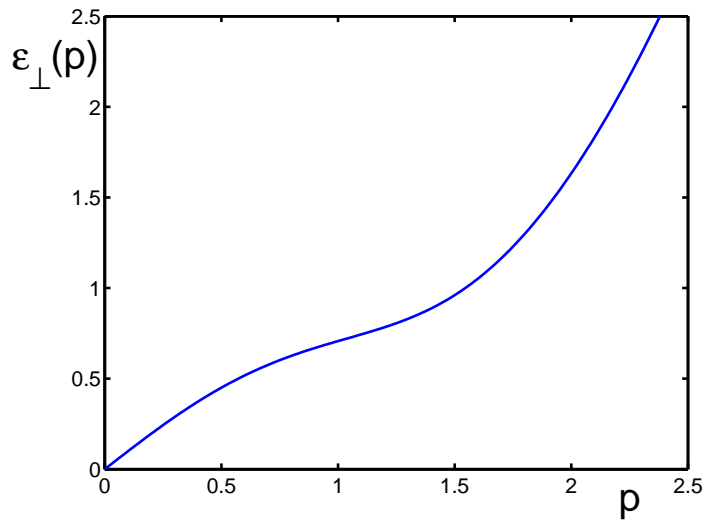} }
\caption{Spectrum $\ep_\perp(p)$ as a function of the dimensionless momentum $p$ for
$\dlt=0.1$ and $\ep_D=7$.
}
\label{fig:Fig.2}
\end{figure}

\begin{figure}[ht!]
\vspace{9pt}
\centerline{
\includegraphics[width=10cm]{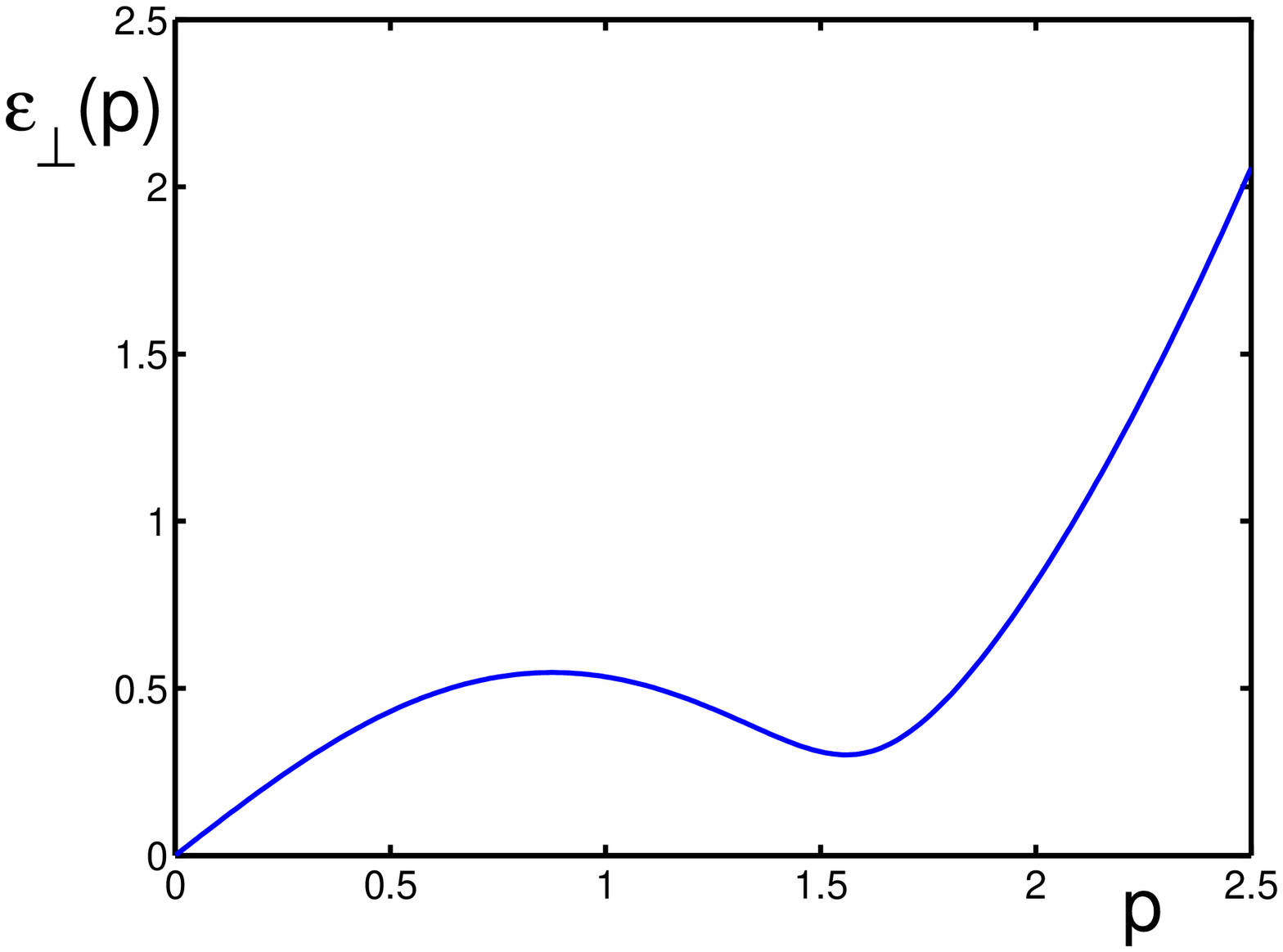} }
\caption{Spectrum $\ep_\perp(p)$ as a function of the dimensionless momentum $p$ for
$\dlt=0.1$ and $\ep_D=8.5$.
}
\label{fig:Fig.3}
\end{figure}

\begin{figure}[ht!]
\vspace{9pt}
\centerline{
\includegraphics[width=10cm]{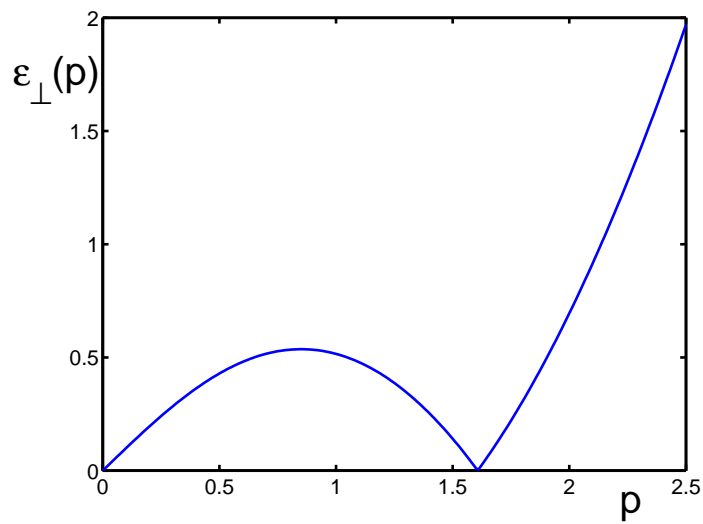} }
\caption{Spectrum $\ep_\perp(p)$ as a function of the dimensionless momentum $p$ for
$\dlt=0.1$ and $\ep_D=8.63845$.
}
\label{fig:Fig.4}
\end{figure}

Bose-Einstein condensates have been realized for $^{52}$Cr with $\mu_0 = 6 \mu_B$,
\cite{Griesmaier_57,Griesmaier_58}, $^{164}$Dy with $\mu_0 = 10 \mu_B$
\cite{Lu_59,Lu_60}, and for $^{168}$Er with $\mu_0 = 7 \mu_B$ \cite{Aikawa_61}.
The spectrum of collective excitations for $^{52}$Cr is studied in \cite{Bismut_62}
showing the spectrum anisotropy. No roton minimum was observed for $^{52}$Cr.
This is because the dipole interactions for $^{52}$Cr are not strong enough. Really,
in this case the scattering length is $a_s = 103 a_B = 0.545 \times 10^{-6}$ cm,
the dipolar length is $a_D = 2.4 \times 10^{-7}$ cm, the peak density is
$\rho \approx 3 \times 10^{14}$ cm$^{-3}$, which corresponds to the mean interatomic
distance $a \approx \rho^{-1/3} \approx 1.5 \times 10^{-5}$ cm, the healing length is
$\xi \approx 1.56 \times 10^{-5}$ cm. Then $a_D/3a_s = 0.147$, which is much smaller
than one.

Theoretically, the possibility of the roton minimum in the spectrum of trapped Bose
gases, interacting through dipolar forces, have been considered for quasi-one-dimensional
\cite{Giovanazzi_63,Dell_64,Mazets_65,Kurizki_66} and quasi-two-dimensional
\cite{Santos_67,Macia_68,Fedorov_69} cases. The instability develops for sufficiently
strong dipolar forces \cite{Goral_70,Goral_71,Lushnikov_72,Dutta_73}, which depends
on the trap shape, and can be connected with the roton instability \cite{Ronen_74}.

The origin of the roton minimum is due to the attractive part of the dipolar interactions
leading to the appearance of the negative term in spectrum (\ref{3.44}).

The Bogolubov approximation is applicable for temperature close to zero. At finite
temperature, one should consider spectrum (\ref{3.28}) taking into account the density
of uncondensed particles $\rho_1$ and the anomalous average $\sigma_1$. It is not
allowed to neglect $\sigma_1$, while retaining $\rho_1$ since these quantities are of
the same order of magnitude
\cite{Yukalov_17,Yukalov_18,Yukalov_29,Yukalov_33,Yukalov_54,Boudjemaa_75}.
Moreover, it is only taking into account the anomalous average $\sigma_1$ makes it
possible to correctly describe the Bose-Einstein condensation as a second-order
phase transition \cite{Yukalov_76} and to achieve for the ground state energy and
condensate density \cite{Yukalov_77} good numerical agreement with Monte Carlo
data \cite{Rossi_78}.

Let us emphasize that, in the Bogolubov approximation, the spectrum of collective
excitations (\ref{3.31}), or (\ref{3.35}), is anisotropic, while the sound velocity (\ref{3.33})
is isotropic. This is because taking account of the screening makes zero the long-wave
limit of $J_p \ra 0$ as $p \ra 0$. Not taking account of the screening is equivalent to
setting unity instead of $J_p$. In that case, the Bogolubov sound velocity would become
anisotropic, but simultaneously thermodynamic potential and other thermodynamic
quantities that have to be scalars, would become explicitly dependent on the angle
$\vartheta_k$, thus loosing their scalar form, which would be senseless \cite{Yukalov_43}.
However, in the HFB approximation, the sound velocity (\ref{3.30}) is anisotropic,
which requires to consider the density of uncondensed particles $\rho_1$ and the
anomalous average $\sigma_1$.

\subsection{Dipolar instability}

Sufficiently strong dipolar interactions can lead to the instability of a Bose-condensed
system \cite{Goral_70,Goral_71,Lushnikov_72,Dutta_73,Santos_79}. Roton instability,
arising at finite momenta of excitations, is one of the possibilities. It may also happen
instability, occurring at the zero momentum. As and example, let us consider the
ultimate situation when the scattering length $a_s$ goes to zero. Varying the scattering
length can be done by employing the Feshbach resonance \cite{Chin_80} that is well
studied for $^{52}$Cr \cite{Werner_81}. Effective dipole-dipole interactions can be
modified by using rotating external fields \cite{Giovanazzi_82}.

In the case of a homogeneous system or for a large trap in the local-density approximation,
the spectrum of collective excitations is given by equation (\ref{3.28}). If $a_s = 0$, then
we get the equation
\be
\label{3.46}
 \ep_k^2 = \left [ \frac{k^2}{2m} + (\rho_1 - \sgm_1) f_k \right ] 
 \left [ \frac{k^2}{2m} + (2\rho_0 + \rho_1 + \sgm_1) f_k \right ]  \;  .
\ee
In the local-density approximation, the quantities $\rho_1$, $\sigma_1$, and $\rho_0$
are functions of ${\bf r}$. In the long-wave limit, the spectrum becomes of a single-particle
type
\be
\label{3.47}
  \ep_k \simeq \frac{k^2}{2m_k} \qquad ( k \ra 0 ) \; ,
\ee
with the effective mass
\be
\label{3.48}
 m_k \equiv 
\frac{m}{ \{ [ 1 + 2m(\rho_1-\sgm_1)AD_k][1+2m(2\rho_0+\rho_1-\sgm_1)AD_k]\}^{1/2} } \;  ,
\ee
where
\be
\label{3.49}
 A \equiv \frac{1+\varkappa b}{5\varkappa^2 e^{\varkappa b} } \; .
\ee
In the Bogolubov approximation, the effective mass reduces to
\be
\label{3.50}
 m_k = \frac{m}{\sqrt{1+4m\rho AD_k} } \;  .
\ee
The mass and also the spectrum become negative for $\vartheta_k = \pi/2$, when
\be
\label{3.51}
 \frac{\rho a_D}{\varkappa^2} > \frac{15e^{\varkappa b}}{16\pi(1+\varkappa b)} \;  .
\ee
If $\varkappa \sim 1/\xi$ and $\varkappa b \ll 1$, then the above inequality reads as
$\rho a_D \xi^2 > 15/(16 \pi)$. With the expression (\ref{3.36}) for the healing length,
we have $a_D/a_s > 7.5$.

Moreover, even for a positive $m_k$, such a quadratic spectrum does not satisfy
the Landau criterion, hence cannot support superfluidity. Recall also that Bose-Einstein
condensation is necessarily accompanied by global gauge symmetry breaking
\cite{Lieb_1,Yukalov_12,Yukalov_17,Yukalov_18,Yukalov_29}. And the collective
excitations in a system with global gauge symmetry breaking exhibit the phonon
behavior in the long-wave limit \cite{Bogolubov_30,Bogolubov_31,Gavoret_83}.
That is, the quadratic spectrum contradicts the presence of Bose-Einstein condensate
or, in other words, signifies the instability of the latter.

The phonon instability at $a_s \ra 0$ leads to the divergence of the isothermal
compressibility
\be
\label{3.52}
 \varkappa_T = \frac{1}{\rho N} \left ( \frac{\prt N}{\prt\mu} \right )_{TV} = 
\frac{{\rm var}(\hat N)}{\rho N T} \;  .
\ee
In the local density approximation, this gives
\be
\label{3.53}
\varkappa_T = \frac{1}{\rho N} \int \frac{\rho(\br)}{\Dlt(\br)} \; d\br \;   ,
\ee
where $\Delta({\bf r})$ is defined in (\ref{2.144}) and, in the Bogolubov approximation,
it is
\be
\label{3.54}
\Dlt(\br) = \rho(\br)\Phi_0 \;   .
\ee
Then the compressibility becomes
\be
\label{3.55}
\varkappa_T = \frac{1}{\rho^2\Phi_0} =  \frac{m}{4\pi\rho^2 a_s} \;  .
\ee
As is evident, when $a_s \ra 0$, the compressibility diverges.

\subsection{Trapped atoms}

In the previous sections, the system stability has been considered for either a
uniform system or for a large trap with a slowly varying trapping potential, for which
the local-density approximation is valid. But when atoms are confined in a trap,
the trapping potential can stabilize atomic system that in the uniform case would be
unstable.

Limiting the consideration by the Bogolubov approximation, we may neglect $\rho_1$
and $\sigma_1$, as compared to $\rho_0 \sim \rho$. Then the condensate function
equation (\ref{2.34}) reduces to the form given by the equation
\be
\label{3.56}
i\; \frac{\prt}{\prt t} \; \eta(\br,t) = \frac{\dlt H^{(0)}}{\dlt\eta^*(\br,t)} \; ,
\ee
which coincides with equation (\ref{2.40}) for the vacuum coherent field.

The trapping potential is usually taken as the harmonic potential. For a trap with
cylindrical symmetry, it reads as
\be
\label{3.57}
 U(\br) = \frac{m}{2}\; \left ( \om_\perp^2 r_\perp^2 + \om_z^2 z^2 \right ) \; ,
\ee
where
\be
\label{3.58}
 \om_\perp \equiv \om_x = \om_y \; , \qquad r_\perp^2 = x^2 + y^2 \;  .
\ee
The geometry of the trap, characterized by the aspect ratio
\be
\label{3.59}
  \al \equiv \frac{\om_z}{\om_\perp} = \left ( \frac{l_\perp}{l_z} \right )^2 \; ,
\ee
with the oscillator lengths
$$
l_\perp \equiv \frac{1}{\sqrt{m\om_\perp}} \; , \qquad 
l_z \equiv \frac{1}{\sqrt{m\om_z}} \;   ,
$$
plays an important role for the system stability.

In equilibrium, the condensate-function equation takes the form
\be
\label{3.60}
 \left\{ -\; \frac{\nabla^2}{2m} + \frac{m\om_\perp^2}{2} \; \left ( 
r_\perp^2 + \al^2 z^2 \right ) + \Phi_0\rho(\br) +
\int \overline D(\br - \br') \rho(\br')\; d\br' \right \} \; \eta(\br) = 
\mu_0\eta(\br) \; ,
\ee
where $\rho({\bf r}) = |\eta({\bf r})|^2$ and the abbreviated notation for the regularized
potential
\be
\label{3.61}
\overline D(\br) \equiv D(\br,b,\varkappa)
\ee
is used.

For uniform systems, the regularization of the dipolar interaction potential is compulsory
for having correct expressions for physical quantities and avoiding senseless conclusions.
But for finite systems, it may happen that this regularization could be not so important.
To decide on this problem, one has to analyze the integral term of equation (\ref{3.60}),
which can be written in two forms,
\be
\label{3.62}
\int \overline D(\br-\br') \rho(\br') \; d\br' = 
\int \overline D_k \rho_k e^{i\bk\cdot\br}\; \frac{d\bk}{(2\pi)^3} \; , 
\ee
involving the Fourier transforms of the regularized potential
$$
\overline D(\br) = \int \overline D_k e^{i\bk\cdot\br}\; \frac{d\bk}{(2\pi)^3} \; , 
\qquad
\overline D_k \equiv D_k(b,\varkappa) \; , 
$$
and of the density
$$
 \rho(\br) = \int \rho_k e^{i\bk\cdot\br}\; \frac{d\bk}{(2\pi)^3} \;  .
$$

The quantity $\overline{D}_k$ tends to zero for $k \gg 1/b$, while $\rho_k$ quickly
diminishes for $k \gg \max_\alpha \{1 / l_\alpha\}$, with $l_\alpha$ being the effective
system length in the $\alpha$-direction. Since $b$ is of the order of $a_s$, one
usually has
\be
\label{3.63}
 b \ll \min_\al \{ l_\al \} \;  .
\ee
In other words,
\be
\label{3.64}
 \frac{1}{b} \gg \frac{1}{ \min_\al \{ l_\al \} }  = \max_\al \left\{
\frac{1}{l_\al} \right \} \; .
\ee
This tells us that the integration in the right-hand side of (\ref{3.62}) is effectively
done over the momenta not much exceeding $\max_\alpha\{1/l_\alpha\}$. Hence
the finiteness of the trap plays the role of an effective regularization at large
momenta or small distances. This implies that the short-range regularization by
means of the cutoff $b$ is not as important, provided no divergences arise.

On the other hand, $\overline{D}({\bf r})$ diminishes for $r \gg r_s \equiv 1/ \varkappa$,
while the density $\rho({\bf r})$ for a finite trap quickly goes to zero for
$r \gg \max_\alpha\{l_\alpha\}$. Therefore for a small trap, with the sizes such that
\be
\label{3.65}
 \max_\al \{l_\al  \} \ll r_s \equiv \frac{1}{\varkappa} \;  ,
\ee
the screening is not crucial, since the trap length effectively limits the integration in the
left-hand side of (\ref{3.62}) ar large distance. Therefore for such small traps the
screening at large distance may be not necessary, since
$$
 \varkappa \ll \frac{1}{\max_\al\{l_\al\} } = 
\min_\al  \left\{ \frac{1}{l_\al} \right \} \; .
$$

The screening lengths in condensed matter are typically of ten Angstroms \cite{Kanth_84}.
The healing length for atomic Bose condensate is $\xi \sim (10^{-5} - 10^{-4})$ cm, while
its typical size is of the order of $(10^{-4} - 10^{-2})$ cm. So, the traps, whose maximal
length is smaller that the screening radius, should be really small. Those calculations for
trapped atoms or molecules, which do not take account  of screening are valid only for
small traps in the sense of condition (\ref{3.65}).

\subsection{Geometric stabilization}

Recall first that a Bose-condensed gas with purely local attractive interactions, when
$a_s < 0$ and $a_D = 0$, is unstable in a uniform system, but can be stabilized
in a trap for the number of particles smaller that a critical number $N_c$ depending
on the shape geometry and the interaction strength. The value of the critical number
can be estimated \cite{Yukalov_85} from the expression
\be
\label{3.66}
 N_c = \sqrt{\frac{\pi}{2} } \; \frac{l_x l_y l_z}{|a_s|(l_x^2+l_y^2+l_z^2)} 
\qquad ( a_s < 0 , \; a_D = 0 ) \; .
\ee

Similarly, the stability of a gas with contact plus dipolar interactions depends on the
trap geometry and the interaction strength
\cite{Goral_70,Goral_71,Lushnikov_72,Dutta_73,Santos_79}. One typically considers
small traps in the sense of condition (\ref{3.65}), which allows for the use of the not
regularized dipolar potential. This potential is partially attractive. For instance, let us
consider a cylindrical trap, with the cylinder axis along $z$, filled by atoms whose
dipoles are oriented along the axis $z$. Then expression (\ref{3.7}) shows that the
atomic interactions in the direction of the axis $z$ are attractive, while the interactions
in the perpendicular direction are repulsive:
$$
D(\br) = - 2\; \frac{d_0^2}{r^3} \qquad ( \vartheta = 0 ) \; ,
$$
\be
\label{3.67}
 D(\br) =  \frac{d_0^2}{r^3} \qquad 
\left ( \vartheta = \frac{\pi}{2} \right ) \;  .
\ee
It is clear that the atomic clouds in pencil-shape traps, where the majority of atoms
interact attractively, should be less stable than the atomic clouds inside pancake-shape
traps, where the majority of atoms interact repulsively.

The energetic stability, at zero temperature and weak interactions, can be studied by
looking for the minima of the energy functional
\be
\label{3.68}
E = \int \eta^*(\br) \left [ -\; \frac{\nabla^2}{2m} + U(\br) \right ] \eta(\br) \; d\br + 
\frac{1}{2} \int \Phi(\br-\br') | \eta(\br)|^2 | \eta(\br')|^2 \; d\br d\br' \;,
\ee
similarly to the stability analysis for different finite quantum systems \cite{Birman_86}.
Here a small trap is assumed in the sense of condition (\ref{3.65}), so that the dipolar
potential is not regularized. Taking the trial condensate function in the Gaussian form
\be
\label{3.69}
 \eta(\br) = \frac{1}{\pi^{3/4}} \; \sqrt{ \frac{N}{L_\perp^2 L_z} } \;
\exp\left\{ -\; \frac{1}{2} \left ( \frac{r_\perp}{L_\perp} \right )^2 \; - \;
\frac{1}{2} \left ( \frac{z}{L_z} \right )^2 \right \} \; ,
\ee
in which $L_\perp$ and $L_z$ are variational parameters describing the effective radius
and length of the atomic cloud, gives the internal energy
\be
\label{3.70}
 E = E_{kin} + E_{tr} + E_{int} \;  ,
\ee
consisting of the kinetic energy
\be
\label{3.71}
 E_{kin} = \frac{N}{4m} \left ( \frac{2}{L_\perp^2} + \frac{1}{L_z^2} \right ) \; ,
\ee
potential trap energy
\be
\label{3.72}
  E_{tr} = \frac{N}{4}\; m \om_\perp^2 \left ( 2 L_\perp^2 + \al^2 L_z^2 \right ) \;  ,
\ee
and the interaction energy
\be
\label{3.73}
 E_{int} = \frac{N^2 a_s}{\sqrt{2\pi}\; m L_\perp^2 L_z}\; [ 1 - \ep_D F(\lbd) ] \;  .
\ee
Here the notations are used:
\be
\label{3.74}
F(\lbd) = \frac{1+2\lbd}{1-\lbd} \; - \; \frac{3\lbd f(\lbd)}{(1-\lbd)^{3/2}}
\ee
and
\be
\label{3.75}
 \lbd \equiv \left ( \frac{L_\perp}{L_z} \right )^2 \;  ,
\ee
in which
\begin{eqnarray}
\nonumber
f(\lbd) = \left \{ \begin{array}{cc}
{\rm artanh} \sqrt{\lbd-1} \; , ~ & ~ \lbd > 1 \\
                                  & \\
\frac{1}{2}\; \ln \; \frac{1+\sqrt{1-\lbd}}{1-\sqrt{1-\lbd}} \; , ~ & ~ \lbd < 1
\end{array} \right. .
\end{eqnarray} 
One should not confuse the variational parameters $L_\perp$ and $L_z$
with the fixed oscillator lengths $l_\perp$ and $l_z$. Hence $\lambda$ is different
from the aspect ratio $\alpha = (l_\perp/l_z)^2$.

Accomplishing the minimization with respect to the parameters $L_\perp$ and $L_z$
results in the energy $E$ being a function of five parameters, $m$, $N$,
$\omega_\perp$, $\alpha$, and $\varepsilon_D$. A discussion on numerical analysis
can be found in \cite{Baranov_20,Baranov_23}. Briefly speaking, the results are as
follows. When $a_s < 0$, the system is stable (metastable) for $N < N_c$. If $a_s > 0$,
the system is stable for weak dipolar interactions with $0 < \varepsilon_D < 1$. When
the dipolar interactions are strong, such that $\varepsilon_D > 1$, the system can be
metastable for $N < N_c$. The value of the critical number $N_c$ depends on the
trap ratio $\alpha$ and the parameter $\varepsilon_D$. Generally, atomic clouds in
pancake-shape traps are more stable than those in pencil-shape traps, provided the
polarization of dipoles is along the axis $z$.

\subsection{Thomas-Fermi approximation}

The shape of the atomic cloud in a small trap can be found by invoking the Thomas-Fermi
approximation \cite{Yi_87,Eberlein_88}, when the kinetic energy can be neglected being
much smaller than the interaction energy. Then equation (\ref{3.60}) gives
\be
\label{3.76}
\rho(\br) = | \eta(\br)|^2 = \rho(0) \left ( 1 - \; \frac{r_\perp^2}{R_\perp^2} \; - \;
\frac{z^2}{R_z^2} \right ) \;   ,
\ee
which is valid in the region
$$
 \frac{r_\perp^2}{R_\perp^2} + \frac{z^2}{R_z^2}  \leq 1 \; ,
$$
where $\rho({\bf r})$ is non-negative, and is zero outside this region. The density in
the center is
\be
\label{3.77}
\rho(0) =  \frac{15N}{8\pi R_\perp^2 R_z}  \; .
\ee
The transverse and longitudinal radii, defining the effective size of the atomic cloud,
are given by the equations
\be
\label{3.78}
R_\perp^5 = \frac{15\Phi_0\sqrt{\lbd}\; N}{4\pi m\om_\perp^2} \; \left \{ 1 +
\ep_D \left [ \frac{3\lbd F(\lbd)}{2(1-\lbd)} \; - \; 1 \right ] \right \}
\ee
and
\be
\label{3.79}
 3\sqrt{\lbd} \; \ep_D \left [ 
\left ( 1 + \frac{\al^2}{2} \right ) \frac{F(\lbd)}{1-\lbd} \; - \; 1 \right ] +
(\ep_D - 1 ) \left ( \lbd -\al^2 \right ) = 0 \; ,
\ee
where
\be
\label{3.80}
 \lbd \equiv \left ( \frac{R_\perp}{R_z} \right )^2 \; , \qquad
 \al \equiv \left ( \frac{l_\perp}{l_z} \right )^2 \;  .
\ee
But the critical number $N_c$ cannot be found in this approximation.

\subsection{Quantum ferrofluid}

A system of atomic or molecular dipoles is somewhat analogous to a collection
of nanosize ferromagnetic particles forming colloidal suspensions \cite{Rosensweig_89}.
Therefore dipolar atomic and molecular systems can be called quantum ferrofluids
\cite{Lahaye_90}. The properties of such systems can be governed in a wide diapason.
The $s$-wave scattering length can be varied by Feshbach resonance, so that the
magnetic dipole-dipole interactions can become comparable in strength with the contact
interaction. Moreover, it is possible to create a purely dipolar quantum gas, reducing the
scattering length $a_s$ to zero \cite{Koch_91}. Such a dipolar gas can be stable in a
pancake trap, provided the number of particles is smaller than the critical number $N_c$
and temperature is close to zero. Dipolar gas in a pancake trap can be stable
(metastable) even for a weak attractive contact interaction with $a_s < 0$, when
$N < N_c$ and $T = 0$ \cite{Koch_91}. However thermal fluctuations make this gas
unstable \cite{Junginger_92}.

An interesting question is: What happens with the strongly dipolar gas, with
$\varepsilon_D \geq 1$ and $a_s \geq 0$, after it becomes unstable? Experiments
with $^{164}$Dy demonstrated that after the system looses stability it forms a metastable
state with one \cite{Schmitt_93,Chomaz_94} or several \cite{Kadau_95,Barbut_96}
self-bound small droplets, whose lifetime being of order $(0.01 - 0.1)$ s. Monte Carlo
simulations \cite{Macia_97,Cinti_98} confirmed the formation of self-bound droplets
of dipolar quantum gas of trapped $^{164}$Dy atoms brought to the regime of instability.
When the number of droplets increases, they form ordered arrays, reminiscent of the
Rosensweig instability of classical ferrofluids in an external magnetic field
\cite{Rosensweig_89}. Theoretically, the appearance of the metastable droplets of dipolar
gas can be explained by taking account of Lee-Huang-Yang corrections
\cite{Petrov_99,Wachtler_100}. Recall that the Lee-Huang-Yang corrections are contained
in the self-consistent mean-field theory
\cite{Yukalov_17,Yukalov_18,Yukalov_29,Yukalov_33,Yukalov_77,Yukalov_101}. These
corrections are just the limiting case of a small gas parameter $\rho^{1/3} a_s$ of this theory.

The system of dipolar bosons, with increasing density or interaction strength, can also
crystallize in three \cite{Jain_102,Osychenko_103} as well as in two \cite{Kurbakov_104} 
dimensions, similarly to classical fluids. The superfluid-crystal quantum phase transition of
a system of purely repulsive dipolar bosons in two dimensions has been studied by Quantum
Monte Carlo simulations at zero temperature \cite{Moroni_104}. It has been found that for all
practical  purposes there happens a conventional first-order phase transition between the
superfluid and crystalline  phases. And the microemulsion scenario for this system has been
ruled out.

Typical size of trapped Bose-Einstein condensates is of the order of $(10^{-4} - 10^{-2})$ cm.
This makes it difficult to get a good resolution of an absorption image {\it in situ}. Therefore
one usually observes absorption images of the atomic cloud after some time of free
expansion from the trap, when the trapping potential has been switched off. The free
expansion is described by rescaling the characteristic lengths of the cloud by
time-dependent factors \cite{Eberlein_88,Giovanazzi_105}.

In nonequilibrium classical ferrofluids, there can exist solitonic excitations \cite{Richter_106}.
In quantum ferrofluids, three-dimensional solitons are unstable, but can arise in lower
dimensional traps \cite{Lahaye_107}.

\subsection{Magnetic induction}

In the process of time-of-flight imaging, when the trapping potential is switched off, atoms
fall down out of the trap due to gravity. At this stage, in addition to studying absorption
images, it is admissible to get more information on the atoms by considering their fall
through a magnetic coil connected to an electric circuit.

Let the electric circuit be characterized by resistance $R$, inductance $L$, and capacity 
$C$. Suppose the coil of the circuit has $n_c$ turns, cross-section area $A_c$, and length 
$l_c$, the coil axis being directed along the unit vector ${\bf e}_c$. If a cloud of atoms, 
with dipole magnetic moments $\mu_0 {\bf e}_d$, passes through the coil, with the number-of 
particle rate $\dot{N}$, this moving atomic cloud induces the electromotive force
\be
\label{3.81}
 E_f = - \; \frac{4\pi c}{cl_c}\;\mu_0 \; \bfe_d \cdot \bfe_c \; \dot{N} \;  ,
\ee
where $c$ is light velocity. The electromotive force generates in the electric circuit
an electric current $j$ obeying the Kirchhoff equation
\be
\label{3.82}
 L \; \frac{dj}{dt} + R j + \frac{1}{C} \int_0^t j(t')\; dt' = E_f \;  .
\ee
The latter can be represented \cite{Yukalov_108,Yukalov_109,Yukalov_110} in the
integral form
\be
\label{3.83}
 j = - \mu_0 \; \frac{\bfe_d \cdot \bfe_c}{n_c A_c} \int_0^t G(t-t') \; dN(t') \;  ,
\ee
in which the transfer function
$$
G(t) \equiv 
\left ( \cos \om' t \; - \; \frac{\gm}{\om'} \; \sin \om' t \right ) e^{-\gm t}
$$
contains the notations
$$
\om' \equiv \sqrt{\om^2 -\gm^2} \; , \qquad \om \equiv \frac{1}{\sqrt{LC}} \; ,
\qquad \gm \equiv \frac{R}{2L} \;  .
$$
Here $\omega$ is the circuit natural frequency and $\gamma$ is the circuit damping.
The circuit is assumed to be of good quality, which implies that $\gamma \ll \omega$.

It is straightforward to find the electric current, when the number of atoms passing
through the coil is given. For example, if the number of atoms passing through the coil
is proportional to time,
$$
N(t) = \Gm t \; ,
$$
then the induced current is
\be
\label{3.84}
j = - \mu_0 \; \frac{\bfe_d\cdot\bfe_c\;\Gm}{n_c A_c\om}\; \sin(\om t) e^{-\gm t} \; .
\ee
Measuring the current makes it possible to find the dipole magnetic moment $\mu_0$.

As another example, let us consider the case, when the number of atoms inside the coil
oscillates as
$$
 N(t) = N_0 + \Dlt N \sin(\om_0 t) \;  ,
$$
with $N_0 \geq \Delta N > 0$. If the oscillation frequency $\omega_0$ is close to the
circuit natural frequency $\omega$, so that the resonance condition
$|\omega - \omega_0| \ll \omega$ holds, then the induced current is
\be
\label{3.85}
j = \mu_0 \; \frac{\bfe_d\cdot\bfe_c\;\om_0}{2n_c A_c\gm}\; \Dlt N
\cos(\om_0 t) \left ( 1 - e^{-\gm t} \right ) \; .
\ee
But if $\omega_0$ is far detuned from $\omega$, then the amplitude of the induced
current diminishes by the factor $\gamma^2 / \omega^2$.

\subsection{Optical lattices}

In the presence of an optical lattice, the effective lattice Hamiltonian can be derived
following the scheme of Sec. 2.12. In the case of lattice band gaps that are much larger
than any other energy scales of the system, such as the interaction strength, temperature,
and Zeeman shifts, atoms remain confined in the lowest energy band of the lattice.
Then, expanding the field operators over Wanier functions, according to (\ref{2.155}),
and considering only the lowest lattice band, we get the extended Hubbard model
$$
\hat H = - \sum_{i\neq j} J_{ij} \hat c_i^\dgr \hat c_j + 
\sum_j h_j \hat c_j^\dgr \hat c_j  + \hat H_Z \; + 
$$
\be
\label{3.86}
+ \;
\frac{1}{2} \; \sum_j U_j \hat c_j^\dgr \hat c_j^\dgr \hat c_j \hat c_j +
\frac{1}{2} \; \sum_{i\neq j} U_{ij} \hat c_i^\dgr \hat c_j^\dgr \hat c_j \hat c_i \; ,
\ee
in which
$$
J_{ij} = - \int w^*(\br-\ba_i) \hat H(\br) w(\br-\ba_j) \; d\br \; , \qquad
h_j = \int w^*(\br-\ba_j) \hat H(\br) w(\br-\ba_j) \; d\br \; ,
$$
\be
\label{3.87}
 U_j \equiv U_{jjjj} \; , \qquad U_{ij} \equiv U_{ijji} + U_{ijij} \;  ,
\ee
where $\hat{H}({\bf r})$ is defined in (\ref{2.154}), and the matrix elements $U_{ijkl}$
are defined in (\ref{2.158}), for instance
\be
\label{3.88}
 U_{ijji} = 
\int |w(\br-\ba_i)|^2 \; \Phi(\br-\br')\; |w(\br'-\ba_j)|^2\; d\br d\br' \;  .
\ee
Here we consider atoms or molecules possessing magnetic moment $\mu_S$.
The Zeeman term $H_Z$ takes account of the external magnetic field ${\bf B}({\bf r})$.
The field operators and the operators $\hat{c}_j$, generally, are spinors with respect to
some internal degrees of freedom.

The magnetic moment $\vec{\mu}_0$ of an atom can be written as
\be
\label{3.89}
 \vec{\mu}_0 = \mu_S\bS \qquad  (\mu_S \equiv g_S \mu_B ) \;  ,
\ee
where $g_S$ is the Land\'e factor, and ${\bf S}$ is the effective atomic spin operator. Then
the Zeeman energy operator reads as
\be
\label{3.90}
 \hat H_Z = - 
\mu_S \int \hat\psi^\dgr(\br)\; \bB(\br) \cdot \bS \; \hat\psi(\br) \; d\br \; .
\ee

Introducing the notation
\be
\label{3.91}
\bB_{ij} \equiv \int w^*(\br-\ba_i) \bB(\br) w(\br-\ba_j) \; d\br
\ee
transforms the Zeeman term (\ref{3.90}) to the form
\be
\label{3.92}
 \hat H_Z = - \mu_S \sum_{ij} \bB_{ij} \cdot 
\left ( \hat c_i^\dgr\; \bS\; \hat c_j \right ) \; .
\ee
If either the external magnetic field is slowly varying or atoms are well localized, then
\be
\label{3.93}
 \bB_{ij} = \bB_j \dlt_{ij} \qquad (\bB_j \equiv \bB_{jj} ) \;  .
\ee
Defining the local spin operator
\be
\label{3.94}
 \bS_j \equiv \hat c_j^\dgr \; \bS \; \hat c_j \;  ,
\ee
we come to the Zeeman term
\be
\label{3.95}
 \hat H_Z = -\mu_S \sum_j \bB_j \cdot \bS_j \;  .
\ee

A similar Hamiltonian characterizing quantum magnetism of the dipolar lattice gas
of $^{52}$Cr, having spin $S = 3$, has been realized in experiment \cite{Paz_111}.

In addition to the linear Zeeman effect, there also exists the so-called quadratic
Zeeman effect that can be due either to the hyperfine structure of the atom, or to
alternating external electromagnetic fields. By applying off-resonance linearly polarized
light, populating the internal spin states with different $m = - S, - S +1, \ldots, S$, it is
possible to exert the quadratic Zeeman shift along the light polarization axis
\cite{Paz_111,Cohen_174,Santos_112,Jensen_175}, generating the term
$$
 q_Z \sum_j \left ( S_j^z \right )^2 \;  ,
$$
where $q_Z$ is independent of the constant magnetic field and the axis $z$ is assumed
to be the light polarization axis. This optically induced quadratic Zeeman shift can be
tailored at high resolution and rapidly adjusted using electro-optics. By employing either
positive or negative detuning the sign of $q_Z$ can be varied.

A similar quadratic shift can be induced by a linearly polarized microwave driving field,
with the driving Rabbi frequency $\Omega$ and detuning $\Delta$ from a hyperfine
transition, creating quadratic Zeeman effect with $q_Z = - \hbar \Omega^2/4 \Delta$,
of the order of $100 \hbar/s$, as is demonstrated for $^{87}$Rb \cite{Gerbier_176,Leslie_126}.

Low-temperature phases and stability of dipolar bosons in optical lattices have been
studied in Refs. \cite{Griesmaier_19,Baranov_20,Pupillo_21,Lahaye_22,Baranov_23,Gadway_24,
Ueda_27,Danshita_113,Baier_114}. Optical lattices can stabilize even a Bose gas with 
attractive local interactions \cite{Muller_115}.

It is worth noting that the considered above optical lattices do not take into account
phonon degrees of freedom. Taking these into account may lead to the phonon instability 
of insulating states in optical lattices, although their lifetime can be rather long for 
treating such insulating optical lattices as metastable objects \cite{Yukalov_116,Yukalov_117}.

\subsection{Deep lattice}

Atomic degrees of freedom can be separated from the spin degrees of freedom, when atoms 
(or molecules) are loaded into a deep optical lattice, where the number of atoms in each 
lattice site is fixed and atoms (or molecules) do not jump between the lattice sites. This 
implies that the Wannier functions are so well localized at the related lattice sites that 
their intersection for different sites is practically zero. As a result, the tunneling 
term becomes negligible,
\be
\label{3.96}
 J_{ij} = 0 \qquad  (i \neq j) \;  .
\ee
With the interaction potential (\ref{3.23}), Hamiltonian (\ref{3.86}) splits into the sum
\be
\label{3.97}
\hat H = \hat H_a + \hat H_s
\ee
of atomic and spin terms. Taking into account that, under the assumed good localization,
$$
 \int | w(\br-\ba_i)|^2 | w(\br-\ba_j)|^2 \;  d\br \cong 0 \qquad ( i\neq j) \;  ,
$$
for the atomic term we have
\be
\label{3.98}
 \hat H_a = \sum_j h_j \hat c_j^\dgr \hat c_j \; + \; 
\frac{1}{2} \sum_j U_j \hat c_j^\dgr c_j^\dgr \hat c_j \hat c_j \;  ,
\ee
where
\be
\label{3.99}
 U_j \equiv \Phi_0 \int | w(\br-\ba_j)|^4 \;  d\br \; .
\ee

The spin part is the sum of the Zeeman term and the interaction term,
\be
\label{3.100}
  \hat H_s = \hat H_Z \; + \; 
\frac{1}{2} \sum_{i\neq j} \sum_{\al\bt} \overline D_{ij}^{\al\bt} S_i^\al S_j^\bt \; .
\ee
Here
$$
\overline D_{ij}^{\al\bt} = \int \overline D^{\al\bt}(\br-\br')
\left [ | w(\br-\ba_i)|^2 | w(\br'-\ba_j)|^2 \right. \; +
$$
\be
\label{3.101}
 \left. + \;
w^*(\br-\ba_i)w^*(\br'-\ba_j) w(\br'-\ba_i)w(\br-\ba_j) \right ] \; d\br d\br' \; ,
\ee
where
$$
\overline D^{\al\bt}(\br-\br') = \Theta(|\br-\br'|-b) D^{\al\bt}(\br-\br') 
\exp\{ -\varkappa|\br-\br'| \} \; , 
$$
$$
D^{\al\bt}(\br-\br') = \mu_S^2 \; \frac{\dlt_{\al\bt}-3n^\al n^\bt}{|\br-\br'|^3} \; , 
\qquad 
\bn \equiv \frac{\br-\br'}{|\br-\br'|} \; .
$$
Note that $\overline{D}_{jj}^{\alpha \beta} = 0$. The exchange term is small, since
$$
 \int w^*(\br-\ba_i)w^*(\br'-\ba_j) w(\br'-\ba_i)w(\br-\ba_j) 
\overline D^{\al\bt}(\br-\br') \; d\br d\br' ~ \cong ~
\dlt_{ij} \overline D^{\al\bt}(0) = 0 \; .
$$
Therefore, for a deep lattice,
\be
\label{3.102}
 \overline D_{ij}^{\al\bt} = \Theta(a_{ij}-b) D_{ij}^{\al\bt} \exp(-\varkappa a_{ij} ) \; ,
\ee
where
\be
\label{3.103}
D_{ij}^{\al\bt} = \frac{\mu_S^2}{a_{ij}^3} \; 
\left ( \dlt_{\al\bt} - 3 n_{ij}^\al n_{ij}^\bt \right )
\ee
is the dipolar tensor and
$$
 a_{ij} \equiv | \ba_{ij}| \; , 
\qquad 
\bn_{ij} \equiv \frac{\ba_{ij}}{a_{ij}} \; ,
\qquad 
\ba_{ij} \equiv \ba_i - \ba_j \;  .
$$
In that way, we come to the spin Hamiltonian
\be
\label{3.104}
\hat H_s = -\mu_S \sum_j \bB_j \cdot \bS_j \; + \;
\frac{1}{2} \sum_{i\neq j} \sum_{\al\bt} \overline D_{ij}^{\al\bt} S_i^\al S_j^\bt \;  .
\ee
Since the linear size of an atom (or molecule) is smaller than the intersite distance
$$
 a \equiv \min_{i\neq j} a_{ij} > b \;  ,
$$
then
\be
\label{3.105}
 \overline D_{ij}^{\al\bt} = D_{ij}^{\al\bt} \exp(-\varkappa a_{ij}) \qquad ( i \neq j) \; .
\ee

So, the lattice provides the short-range regularization for dipolar interactions. If the 
length of the considered specimen is much smaller than the screening radius, then the 
long-range exponential regularization can also be omitted, since the system length plays 
the role of an effective regularization parameter.

When the lattice is deep, such that the number of atoms in each lattice site is fixed,
say being $\nu_j$, this can be formulated as the condition
$$
 \hat c_j^\dgr \hat c_j = \nu_j \;  .
$$
Then the atomic Hamiltonian (\ref{3.98}) becomes a constant, and, hence, does not
influence the spin part. That is, the total Hamiltonian (\ref{3.97}) reduces to the spin
Hamiltonian (\ref{3.104}). In a deep lattice, the occupation number of lattice sites
$\nu_j$ can be made the same for all sites and, thus, equal to the lattice filling factor,
$$
 \nu_j ~ \ra ~ \nu \equiv \frac{N}{N_L} \;  .
$$

\subsection{Spin dynamics}

The Hamiltonian of dipolar atoms (or molecules) in a deep optical lattice is characterized
by the spin term (\ref{3.104}). If these atoms would possess hyperfine structure, due to
nonzero nuclear spin, then there would also exist the quadratic Zeeman effect \cite{Schiff_118}
caused by a constant magnetic field. But even, if the nuclear spin is zero and hyperfine
structure is absent, quadratic Zeeman effect can be induced by applied laser beams
\cite{Paz_111,Cohen_174,Santos_112,Jensen_175} or microwave dressing
\cite{Gerbier_176,Leslie_126}. Then the spin Hamiltonian, taking account of the
alternating-field quadratic Zeeman effect, reads as
\be
\label{3.106}
 \hat H = -\mu_S \sum_j \bB_j \cdot \bS_j \; + \; q_Z \sum_j ( S_j^z)^2 \; + \;
\frac{1}{2} \sum_{i\neq j} \sum_{\al\bt} \overline D_{ij}^{\al\bt} S_i^\al S_j^\bt \; .
\ee
The coefficient $q_Z$ here is a function of the applied intensity and the detuning, but
independent of the magnetic field \cite{Paz_111,Cohen_174,Santos_112,Jensen_175}.
For example, for $^{52}$Cr, that does not possess hyperfine structure, this coefficient
can reach \cite{Santos_112}
$$
 q_Z \; \sim \; \pm 0.01 \mu_B G ~ \sim ~ \pm 10^{-22} {\rm erg} ~ \sim ~
\pm 10^5 \; \frac{\hbar}{s} \;  ,
$$
although usually it is less than $100 - 400 \hbar$/s \cite{Paz_111}. Anyway, this is a rather
large quantity comparable or even larger than the characteristic dipolar interaction energy
$\rho \mu_S^2$, where $\rho$ is the average atomic density. The mean interatomic distance
in optical lattices is of the order $a \sim (10^{-5} - 10^{-4})$ cm. Then
$\rho \sim (10^{12} - 10^{15})$ cm$^{-3}$. Taking $\mu_S \sim 10 \mu_B$ yields
$$
 \rho \mu_S^2 ~ \sim ~ \left ( 10^{-26} - 10^{-23} \right ) {\rm erg} ~ \sim ~
\left ( 10 - 10^4 \right ) \frac{\hbar}{s} \;  .
$$

Hamiltonian (\ref{3.106}) has the structure similar to that of the system of magnetic
nanomolecules and magnetic nanoclusters
\cite{Yukalov_110,Yukalov_119,Yukalov_120,Yukalov_121,Yukalov_122,Kharebov_123},
with the quadratic Zeeman term analogous to the single-site anisotropy. Therefore the
spin dynamics for this Hamiltonian can be studied in the same way as in the cited papers.

To more efficiently regulate the motion of spins, it is possible to connect the sample
with a resonant electric circuit creating a magnetic feedback field  $H$ described by the
Kirchhoff equation
\be
\label{3.107}
 \frac{dH}{dt} + 2\gm H + \om^2 \int_0^t H(t') \; dt' = - 4\pi \eta_c \;
\frac{dm_x}{dt} \;  ,
\ee
in which $\eta_c \equiv V/V_c$ is the coil filling factor, $V$ being the sample volume,
$V_c$, the coil volume, $\gamma$ is the circuit attenuation, $\omega$, the circuit natural
frequency, and
\be
\label{3.108}
m_x = \frac{1}{V} \sum_{j=1}^{N_L} \mu_S \lgl S_j^x \rgl \;   .
\ee
In the presence of an external magnetic field $B_0$, the total magnetic field becomes
\be
\label{3.109}
\bB = B_0 \bfe_z + H \bfe_x \;   .
\ee
The evolution equations are written for the variables
\be
\label{3.110}
 u \equiv \frac{1}{SN_L} \sum_{j=1}^{N_L} \lgl S_j^- \rgl \; , \qquad 
 w \equiv \frac{1}{SN_L^2} \sum_{i\neq j}^{N_L} \lgl S_i^+ S_j^- \rgl \; , \qquad
 s \equiv \frac{1}{SN_L} \sum_{j=1}^{N_L} \lgl S_j^z \rgl \; ,
\ee
where $S^\pm \equiv S^x \pm iS^y$. The analysis of spin dynamics can be done as in papers
\cite{Yukalov_110,Yukalov_119,Yukalov_120,Yukalov_121,Yukalov_122,Kharebov_123}.

It is important to stress that in order to organize coherent spin motion, that is necessary for
realizing fast spin reversal, the presence of the resonant electric circuit is compulsory. If
this circuit is absent, then dipolar interactions destroy spin coherence and do not allow for
their coherent motion \cite{Yukalov_124,Yukalov_125}.

As a more general case, it is possible to consider a mixture of two different atomic species,
with spins $S$ and $I$, in addition to dipolar interactions, having exchange interactions.
The total spin Hamiltonian for such a mixture has the structure
$$
\hat H = \hat H_S + \hat H_I + \hat H_{SI} \;   ,
$$
with the Hamiltonian for $S$ spins
$$
 \hat H_S = \sum_i \hat H_i^S \; + \; \frac{1}{2} \sum_{i\neq j} \hat H_{ij}^S \; \qquad
\hat H_i^S = -\mu_S \bB \cdot \bS_i + q_Z^S \left ( S_i^z \right )^2 \; ,
$$
$$
\hat H_{ij}^S = \sum_{\al\bt} \overline D_{ij}^{\al\bt} S_i^\al S_j^\bt \; - \;
2J_{ij}^S \; \bS_i \cdot \bS_j 
$$
and for $I$ spins
$$
 \hat H_I = \sum_j \hat H_j^I \; + \; \frac{1}{2} \sum_{i\neq j} \hat H_{ij}^I \; \qquad
\hat H_j^I = -\mu_I \bB \cdot \bI_j + q_Z^I \left ( I_j^z \right )^2 \; ,
$$
$$
\hat H_{ij}^I = \sum_{\al\bt} C_{ij}^{\al\bt} I_i^\al I_j^\bt \; - \;
2J_{ij} \; \bI_i \cdot \bI_j \; .
$$
The dipolar interactions between $S$ spins are defined in (\ref{3.105}). And for $I$ spins,
the dipolar interactions are given by the dipolar tensor
$$
 C_{ij}^{\al\bt} = \frac{\mu_I^2}{a_{ij}^3} \; 
\left ( \dlt_{\al\bt} - 3 n_{ij}^\al n_{ij}^\bt \right ) \exp(-\varkappa_I a_{ij} ) \; .
$$
Interactions between the spins of different species are described by the Hamiltonian
$$
 \hat H_{SI} = \sum_i A \; \bS_i \cdot \bI_i \; + \; 
\sum_{i\neq j} \sum_{\al\bt} A_{ij}^{\al\bt} S_i^\al I_j^\bt \;  ,
$$
with
$$
 A_{ij}^{\al\bt} = \frac{\mu_S \mu_I}{a_{ij}^3} \; 
\left ( \dlt_{\al\bt} - 3 n_{ij}^\al n_{ij}^\bt \right ) \exp(-\varkappa_{SI} a_{ij} ) \;  .
$$
The transverse magnetization, entering the Kirchhoff equation (\ref{3.107}) is
$$
 m_x = \frac{\mu_S}{V} \sum_i \lgl S_i^x \rgl \; + \; 
\frac{\mu_I}{V} \sum_j \lgl I_j^x \rgl \;  .
$$

The possibility of the efficient manipulation of spin dynamics of dipolar gases
in deep optical lattices can be used for quantum information processing.

\section{Spinor interaction potentials}

\subsection{Hyperfine structure}

Many atoms (and molecules), having a nonzero nuclear spin, exhibit hyperfine
structure of their spectra \cite{Cowan_127,Woodgate_128,Brown_129,Demtroder_130}.
The difference between two adjacent hyperfine Zeeman levels is caused by the
influence on the nuclear spin ${\bf I}$ (total nuclear angular momentum) of the atomic
electrons.The total angular momentum of an atom ${\bf F}$ is the sum
$$
\bF = \bJ + \bI
$$
of the total angular momentum of all electrons ${\bf J}$ and of the nuclear spin ${\bf I}$.
The total angular momentum of electrons
$$
\bJ = \bL + \bS
$$
is formed by the electron angular momentum ${\bf L}$ and the total electron spin ${\bf S}$,
$$
 \bL = \sum_i \bL_i \; , \qquad  \bS = \sum_i {\bf s}_i \; .
$$
The hyperfine energy splitting is
$$
 \Dlt W = A_{hyp} \lgl \bI \cdot \bJ \rgl \;  ,
$$
where $A_{hyp}$ is the hyperfine structure constant that is usually determined from
experiments. Rewriting ${\bf J} \cdot {\bf I}$ as
$$
 \bJ \cdot \bI = \frac{1}{2} \; \left ( \bF^2 - \bI^2 - \bJ^2 \right )
$$
and taking into account that
$$
\bF^2 = F(F+1) \; , \qquad  \bJ^2 = J(J+1) \; , \qquad \bI^2 = I(I+1) \; ,
$$
we have
$$
  \Dlt W = \frac{1}{2} \; A_{hyp} ( F(F+1) - J(J+1) - I(I+1) ) \;  .
$$

The total atomic angular momentum is a vector matrix
$$
\bF = F^x \bfe_x + F^y \bfe_y + F^z \bfe_z \;   ,
$$
with the matrix components
$$
 F^\al = [ F_{mn}^\al ] \qquad ( \al = x,y,z ) \;  ,
$$
whose indices take the values $m \equiv m_F = - F, - F +1, \ldots, F$.

The hyperfine structure of different atoms has been described in a number of publications
\cite{Cowan_127,Woodgate_128,Brown_129,Demtroder_130,Frosch_131,Murukawa_132,
Cohen_133,Stevens_134,Walker_135,Reinhardt_136,Frosch_137,Papa_138}. The trapping
of spinor atoms is accomplished in optical traps \cite{Barret_156}.

\subsection{Multicomponent versus fragmented}

Spinor Bose gases can condense. One sometimes name spinor condensates as
fragmented. This however is not correct. A spinor condensate is multicomponent.

By the definition of  Pollock \cite{Pollock_139} and Nozieres and Saint James
\cite{Nozieres_140}, a fragmented condensate is a condensate consisting of several
coexisting condensates distinguished by the collective quantum index labelling the
natural orbitals. The natural orbitals are the eigenfunctions of the first-order density
matrix \cite{Coleman_141} satisfying the eigenproblem
\be
\label{4.1}
 \int \rho(\br,\br') \vp_k(\br') \; d\br' = n_k \vp_k(\br) \;  ,
\ee
where $k$ is the multi-index of collective states and $n_k$ is the occupation number
of the $k$-th state. It may happen that there are several collective states
$k_1, k_2, \ldots$ for which, in thermodynamic limit, one has
\be
\label{4.2}
 \lim_{N\ra \infty} \; \frac{n_{k_i}}{N} > 0 \qquad ( i = 1,2,\ldots ) \;  .
\ee
Then one says that this is a fragmented condensate. If there is just a single
state labelled by $k_0$ satisfying the above limit, this corresponds to the standard
case of a Bose-Einstein condensate.

A spinor gas is characterized by a spinor field operator
$$
\hat\psi(\br) = [ \hat\psi_m(\br) ] \qquad ( m = -F,-F+1,\ldots,F ) \;   ,
$$
with the index $m$ enumerating the $2 F + 1$ hyperfine components. For each
of the components, it is straightforward to define the first-order density matrix
\be
\label{4.3}
\rho_m(\br,\br') ~ \equiv ~ \lgl \hat\psi_m^\dgr(\br') \hat\psi_m(\br) \rgl
\ee
and to study the related eigenproblem
$$
\int \rho_m(\br,\br')\vp_{mk}(\br') \; d\br' = n_{mk} \vp_{mk}(\br) \;  .
$$
If for an $m$-th component there exists the limit
\be
\label{4.4}
 \lim_{N\ra \infty} \; \frac{n_{mk_m}}{N} > 0 \qquad ( - F \leq m \leq F ) \; ,
\ee
then this component is said to be condensed, with the $n_{m k_m}$ number of
atoms in the condensate.

The total first-order density matrix is
\be
\label{4.5}
 \rho(\br,\br')  ~ \equiv ~ \lgl \hat\psi^\dgr(\br') \hat\psi(\br) \rgl 
= \sum_m \rho_m(\br,\br') \; .
\ee
But this density matrix does not satisfy eigenproblem (\ref{4.1}), instead we have
$$
 \sum_m \int \rho_m(\br,\br') \vp_{mk}(\br') \; d\br' = 
\sum_m n_{mk} \vp_{mk}(\br) \; ,
$$
which is not an eigenproblem.

When the density matrix can be written as a sum, like in (\ref{4.5}), this does not
mean that there is fragmentation. In other words, using the representation
$$
\rho_m(\br,\br') = \sum_p n_{mp} \vp_{mp}(\br) \vp^*_{mp}(\br') \;   ,
$$
we can write
$$
\int \rho(\br,\br') \vp_{nk}(\br') \; d\br' =  
\sum_{mp} n_{mp} \vp_{mp}(\br)  (\vp_{mp},\vp_{nk} ) \; .
$$
But this is not an eigenproblem, since the function $\varphi_{nk}$ is not a natural
orbital. Even in the simplified case, that is often met in dealing with spinor gases,
when
$$
 \vp_{mp}(\br) = \zeta_m \vp_p(\br) \; , \qquad ( \vp_k, \vp_p) = \dlt_{kp} \; ,
$$
we get
$$
 \int \rho(\br,\br') \vp_{nk}(\br') \; d\br' = 
\sum_m n_{mk} \vp_{mk}(\br) \zeta_m \zeta^*_n \;  ,
$$
which again is not an eigenproblem.

The core of the trouble is that the index $m$ is not an index of a collective state for
an $N$-particle system, but it is a single-particle index enumerating internal hyperfine
states of separate atoms. Therefore spinor condensates are multicomponent, but not
fragmented.

\subsection{Spinor Hamiltonian}

The Hamiltonian of a system of spinor atoms can be written
\cite{Pethick_3,Kurn_25,Kurn_26,Ueda_27} as a sum of three terms
\be
\label{4.6}
 \hat H = \hat H_0 + \hat H_Z + \hat H_{int} \;  .
\ee
The first term
\be
\label{4.7}
\hat H_0 = \int \hat\psi^\dgr(\br) \left [ - \; 
\frac{\nabla^2}{2m} + U(\br) \right ] \hat\psi(\br) \; d\br
\ee
is the single-atom Hamiltonian not containing the angular moment ${\bf F}$.

The second term is the Zeeman energy operator
\be
\label{4.8}
\hat H_Z = \hat H_{LZ} + \hat H_{QZ}
\ee
including the linear and quadratic Zeeman effects. The linear Zeeman term is
\be
\label{4.9}
 \hat H_{LZ} = - \mu_F 
\int \hat\psi^\dgr(\br) \; \bB \cdot \bF \; \hat\psi(\br) \; d\br \;  ,
\ee
where $\mu_F = - \mu_B g_F$, with $g_F$ being the Land\'e factor. The external
magnetic field, in general, can be a function of spatial and time variables,
${\bf B} = {\bf B}({\bf r}, t)$. And the scalar product ${\bf B} \cdot {\bf F}$ implies
the matrix
$$
 \bB \cdot \bF =\left [ \sum_\al B^\al F_{mn}^\al \right ] \;  .
$$

The quadratic Zeeman effect is described by the Hamiltonian
\be
\label{4.10}
 \hat H_{QZ} = \mp \int \hat\psi^\dgr(\br) \; 
\frac{(\mu_F \; \bB \cdot \bF)^2}{\Dlt W(1+2I)^2} \; \hat\psi(\br) \; d\br \;  ,
\ee
where $\Delta W$ is the hyperfine energy splitting. The sign minus or plus corresponds 
to the relative alignment (parallel or antiparallel) of the nuclear and electronic spin
projections in the considered atom.

The interaction part of the Hamiltonian
\be
\label{4.11}
\hat H_{int} = \hat H_F + \hat H_D
\ee
consists of two terms describing local interactions of atoms with angular momentum
${\bf F}$ and their nonlocal dipolar interactions. Keeping in mind rotationally symmetric
pair collisions in the $s$-wave approximation, the angular momentum of the pair $f$
is to be even \cite{Kurn_25,Kurn_26}. The latter conclusion holds both for bosons
and fermions. The binary collisions of atoms, of angular momentum $F$ each, with
the total angular momentum of the pair $f = 0, 2, \ldots, 2F$, are characterized by the
interaction potential
\be
\label{4.12}
 \Phi_F(\br) = \frac{4\pi}{m} \; \dlt(\br) \sum_f a_f \hat P_f \;  ,
\ee
in which $a_f$ is the scattering length for collisions between two atoms with the total
angular momentum of the pair $f$ and $\hat{P}_f$ is a projection operator onto the
state with the even angular momentum $f = 0, 2, \ldots, 2F$. For atoms with the
angular momentum $F$, there are $2 F + 1$ components corresponding to the angular
momentum projections $- F, - F + 1, \ldots, F$. In the rotationally symmetric $s$- wave
approximation, the system of $2 F + 1$ components is fully described by $F + 1$
scattering lengths $a_f$.

The Hamiltonian of local interactions of atoms with angular momentum $F$ each has
the form
\be
\label{4.13}
 \hat H_F = \frac{1}{2} \sum_{klmn} \int \hat\psi^\dgr_k(\br) \hat\psi^\dgr_l(\br) 
\Phi_{klmn} \hat\psi_m(\br) \hat\psi_n(\br) \; d\br \; ,
\ee
where $\Phi_{klmn}$ is a matrix element of the potential (\ref{4.12}).

As soon as there are nonzero angular momenta, they induce the related dipolar
interactions with the Hamiltonian
\be
\label{4.14}
  \hat H_D = \frac{\mu_F^2}{2} \sum_{klmn} \int \hat\psi^\dgr_k(\br) \hat\psi^\dgr_l(\br') 
\hat\psi_m(\br') \hat\psi_n(\br) \overline D_{klmn}(\br-\br')\; d\br d\br' \; ,
\ee
in which the regularized dipolar interactions are given by the factor $\bar{D}_{klmn}$
that should include the short-range regularization as well as long-range screening,
similarly to dipolar interactions in the previous chapter \cite{Yukalov_43,Yukalov_45}.
As in the previous chapter, the regularized dipolar interaction potential can be written 
as
$$
\overline D_{klmn}(\br) = \Theta(r-b_F) D_{klmn}(\br) \exp(-\varkappa_F r) \; , 
$$
\be
\label{4.15}
D_{klmn}(\br) = 
\frac{(\bF_{kn}\cdot\bF_{lm})-3(\bF_{kn} \cdot\bn)(\bF_{lm}\cdot\bn)}{r^3} \;   ,
\ee
with the appropriate short-range cutoff $b_F$, a long-range screening parameter
$\varkappa_F$, and  the bare dipolar interaction potential $D_{klmn}({\bf r})$,
where ${\bf n} = {\bf r}/ r$. The screening parameter $\varkappa_F$ can be of the
order of the inverse spinor healing length \cite{Baraban_181} $\xi_F \sim 10^{-4}$ cm.

Since
$$
\overline D_{klmn}(0) = 0\;  ,
$$
the order of the field operators in (\ref{4.14}) can be changed, so that to get the
expression
$$
 \hat H_D = \frac{\mu_F^2}{2} \sum_{klmn} \int \hat\psi^\dgr_k(\br) \hat\psi_n(\br) 
\overline D_{klmn}(\br-\br')\hat\psi_l^\dgr(\br') \hat\psi_m(\br')  \; d\br d\br' \;  .
$$

The dipolar interactions are usually smaller than the local spinor interactions.

\subsection{Grand Hamiltonian}

The grand Hamiltonian for a multicomponent system is defined similarly to the
case of a single-component system. The appearance of the condensate in the
$m$-th component is taken into account by the Bogolubov shift of the field
operator
\be
\label{4.16}
 \hat\psi_m(\br) = \eta_m(\br) + \psi_m(\br) \;  ,
\ee
where $\eta_m$ is the condensate function and $\psi_m$ is the field operator of
uncondensed atoms in the $m$-th component. The quantities $\eta_m$ and $\psi_m$
are independent variables, being orthogonal to each other,
\be
\label{4.17}
 \int \eta_m^*(\br)\psi_m(\br) \; d\br = 0 \; .
\ee
The condensate function is the order parameter for the $m$-th component, which
requires that
\be
\label{4.18}
 \eta_m(\br) = \lgl  \hat\psi_m(\br) \rgl \; , \qquad  
\lgl \psi_m(\br) \rgl = 0 \; .
\ee
The number of condensed atoms in the $m$-th component is
\be
\label{4.19}
 N_{0m} = \int | \eta_m(\br) |^2 \; d\br \;  .
\ee
And the number of uncondensed atoms in this component is
\be
\label{4.20}
N_{1m} = \lgl \hat N_{1m} \rgl \; , \qquad
\hat N_{1m} = \int \psi_m^\dgr(\br)\psi_m(\br) \; d\br \;  .
\ee
Condition (\ref{4.18}) can be represented as the equality
\be
\label{4.21}
 \lgl  \hat\Lbd_m \rgl = 0 \;  ,
\ee
in which the operator
$$
 \hat\Lbd_m = \int \left [ \lbd_m(\br) \psi_m^\dgr(\br) + 
\lbd_m^* \psi_m(\br) \right ] \; d\br
$$
eliminates in the grand Hamiltonian the terms linear in the operators $\psi_m$.

Thus the grand Hamiltonian reads as
\be
\label{4.22}
  H = \hat H - \sum_m \left ( \mu_{0m} N_{0m} + \mu_{1m} \hat N_{1m}
+ \hat\Lbd_m \right ) \; ,
\ee
where $\mu_{0m}$ and $\mu_{1m}$ are the Lagrange multipliers guaranteeing the
normalization conditions (\ref{4.19}) and (\ref{4.20}).

The condensate-function equations and the equations of motion for the operators
of uncondensed atoms are defined as
\be
\label{4.23}
 i\; \frac{\prt}{\prt t} \; \eta_m(\br,t) = 
\left \lgl \frac{\dlt H}{\dlt\eta_m^*(\br,t)} \right \rgl \; , 
\qquad
i\; \frac{\prt}{\prt t} \; \psi_m(\br,t) = \frac{\dlt H}{\dlt\psi_m^\dgr(\br,t)}  \; .
\ee

Not all chemical potentials $\mu_{0m}$ and $\mu_{1m}$ are independent. The
components are in chemical equilibrium with each other \cite{Kubo_177}, so that
for the variation of the Gibbs thermodynamic potential one has
$$
 \dlt G = \sum_m \left (  \frac{\dlt G}{\dlt N_{0m}} \; \dlt N_{0m} +
\frac{\dlt G}{\dlt N_{1m}} \; \dlt N_{1m} \right ) = 0 \; .
$$
With the equalities
$$
\mu_{0m} = \frac{\dlt G}{\dlt N_{0m}} \; , \qquad  
\mu_{1m} = \frac{\dlt G}{\dlt N_{1m}} \;  ,
$$
this translates into
$$
 \dlt G = \sum_m ( \mu_{0m} \dlt N_{0m} + \mu_{1m} \dlt N_{1m} ) = 0 \; .
$$
But the variation with respect to the numbers of atoms in the components assumes
the constraints for the given total numbers of condensed and uncondensed atoms
$$
N_0 = \sum_m N_{0m} \; , \qquad  N_1 = \sum_m N_{1m} \;   ,
$$
which implies that
$$
 \dlt N_{00} + \sum_{m\neq 0} \dlt N_{0m} = 0 \; , \qquad 
 \dlt N_{10} + \sum_{m\neq 0} \dlt N_{1m} = 0 \; .
$$
Thus we come to the equation
$$
\sum_{m\neq 0} ( \mu_{0m} - \mu_{00} ) \dlt N_{0m} + 
( \mu_{1m} - \mu_{10} ) \dlt N_{1m} = 0 \; .
$$

Another constraint for an equilibrium system is the conservation of the $z$-component
of the angular momentum, which implies
$$
 \dlt N_{0m} - \dlt N_{0,-m} = 0 \; , \qquad  
\dlt N_{1m} - \dlt N_{1,-m} = 0 \qquad (m\neq 0) \; .
$$
Taking this into account yields
\be
\label{4.24}
 \sum_{m>0} \left [ (\mu_{0m} + \mu_{0,-m} - 2\mu_{00} ) \dlt N_{0m} +
(\mu_{1m} + \mu_{1,-m} - 2\mu_{10} ) \dlt N_{1m} \right ] = 0 \; .
\ee
From here it follows that
\be
\label{4.25}
 \mu_{0m} + \mu_{0,-m} = 2\mu_{00} \; , \qquad  
\mu_{1m} + \mu_{1,-m} = 2\mu_{10} \qquad (m\neq 0) \;   .
\ee

Additional two conditions are defined by the fixed total number of particles $N$ in
the system and by the minimization of thermodynamic potential with respect to $N_0$.

\subsection{Atoms with $F = 1$}

The angular momentum $F$ for atoms with hyperfine structure is often called
hyperfine spin. As an example, the hyperfine spin ${\bf F}$ for $F = 1$ has the
components
\begin{eqnarray}
\label{4.26}
F^x = \frac{1}{\sqrt{2}}\left [ \begin{array}{ccc}
0 & 1 & 0 \\
1 & 0 & 1 \\
0 & 1 & 0 \end{array}
\right ] \; , \quad
F^y = \frac{i}{\sqrt{2}}\left [ \begin{array}{ccc}
0 & -1 & 0 \\
1 & 0 & -1 \\
0 & 1 & 0 \end{array}
\right ] \; , \quad
F^z = \frac{1}{2}\left [ \begin{array}{ccc}
1 & 0 & 0 \\
0 & 0 & 0 \\
0 & 0 & -1 \end{array}
\right ] \; .
\end{eqnarray}
The field operator is the three-component spinor labelled by the index $m = -1, 0, 1$,
\begin{eqnarray}
\hat\psi = \left [ \begin{array}{l}
\hat\psi_1 \\
\hat\psi_0 \\
\hat\psi_{-1} \end{array}
\right ] \; , \qquad \hat\psi_m = \hat\psi_m(\br,t) \; .
\end{eqnarray}

The Hamiltonian part, corresponding to the local spinor interactions, reads as
$$
\hat H_F = \frac{1}{2} \; c_0 \sum_{mn} \int \hat\psi_m^\dgr(\br) \hat\psi_n^\dgr(\br) 
\hat\psi_n(\br) \hat\psi_m(\br) \; d\br \; +
$$
\be
\label{4.28}
 + \; 
\frac{1}{2}\; c_2 \sum_\al \sum_{klmn} F_{kn}^\al F_{lm}^\al 
\int \hat\psi_k^\dgr(\br) \hat\psi_l^\dgr(\br) \hat\psi_m(\br)\hat\psi_n^(\br) \; d\br \; ,
\ee
where
\be
\label{4.29}
 c_0 ~ \equiv ~ \frac{4\pi}{3m} \; (2a_2 + a_0 ) \; , \qquad 
 c_2 ~ \equiv ~ \frac{4\pi}{3m} \; (a_2 - a_0 ) \; .
\ee
Explicitly, this takes the form
$$
\hat H_F = \frac{1}{2} \int \left \{ 
c_0 \hat\psi_0^\dgr \hat\psi_0^\dgr \hat\psi_0 \hat\psi_0 + (c_0 + c_2) \left [
\hat\psi_1^\dgr \hat\psi_1^\dgr \hat\psi_1 \hat\psi_1 + 
\hat\psi_{-1}^\dgr \hat\psi_{-1}^\dgr \hat\psi_{-1} \hat\psi_{-1} +
2\hat\psi_1^\dgr \hat\psi_0^\dgr \hat\psi_1 \hat\psi_0 +
2\hat\psi_{-1}^\dgr \hat\psi_0^\dgr \hat\psi_{-1} \hat\psi_0 \right ] + \right.
$$
\be
\label{4.30}
 + \left. 
2(c_0 - c_2) \hat\psi_1^\dgr \hat\psi_{-1}^\dgr \hat\psi_1 \hat\psi_{-1} +
2c_2 \left ( \hat\psi_0^\dgr \hat\psi_0^\dgr \hat\psi_1 \hat\psi_{-1} +
\hat\psi_1^\dgr \hat\psi_{-1}^\dgr \hat\psi_0 \hat\psi_0 \right ) \right \} \; d\br \; .
\ee
One calls the interactions ferromagnetic, when $c_2 < 0$, and antiferromagnetic
when $c_2 > 0$.

Examples of the Bose atoms with the hyperfine angular momentum $F = 1$ and
their scattering lengths are: lithium \cite{Kurn_26},
$$
a_0 = 23.9 a_B \; , \qquad a_2 = 6.8 a_B \qquad (\; ^7{\rm Li} \; ) \;   ,
$$
sodium \cite{Crubellier_142},
$$
a_0 = (50.0 \pm 1.6) a_B \; , \qquad a_2 = (55.0 \pm 1.7) a_B 
\qquad (\; ^{23}{\rm Na} \; ) \;   ,
$$
potassium \cite{Falke_143,Lysebo_144},
$$
 a_0 = (68.5 \pm 0.7) a_B \; , \qquad a_2 = (63.5 \pm 0.6) a_B 
\qquad (\; ^{41}{\rm K} \; ) \;  ,
$$
and rubidium \cite{Klausen_145,Kempen_146},
$$
a_0 = (101.8 \pm 0.2) a_B \; , \qquad a_2 = (100.4 \pm 0.1) a_B 
\qquad (\; ^{87}{\rm Rb} \; ) \;   .
$$

\subsection{Equilibrium properties}

Bose-Einstein condensation of a spinor gas of sodium atoms $^{23}$Na with $F = 1$
was recently experimentally studied in Ref. \cite{Frapolli_147} taking into account the
linear and quadratic Zeeman effects. Dipolar interactions here can be neglected, since
$\mu_F^2/|c_2| = 0.007$. Therefore the longitudinal magnetization
$M_z = \langle F^z \rangle / N$ along the external magnetic field direction is conserved
by local interatomic interactions. The linear Zeeman effect thus becomes irrelevant for
equilibrium states, since it is proportional to a conserved quantity. Bose-Einstein
condensation was found to occur in steps, when different Zeeman components condense
one at a time. The sequence of the condensation transitions essentially depends on the
quadratic Zeeman energy
$$
 E_{QZ}  = \frac{(\mu_F B)^2}{\Dlt W(1+2I)^2} \;  .
$$
Here $\mu_F^2/ \Delta W \approx 277 \hbar / {\rm s} {\rm G}^2$ \cite{Jacob_165}.
For example, for $E_{QZ} \approx 56 \hbar/s$ and a highly magnetized sample, the
component $m_F = +1$ condenses first at a critical temperature $T_1$, followed by
the component $m_F = 0$ at a lower temperature $T_0 < T_1$. For low magnetization,
the condensation sequence is reversed: first the component $m_F = 0$ condenses
at $T_0$, after which the component  $m_F = +1$ follows at $T_1 < T_0$. For
$E_{QZ} \approx 434 \hbar/s$, there occurs only one sequence for any magnetization,
with the component $m_F = +1$ condensing first and $m_F = 0$, second at a lower
temperature. At the low field, when $E_{QZ} \approx 18 \hbar/s$, for high values of
$M_z$, there happens the condensation of the component $m_F = -1$, while the
$m_F = 0$ component remains uncondensed.

Full theoretical investigations of finite temperature properties for spinor gases seem
to be absent. Starting from papers \cite{Ohmi_148,Ho_149}, one usually considers
the case of zero temperature and weak interactions, when all the system is assumed
to be condensed. Finite temperature properties of spinor gases with $F = 1$ were
discussed in \cite{Isoshima_150,Kao_151,Lang_152} without taking into account
interactions and in \cite{Zhang_153,Tao_154,Kawaguchi_155} neglecting anomalous
averages that, however, can be crucially important for condensed systems.

The ground-state properties of spinor gases and their collective excitations at zero
temperature have been studied for the angular momenta
$F = 1$ ($^7$Li, $^{23}$Na, $^{41}$K, $^{87}$Rb)
\cite{Ohmi_148,Ho_149,Ho_156,Ueda_182,Zhang_157,Yi_158,Gu_183,Kawaguchi_184,
Yi_185,Szabo_159,Murata_160,Kjall_161,Cherng_186,Pietila_162,Makela_163,
Shlyapnikov_164,Symes_166},
$F = 2$ ($^{87}$Rb, $^{85}$Rb, $^{23}$Na) \cite{Koashi_167,Klausen_145,Zheng_169},
and $F = 3$ ($^{52}$Cr) \cite{Santos_170,Diener_171,Makela_172}. The Lee-Huang-Yang
corrections to the Bogolubov approximation for gases with $F = 1$ and $F = 2$ are
considered in Ref. \cite{Uchino_173}.

As an example, let us consider the ground state of $F = 1$ atoms, neglecting dipolar
interactions and ignoring the linear Zeeman effect \cite{Kurn_26,Uchino_173}. Assume
that the external magnetic field is directed along the axis $z$, so that the quadratic
Zeeman term (\ref{4.10}) can be presented as
\be
\label{4.31}
\hat H_{QZ} = \int \hat\psi^\dgr(\br) q_B 
\left ( F^z \right )^2 \hat\psi(\br) \; d\br \;  ,
\ee
with
$$
 q_B = \mp \frac{\mu_F^2 B^2}{\Dlt W(1+2I)^2} \;  .
$$

If we assume that all atoms are condensed and use the single-mode approximation
\be
\label{4.32}
  \hat\psi(\br) ~ \approx ~ \sqrt{\rho(\br) } \; \hat\zeta \; ,
\ee
where $\rho({\bf r})$ is average density and $\hat{\zeta} = [\zeta_n]$ is a column
normalized to unity,
$$
 || \hat \zeta ||  = \sqrt{\sum_n | \zeta_n|^2 } ~ = ~ 1 \; ,
$$
then, depending on the values of the parameters $c_2$ and $q_B$, there exist the
following ground-state phases.

{\it Ferromagnetic phase}, when $| \langle {\bf F} \rangle | = 1$, corresponds to one
of the states
\begin{eqnarray}
\label{4.33}
\hat\zeta_+ = \left [ \begin{array}{c}
1 \\
0 \\
0 \\ \end{array} \right ] \; , \qquad
\hat\zeta_- = \left [ \begin{array}{c}
0 \\
0 \\
1 \\ \end{array} \right ] \qquad ( c_2 < 0, \; \; q_B < 0 ) \; .
\end{eqnarray}

{\it Polar phase}, longitudinal, $\hat{\zeta}_0$, or transverse, $\hat{\zeta}_\perp$,
for which $| \langle {\bf F} \rangle | = 0$, is described by the states
\begin{eqnarray}
\label{4.34}
\hat\zeta_0 = \left [ \begin{array}{c}
0 \\
1 \\
0 \\ \end{array} \right ] \quad (c_2>0,\; \; q_B>0) \; , \qquad
\hat\zeta_\perp = \frac{1}{\sqrt{2}} \left [ \begin{array}{c}
1 \\
0 \\
e^{i\vp} \\ \end{array} \right ] \quad ( c_2 > 0, \; \; q_B < 0 ) \; .
\end{eqnarray}
This phase sometimes is named as antiferromagnetic.

{\it Broken-axisymmetry phase}, with the state
\begin{eqnarray}
\label{4.35}
\hat\zeta = \frac{1}{\sqrt{2}} \left [ \begin{array}{c}
\sin\vartheta \\
\sqrt{2}\;\cos\vartheta \\
\sin\vartheta \\ \end{array} \right ] \qquad ( c_2 < 0, \; \; 0< q_B < -2c_2\rho ) \; ,
\end{eqnarray}   
where $\rho = N/V$ is the average atomic density.

\subsection{Stratification of components}

A spinor gas is a mixture of several components. But a system of components
requires special conditions to represent a real mixture in space. If two components
interact through a local potential
\be
\label{4.36}
\Phi_{ij}(\br) = \Phi_{ij}\dlt(\br) \; , \qquad 
\Phi_{ij} ~\equiv ~ \int \Phi_{ij}(\br) \; d\br \;  ,
\ee
their mixture is stable at zero temperature, provided the stability condition for
the interaction strengths is valid:
\be
\label{4.37}
  \Phi_{ij} < \sqrt{\Phi_{ii}\Phi_{jj} } \qquad (stability, \; T = 0 ) \; ,
\ee
where $i \neq j$. The derivation of this condition can be found, e.g., in \cite{Yukalov_29}.

For example, in a spinor gas with $F = 1$, there are three components labeled
by $m_F = -1, 0, 1$. Consider first the two components, $m_F = 0$ and $m_F = \pm 1$,
neglecting the dipolar interactions. As follows from Hamiltonian (\ref{4.30}), the
interaction strengths between the atoms of the $m_F = 0$ component and between the
atoms of $m_F = \pm 1$ component, respectively,  are
$$
\Phi_{00} =  c_0 \; , \qquad \Phi_{11} = \Phi_{-1,-1} = c_0 + c_2 \;   ,
$$
while the interaction strength between the atoms of different components, $m_F = 0$
and $m_F = \pm 1$, is
$$
 \Phi_{01} = \Phi_{0,-1} = c_0 + c_2 \;  .
$$
If the system is antiferromagnetic (polar), where $c_2 > 0$, then
$$
c_0 + c_2 > \sqrt{c_0(c_0 + c_2)} \qquad ( c_2 > 0 ) \;   ,
$$
which implies
\be
\label{4.38}
 \Phi_{01} > \sqrt{\Phi_{00}\Phi_{11} } \qquad (stratification, \; T = 0 ) \;   .
\ee
Therefore such a system at zero temperature stratifies into spatially separated
components $m_F = 0$ and $m_F = \pm 1$. But these components cannot be
mixed in an equilibrium system.

The situation is different for the mixture of $m_F = 1$ and  $m_F = -1$ components,
for which the corresponding interaction strengths are
$$
 \Phi_{11} = \Phi_{-1,-1} = c_0 + c_2 \; , \qquad \Phi_{-11} = c_0 - c_2 \; .
$$
For a polar system, where $c_2 > 0$, one has
$$
 c_0 - c_2 < c_0 + c_2 \qquad (c_2 > 0 ) \;  ,
$$
meaning that
\be
\label{4.39}
 \Phi_{-11} < \sqrt{\Phi_{11}\Phi_{-1,1} } \qquad (stability, \; T = 0 ) \;  .
\ee
This mixture at zero temperature is stable.

These effects have been observed for the $m_F$ components of sodium, for which
$c_2 > 0$ \cite{Stenger_178,Miesner_179}.

At finite temperature, the stability condition reads \cite{Yukalov_29} as
\be
\label{4.40}
 \Phi_{ij} < \sqrt{\Phi_{ii}\Phi_{jj} } \; + \; \frac{TV}{N_iN_j} \; \Dlt S_{mix}
\qquad ( T > 0 ) \;  ,
\ee
where $i \neq j$, $N_i$ is the number of atoms in the $i$-th component, and
$$
\Dlt S_{mix} = - N_i \ln\; \frac{N_i}{N} \; - \; N_j\ln \; \frac{N_j}{N}
$$
is the entropy of mixing. Hence at finite temperature it is easier to mix different
components.

\subsection{Nonequilibrium properties}

One usually considers the equations of motion of different components, neglecting
the anomalous averages. For $F =1$, this has been done in
\cite{Pu_187,Schmaljohann_188,Cheng_189,Plimak_190,Black_191,Gawryluk_192,
Hoshi_193,Swislocki_194,Deuretzbacher_195}. The evolution of the components of the
spinor gas with $F = 2$ at zero temperature has been studied in Refs.
\cite{Schmaljohann_188,Swislocki_194,Schmaljohann_196,Kronjager_197,Klempt_198}.
Population dynamics of an $F = 1$ Bose gas was also considered for temperatures
above the Bose-Einstein condensation transition \cite{Endo_199,Endo_200}.

The process of population dynamics is dominated by the dipolar interactions of magnetic
spins rather than by the spin-mixing contact potential. This is because the Hamiltonian part,
containing the local hyperfine spin interactions, is invariant with respect to spin rotations
and can be expressed through integrals of motion. For instance, let us consider the
interaction Hamiltonian (\ref{4.28}) for $F = 1$ and employ the single-mode approximation
in the form
\be
\label{4.41}
 \hat\psi_m(\br) = \vp(\br) \hat a_m \qquad ( m = -1,0,1) \;  ,
\ee
where $\hat{a}_m$ is a Bose field operator not depending on the spatial variable and
$\varphi({\bf r})$ is a spin-independent mode function normalized to one,
$$
\int | \vp(\br)|^2 \; d\br = 1\;   .
$$
This form of the single-mode approximation assumes that not all atoms are condensed,
since otherwise $\hat{a}_m$ could not be a Bose operator.

In terms of the operators $\hat{a}_m$, the operator of the total number of atoms is
\be
\label{4.42}
 \hat N = \sum_m \hat a_m^\dgr \hat a_m \; .
\ee
It is possible to introduce the {\it effective total spin operator}
\be
\label{4.43}
 \bS ~ \equiv ~ \sum_{mn} \hat a_m^\dgr \; \bF_{mn} \; \hat a_n \;  .
\ee
Explicitly,  the spin components for $F = 1$ are
$$
S^+ = \sqrt{2}\; \left ( \hat a_1^\dgr \hat a_0 + \hat a_0^\dgr \hat a_{-1}\right ) \; ,
\qquad
S^- = \sqrt{2}\; \left ( \hat a_0^\dgr \hat a_1 + \hat a_{-1}^\dgr \hat a_1\right ) \; ,
\qquad
S^z =  \hat a_1^\dgr \hat a_1 - \hat a_{-1}^\dgr \hat a_{-1} \; ,
$$
where $S^{\pm} \equiv S^x \pm iS^y$. The spin operator satisfies the standard
spin (or angular momentum) algebra, with the commutation relations
$$
 [ S^+,\; S^-] = 2S^z \; , \qquad [ S^z,\; S^\pm] = \pm S^\pm \; .
$$
For the squared spin operator
$$
\bS^2 = S^+ S^- + \left ( S^z \right )^2 - S^z \;   ,
$$
one has
$$
 [ S^\pm,\; \bS^2] = [ S^z,\; \bS^2] = 0 \; .
$$

Using the commutation relations for $\hat{a}_m$, we have the equality
\be
\label{4.44}
\sum_{mnk} \hat a_m^\dgr F_{mn}^\al F_{nk}^\bt \hat a_k = S^\al S^\bt -
\sum_{mnkl} \hat a_m^\dgr \hat a_k^\dgr \hat a_l \hat a_n F_{mn}^\al F_{kl}^\bt \;  .
\ee
Also, we use the property of the angular momentum operators
\be
\label{4.45}
 \sum_k \bF_{mk} \cdot \bF_{kn} = \sum_{\al k } F_{mk}^\al F_{kn}^\al =
 F(F+1) \dlt_{mn} \; .
\ee
In that way, employing the single-mode approximation (\ref{4.41}), the Hamiltonian
$H_F$ in (\ref{4.28}), for $F = 1$, reduces
\cite{Pethick_3,Law_202,Goldstein_203,Pu_204} to the simple form
\be
\label{4.46}
 \hat H_F = \frac{1}{2} \; \overline c_0 \; \hat N (\hat N - 1) +
\frac{1}{2} \; \overline c_2 \; (\bS^2 - 2\hat N) \; ,
\ee
in which
$$
 \overline c_0 ~ \equiv ~ c_0 \int | \vp(\br)|^4 \; d\br \; , \qquad
 \overline c_2 ~ \equiv ~ c_2 \int | \vp(\br)|^4 \; d\br \; .
$$
This Hamiltonian is expressed through the integrals of motion. It does not influence
the evolution of the effective spin, since $[{\bf S},\hat{N}] = 0$. The spin evolution is
governed by the dipolar interactions and the Zeeman terms. The linear Zeeman term
(\ref{4.9}) takes the form
\be
\label{4.47}
 \hat H_{LZ} = -\mu_F \; \bB_{eff} \cdot \bS \;  ,
\ee
in which
\be
\label{4.48}
 \bB_{eff} ~ \equiv ~ \int \bB(\br) | \vp(\br)|^2 \; d\br \;  .
\ee
And the quadratic Zeeman term (\ref{4.10}) that can be rewritten as
\be
\label{4.49}
 \hat H_{QZ} = 
Q_Z \int \hat\psi^\dgr(\br)\; (\bB \cdot \bF)^2\; \hat\psi(\br)\; d\br \; ,
\ee
where
$$
  Q_Z ~ \equiv ~ \mp \; \frac{\mu_F^2}{\Dlt W (1 + 2I)^2} \; ,
$$
becomes
\be
\label{4.50}
  \hat H_{QZ} = Q_Z \sum_{mnk} \sum_{\al\bt} B^{\al\bt} \;
\hat a_m^\dgr \; F_{mn}^\al F_{nk}^\bt \; \hat a_k \;  ,
\ee
with the notation
$$
 B^{\al\bt} ~ \equiv ~ \int B^\al(\br) B^\bt(\br) | \vp(\br)|^2 \; d\br \;  .
$$

Spinor condensates exhibit a rich variety of nonuniform spin structures, such as
magnetic domains and various textures
\cite{Saito_205,Gawryluk_206,Gu_207,Takahashi_208,Matuszhewski_209}. Different
topological excitations can arise, e.g., integer and fractional vortices, monopoles,
skyrmions, and knots \cite{Kurn_26,Marzlin_210,Khawaja_211,Ruostekoski_212}.

An interesting question is the dynamics of equilibration and thermalization of spinor
gases prepared in a strongly nonequilibrium initial state. This problem is of general
interest for finite quantum systems and has been reviewed in
\cite{Dziarmaga_213,Polkovnikov_214,Yukalov_215}. For spinor gases, these topics
are discussed in review \cite{Kurn_26}.

\subsection{Optical lattices}

The field operators of atoms in an optical lattice allow for the expansion in Wannier
functions
\be
\label{4.51}
 \hat\psi_m(\br) = \sum_j \hat c_{jm} w(\br-\ba_j) \;  ,
\ee
where the single-band approximation is assumed, $m = - F, - F + 1, \ldots, F$ is the
hyperfine index, $j = 1, 2, \ldots, N_L$ is the lattice-site index, and ${\bf a_j}$ is a
lattice vector. The Wannier functions are taken to be independent of the hyperfine index.

Substituting this expansion into Hamiltonian (\ref{4.6}), for concreteness, we shall keep
in mind atoms with the angular momentum $F = 1$. The related hyperfine spin is given
explicitly in equation (\ref{4.26}) or can be represented in the compact form through its
components
$$
F_{mn}^x = \frac{1}{\sqrt{2}} \; ( \dlt_{mn-1} + \dlt_{mn+1} ) \; ,
$$
\be
\label{4.52}
F_{mn}^y = \frac{i}{\sqrt{2}} \; ( \dlt_{mn-1} - \dlt_{mn+1} ) \; , \qquad
F_{mn}^z = m\dlt_{mn} \qquad (m,n = -1,0,1) \;  .
\ee
Respectively, the ladder components are
\be
\label{4.53}
 F_{mn}^\pm ~ \equiv ~ F_{mn}^x \pm i F_{mn}^y = \sqrt{2}\; \dlt_{m,n\pm 1} \; .
\ee

The first term, defined in equation (\ref{4.7}), becomes
\be
\label{4.54}
 \hat H_0 = - \sum_{i\neq j} \sum_m J_{ij} \hat c_{im}^\dgr \hat c_{jm} + 
\sum_j \sum_m h_j \hat c_{jm}^\dgr \hat c_{jm} \;  ,
\ee
with the tunneling parameter
\be
\label{4.55}
J_{ij} = - \int w^*(\br -\ba_i) \left [ - \; \frac{\nabla^2}{2m} + U(\br) \right ]
w(\br-\ba_j) \; d\br
\ee
and the local-energy offset due to the external potential, including the lattice and
confining potentials,
\be
\label{4.56}
  h_j = \int w^*(\br -\ba_j) \left [ - \; \frac{\nabla^2}{2m} + U(\br) \right ]
w(\br-\ba_j) \; d\br \; .
\ee

The linear Zeeman term (\ref{4.9}) is
\be
\label{4.57}
 \hat H_{LZ} = - \mu_F \sum_{ij} 
\sum_{mn} \bB_{ij} \cdot \bF_{mn} \hat c_{im}^\dgr \hat c_{jn} \;  ,
\ee
where
\be
\label{4.58}
 \bB_{ij} ~ \equiv ~ \int w^*(\br -\ba_i) \bB(\br)  w(\br -\ba_j)\; d\br \;  .
\ee
The quadratic Zeeman term (\ref{4.10}) transforms into
\be
\label{4.59}
 \hat H_{QZ} = Q_Z \sum_{ij} \sum_{mnk} 
\sum_{\al\bt} B_{ij}^{\al\bt}\; \hat c_{im}^\dgr \; F_{mn}^\al F_{nk}^\bt \; \hat c_{jk} \;  ,
\ee
with the notation
\be
\label{4.60}
 B_{ij}^{\al\bt} ~ \equiv ~ 
\int w^*(\br -\ba_i) B^\al(\br) B^\bt(\br) w(\br -\ba_j)\; d\br \;  .
\ee

Taking into account that the hyperfine spin-mixing interaction is short-range, it is
admissible to retain in Hamiltonian (\ref{4.28}) only the single-site terms, getting
\be
\label{4.61}
 \hat H_F = \frac{\overline c_0}{2} 
\sum_j \sum_{mn} \hat c_{jm}^\dgr \hat c_{jn}^\dgr \hat c_{jn} \hat c_{jm} \; + \;
\frac{\overline c_2}{2} \sum_j \sum_{mnkl} 
\hat c_{jm}^\dgr \hat c_{jk}^\dgr\; \bF_{mn}\cdot\bF_{kl}\; \hat c_{jl} \hat c_{jn} \; ,
\ee
where
\be
\label{4.62}
\overline c_0 ~ \equiv ~ c_0 \int | w(\br-\ba_j)|^4 \; d\br \; , \qquad 
\overline c_2 ~ \equiv ~ c_2 \int | w(\br-\ba_j)|^4 \; d\br \;  .
\ee

The dipolar Hamiltonian (\ref{4.14}) takes the form
\be
\label{4.63}
 \hat H_D = \frac{\mu_F^2}{2} \sum_{ijfg} \sum_{klmn} \overline D_{klmn}^{ijfg}
\hat c_{ik}^\dgr \hat c_{jl}^\dgr \hat c_{fm} \hat c_{gn} \;  ,
\ee
with the notation
\be
\label{4.64}
 \overline D_{klmn}^{ijfg} ~ \equiv ~ \int w^*(\br-\ba_i) w^*(\br'-\ba_j)
w(\br'-\ba_f) w(\br-\ba_g) \overline D_{klmn}(\br-\br') \; d\br d\br' \; ,
\ee
in which $\overline D_{klmn}$ is defined in equation (\ref{4.15}).

Choosing the Wannier functions to be well localized \cite{Marzari_47} makes
it possible to get
$$
\overline D_{klmn}^{ijfg} = \dlt_{ig} \dlt_{jf} \overline D_{klmn}^{ij} \; , 
$$
\be
\label{4.65}
\overline D_{klmn}^{ij} ~ \equiv ~ 
\int | w(\br-\ba_i)|^2 | w(\br'-\ba_j)|^2 \overline D_{klmn}(\br-\br') \; d\br d\br' \; .
\ee

Depending on the lattice depth, thermodynamic parameters, and atomic interactions,
the system can be in an insulating state or in superfluid state
\cite{Cheng_216,Rizzi_217,Pai_218,Wagner_219}.

\subsection{Insulating lattice}

If the optical lattice is sufficiently deep, such that there are no jumps of atoms between
different lattice sites, when $J_{ij} = 0$, the system is in an insulating state. Then
Hamiltonian (\ref{4.54}) reads as
\be
\label{4.66}
\hat H_0 = \sum_j h_j \hat n_j \;   ,
\ee
with the operator
\be
\label{4.67}
\hat n_j ~ \equiv ~ \sum_m \hat c_{jm}^\dgr \hat c_{jm}
\ee
for the number of atoms in a $j$-th lattice site.

Assuming that the external magnetic field is slowly varying in space, as compared to
the variation of well-localized Wannier functions, we have
\be
\label{4.68}
  \bB_{ij} ~ \cong ~ \dlt_{ij}\bB(\ba_j) ~ \equiv ~ \dlt_{ij}\bB_j \; .
\ee
In particular, when the external magnetic filed is constant in space, then we have
the exact equality
$$
 \bB_{ij} = \dlt_{ij}\bB \qquad (\bB = const) \;  .
$$
In this case, the linear Zeeman term (\ref{4.57}) can be written as
\be
\label{4.69}
  \hat H_{LZ} = -\mu_F \sum_j \bB_j \cdot \bS_j \; ,
\ee
where the effective spin operator is
\be
\label{4.70}
 \bS_j ~ \equiv ~ \sum_{mn}   \hat c_{jm}^\dgr \; \bF_{mn} \; \hat c_{jn} \; .
\ee

Under the condition that the external magnetic field is slowly varying, as compared
to the variation of Wannier functions, from notation (\ref{4.60}), we find
\be
\label{4.71}
 B_{ij}^{\al\bt}  ~ \cong ~ \dlt_{ij} B_i^\al B_j^\bt \;  .
\ee
For a constant in space magnetic field, one has exactly
$$
B_{ij}^{\al\bt} = \dlt_{ij} B^\al B^\bt  \qquad ( \bB = const ) \;   .
$$
Then the quadratic Zeeman term (\ref{4.59}) takes the form
\be
\label{4.72}
\hat H_{QZ} = Q_Z \sum_j \sum_{mnk} (\bB_j\cdot\bF_{mn})  (\bB_j\cdot\bF_{nk}) 
\hat c_{jm}^\dgr \hat c_{jk} \; .
\ee

Using the commutation relations of the operators $\hat{c}_{jm}$, we have the equality
\be
\label{4.73}
\sum_{mnk}  \hat c_{jm}^\dgr \; F_{mn}^\al F_{nk}^\bt \; \hat c_{jk} = 
S_j^\al S_j^\bt - \sum_{mnkl} \hat c_{jm}^\dgr \hat c_{jk}^\dgr \;
F_{mn}^\al F_{kl}^\bt \; \hat c_{jl} \hat c_{jn} \; .
\ee
We may notice that the term
\be
\label{4.74}
S_j^\al S_j^\bt = \sum_{mnkl} 
\hat c_{jm}^\dgr \; F_{mn}^\al\; \hat c_{jn} \hat c_{jk}^\dgr\; F_{kl}^\bt \;\hat c_{jl} 
\ee
describes a consecutive destruction and creation of a single atom. While the second
term in the right-hand site of equality (\ref{4.73}) corresponds to the simultaneous
destruction and creation of a pair of atoms. Such two-particle processes are usually
less important than the single-particle ones and can be neglected. For instance, if the
number of atoms in a lattice site is strictly one, then $\hat{c}_{jm} \hat{c}_{jn} = 0$.
In that way, for the quadratic Zeeman term, we can write
\be
\label{4.75}
 \hat H_{QZ} = Q_Z \sum_j (\bB_j\cdot\bS_j)^2 \; ,
\ee
where
\be
\label{4.76}
 Q_Z = \mp \; \frac{\mu_F^2}{\Dlt W(1+2I)^2} \;  .
\ee
Here $\Delta W$ is the hyperfine energy splitting.

Using equality (\ref{4.73}) transforms the spin-mixing Hamiltonian (\ref{4.61}) to the form
\be
\label{4.77}
 \hat H_F = \frac{\overline c_0}{2} \sum_j \hat n_j(\hat n_j -1 ) + 
\frac{\overline c_2}{2} \sum_j \left ( \bS_j^2 - 2\hat n_j \right ) \; .
\ee

The dipolar part (\ref{4.63}) for a deep insulating lattice, taking into account that the
dipolar potential is regularized by a short-range cutoff, because of which
$\bar{D}_{klmn}^{jj} = 0$, can be written as
\be
\label{4.78}
 \hat H_D = \frac{\mu_F^2}{2} \sum_{i\neq j} \sum_{klmn} \overline D_{klmn}^{ij}
 \hat c_{ik}^\dgr \hat c_{in} \hat c_{jl}^\dgr \hat c_{jm} \; .
\ee
Employing the same procedure as in Chapter 3, Hamiltonian (\ref{4.78}) can be
represented as
\be
\label{4.79}
 \hat H_D = \frac{1}{2} 
\sum_{i\neq j} \sum_{\al\bt} \overline D_{ij}^{\al\bt} S_i^\al S_j^\bt \;  ,
\ee
with the interaction potential
\be
\label{4.80}
\overline D_{ij}^{\al\bt} = \mu_F^2 \int |w(\br-\ba_i)|^2 |w(\br'- \ba_j)|^2 
\overline D^{\al\bt}(\br-\br') \; d\br d\br' \;  ,
\ee
in which
$$
 \overline D^{\al\bt}(\br) = \Theta(r-b_F) D^{\al\bt}(\br)\exp(-\varkappa_F r) \; ,
$$
$$
D^{\al\bt}(\br) = \mu_F^2 \; \frac{\dlt_{\al\bt}-3n^\al n^\bt}{r^3} 
\qquad
\left ( \bn \equiv \frac{\br}{r} \right ) \;  .
$$
For well localized atoms, potential (\ref{4.80}) simplifies to the expression
\be
\label{4.81}
 \overline D_{ij}^{\al\bt} = 
\Theta(a_{ij}-b_F) D_{ij}^{\al\bt}\exp(-\varkappa_F|\ba_{ij}|) \;  ,
\ee
with the dipolar tensor
\be
\label{4.82}
D_{ij}^{\al\bt} = 
\frac{\mu_F^2}{a_{ij}^3} \; (\dlt_{\al\bt}-3n_{ij}^\al n_{ij}^\bt) \; ,
\ee
where
$$
 \bn_{ij} ~ \equiv ~ \frac{\ba_{ij}}{|\ba_{ij}|} \; , \qquad
\ba_{ij} ~ \equiv ~ \ba_i - \ba_j \;  .
$$

The short-range cutoff $b_F$ is of the order of the effective size of an atom, which is
usually smaller than the lattice spacing. Hence the factor $\Theta(a_{ij} - b_F)$ can be
omitted, being effectively taken into account by the summation with $i \neq j$. The
screening parameter $\varkappa_F$ is of the order of the inverse spin healing length.
The latter can be defined by equating the effective kinetic energy $\hbar^2 / 2 m \xi_F^2$
with the effective potential energy of spin interactions $\rho \mu_F^2$, which gives
the spin healing length
$$
 \xi_F = \frac{\hbar}{\sqrt{2m\rho\mu_F^2}} \;  .
$$
The typical lattice spacing for optical lattices is $a \sim (10^{-5} - 10^{-4})$ cm, with the
average density $\rho \sim (10^{12} - 10^{15})$ cm$^{-3}$. Assuming $\mu_F \sim 10 \mu_B$
gives the typical energy $\rho \mu_F^2 \sim (10^{-26} - 10^{-23})$ erg. Then the spin healing
length is $\xi_F \sim (10^{-5} - 10^{-3})$ cm, which is close to the mean interatomic distance.
Therefore taking into account the dipolar interactions of only nearest neighbors is a good
approximation.

Summarizing, Hamiltonian (\ref{4.6}) can be considered as being composed of three
parts
\be
\label{4.83}
 \hat H = \hat H_L + \hat H_Z + \hat H_D \;  .
\ee
The first term $\hat{H}_L = \hat{H}_0 + \hat{H}_F$,
\be
\label{4.84}
 \hat H_L = \sum_j \left [ h_j \hat n_j + 
\frac{\overline c_0}{2} \; \hat n_j ( \hat n_j - 1 )
+  \frac{\overline c_0}{2} \; \left ( \bS_j^2 - 2\hat n_j \right ) \right ] \; ,
\ee
includes local spinor interactions. This term is invariant with respect to spin rotations
and, since
$$
[\bS_j , \; \hat n_j]  = [\bS_j , \; \hat H_L]  = 0 \; ,
$$
it does not influence spin motion.

Spin motion is governed by the Zeeman - effect part
\be
\label{4.85}
\hat{H}_Z = \sum_j \left [ - \mu_F \bB_j\cdot\bS_j + Q_Z (\bB_j \cdot \bS_j)^2
+q_Z (S_j^z)^2 \right ]
\ee
and by the dipolar Hamiltonian (\ref{4.79}). The Zeeman term consists of the linear
Zeeman part, the quadratic Zeeman part caused by an external magnetic field and
by the alternating-current quadratic Zeeman term.

The Hamiltonian similar to expression (\ref{4.83}), but without the quadratic Zeeman
term, has been derived in Refs. \cite{Pu_220,Gross_221,Zhang_222,Pu_223}.
However, the quadratic Zeeman effect is an essential phenomenon in the physics
of spinor atoms.

The alternating-current quadratic Zeeman term, due to the alternating-current Stark shift,
can be induced either by applying off-resonance linearly polarized light, populating the
internal spin states and exerting the quadratic shift along the light polarization axis
\cite{Paz_111,Cohen_174,Santos_112,Jensen_175}, or by a linearly polarized
microwave driving field \cite{Gerbier_176,Leslie_126,Bookjans_224}. Here we assume
that the light polarization axis is the axis $z$. The coefficient $q_Z$ is independent of
the non-driving magnetic field and depends only on the intensity and detuning of the
quasi-resonance driving field. The alternating-field induced quadratic Zeeman shift can
be tailored at high resolution and rapidly adjusted using electro-optics. By employing
either positive or negative detuning the sign of $q_Z$ can be varied. The quasi-resonance
quadratic Zeeman effect makes it possible to get $q_Z$ of the order $(100 - 10^5) \hbar/$s.

\subsection{Spin dynamics}

Considering weak deviations of spins from their equilibrium positions, one can study such
phenomena as spin echo \cite{Yasunaga_225} and ferromagnetic resonance
\cite{Yasunaga_226} in spinor gases. In these studies, one usually does not take into
account the quadratic Zeeman effect.

Here we shall describe how one could consider strongly nonequilibrium spin states and
spin motion in the presence of the quadratic Zeeman effect. Also, we take into account
the existence of a longitudinal and transverse external magnetic fields, so that the total
magnetic field at site $j$ is
\be
\label{4.86}
\bB_j = B_0\bfe_z + H\bfe_x \;   .
\ee
The transverse field can be created by a resonant magnetic coil, as in equation (\ref{3.107}),
which will allow for the efficient regulation of spin motion.

From the spin-operator components $S_j^\alpha$, we pass to the ladder operators using
the relations
$$
 S_j^x = \frac{1}{2} \left ( S_j^+ + S_j^- \right ) \; , \qquad 
 S_j^y = -\;\frac{i}{2} \left ( S_j^+ - S_j^- \right ) \;  .
$$

Then the Zeeman Hamiltonian (\ref{4.85}) becomes
$$
\hat H_Z = \sum_j \left \{ - \mu_F B_0 S_j^z -\; 
\frac{\mu_F H}{2} \; \left ( S_j^+ + S_j^- \right ) + 
( Q_Z B_0^2 + q_Z) ( S_j^z)^2 + \right.
$$
\be
\label{4.87}
\left.
  + \frac{Q_ZH^2}{4} \left [ ( S_j^+)^2 + ( S_j^-)^2
+ S_j^+ S_j^- +  S_j^-  S_j^+ \right ] + \frac{Q_ZB_0H}{2 } \left [ ( S_j^+ +  S_j^-)S_j^z +
S_j^z ( S_j^+ +  S_j^-) \right ] \right \} \; ,
\ee
where $\mu_F = - g_F \mu_B$ and the equality
$$
\bB_j \cdot \bS_j = \frac{H}{2} ( S_j^+ +  S_j^-) + B_0 S_j^z
$$
is used.

The dipolar term (\ref{4.79}) can be written in the form
\be
\label{4.88}
 \hat H_D = \frac{1}{2} \sum_{i\neq j} \left \{ 
\frac{a_{ij}}{2} \left ( 2 S_i^z S_j^z - S_i^+ S_j^- \right ) + 
b_{ij}  S_i^+ S_j^+ + b_{ij}^*  S_i^- S_j^- + 2c_{ij}  S_i^+ S_j^z
+ 2c_{ij}^*  S_i^- S_j^z \right \} \;  ,
\ee
in which
$$
a_{ij} ~ \equiv ~ \overline D_{ij}^{zz} \; , \qquad
b_{ij} ~ \equiv ~ \frac{1}{4} \left ( \overline D_{ij}^{xx} - \overline D_{ij}^{yy} 
-2i \overline D_{ij}^{xy} \right ) \; , \qquad
c_{ij} ~ \equiv ~ \frac{1}{2} \left ( \overline D_{ij}^{xz} - i \overline D_{ij}^{yz}
\right ) \; .
$$

The equations of motion
$$
i\; \frac{d}{dt} \; S_j^\pm =  [ S_j^\pm , \; H] \; , \qquad 
i\; \frac{d}{dt} \; S_j^z =  [ S_j^z , \; H] \; ,
$$
are derived by employing the commutation relations
$$
[ S_i^+ , \;  S_j^-]  = 2\dlt_{ij} S_j^z \; , \qquad 
[ S_i^z , \;  S_j^\pm]  = \pm \dlt_{ij} S_j^\pm \; .
$$
Recall that in the Hamiltonian (\ref{4.83}) only the Zeeman (\ref{4.87}) and dipolar
(\ref{4.88}) terms contribute to the equations of motion, since the spin-mixing part
$H_L$ commutes with the spin operator.

In what follows, we shall need the notations for the Zeeman frequency
\be
\label{4.89}
\om_0 ~ \equiv ~ - \mu_F B_0 > 0
\ee
and for the quadratic Zeeman parameter
\be
\label{4.90}
  D ~ \equiv ~ Q_Z B_0^2 + q_Z \; .
\ee

Dipolar interactions enter the evolution equations through the local fluctuating fields
\be
\label{4.91}
\xi_D ~ \equiv ~ \sum_j \left ( a_{ij} S_j^z + c_{ij}^* S_j^- + c_{ij} S_j^+ \right )
\ee
and
\be
\label{4.92}
 \vp_D ~ \equiv ~ \sum_j \left ( \frac{a_{ij}}{2}\; S_j^- 
-2 c_{ij} S_j^z -2 b_{ij} S_j^+ \right ) \; .
\ee
Here the index $j$ runs over all lattice sites. The self-interaction term is zero because
the dipolar interaction potential is regularized, such that $\overline D_{jj}^{\al\bt}=0$.

Since for a macroscopic sample, for which boundary effects are negligible,
$$
 \sum_j \overline D_{ij}^{\al\bt} = 0 \;  ,
$$
we have
$$
 \sum_j a_{ij} =  \sum_j b_{ij} = \sum_j c_{ij} = 0 \; .
$$
Therefore, assuming that the average $\langle S_j^\alpha \rangle$ does not depend
on the site index $j$, we see that the fluctuating dipolar fields on the average are
zero centered.
$$
  \lgl \xi_D \rgl = \lgl \vp_D \rgl = 0 \; .
$$

Also, we define the effective force
\be
\label{4.93}
 f  ~ \equiv ~ - i ( \mu_F H + \vp_D ) \;  .
\ee

Finally, we come to the Heisenberg equations of motion for the ladder operator,
$$
\frac{d S_j^-}{dt} = - i (\om_0 +\xi_D) S_j^- + f S_j^z -
iD ( S_j^- S_j^z + S_j^z S_j^- ) +
$$
$$
 + \frac{i}{2} \; Q_Z H^2 \left [ (S_j^+ + S_j^-) S_j^z
 + S_j^z (S_j^+ + S_j^-) \right ] -
$$
\be
\label{4.94}
 - \; 
\frac{i}{2} \; Q_Z B_0 H \left [ S_j^+  S_j^- + S_j^- S_j^+ -4(S_j^z)^2
 + 2 ( S_j^-)^2 \right ] \; ,
\ee
and for the $z$-component of the spin operator,
$$
\frac{d S_j^z}{dt} = -\; \frac{1}{2} ( f^+ S_j^- + S_j^+ f) + 
\frac{i}{2} \; Q_Z H^2 \left [ (S_j^-)^2 - (S_j^+)^2 \right ] +
$$
\be
\label{4.95}
  + \; 
\frac{i}{2} \; Q_Z B_0 H \left [ (S_j^- - S_j^+) S_j^z + S_j^z (S_j^- - S_j^+) \right ] \;  .
\ee
The equation for $S_j^+$ follows from the Hermitian conjugation of equation (\ref{4.94}).

\subsection{Stochastic quantization}

Averaging the equations of motion (\ref{4.94}) and (\ref{4.95}), we consider the following
functions: the {\it transition function}
\be
\label{4.96}
  u ~ \equiv ~ \frac{1}{S N_L} \sum_{j=1}^{N_L} \lgl S_j^- \rgl \; ,
\ee
describing the mean spin rotation of the transverse spin component, the
{\it coherence intensity}
\be
\label{4.97}
w ~ \equiv ~ \frac{1}{S N_L(N_L-1)} \sum_{i\neq j}^{N_L} \lgl S_i^+ S_j^- \rgl \;   ,
\ee
showing the level of coherence in the spin motion, and the {\it spin polarization}
\be
\label{4.98}
 s ~ \equiv ~ \frac{1}{S N_L} \sum_{j=1}^{N_L} \lgl S_j^z \rgl \;  ,
\ee
defining the average magnetic polarization per atom. Here $S$ implies the maximal
eigenvalue of $S_j^z$.

In the process of averaging the equations of motion, there is the necessity of
decoupling the spin-spin correlation functions. The natural decoupling seems to
be the mean-field approximation
\be
\label{4.99}
 \lgl S_i^\al S_j^\bt \rgl  = \lgl S_i^\al \rgl \lgl S_j^\bt \rgl 
\qquad (i \neq j) \; .
\ee
There are, however, some problems when using this decoupling. The first problem is
that decoupling (\ref{4.99}) can be used for different lattice sites, but it is not valid
for coinciding sites, if $S$ is arbitrary. This is because there happen such expressions
$$
S_j^\al S_j^\bt + S_j^\bt S_j^\al = 0 \qquad \left ( S = \frac{1}{2} \right ) \; ,
$$
that become zero for $S = 1/2$. The general decoupling, taking into account this case
reads \cite{Yukalov_110,Yukalov_119,Yukalov_227} as
\be
\label{4.100}
\lgl  S_j^\al S_j^\bt + S_j^\bt S_j^\al   \rgl  =
\left ( 2 - \; \frac{1}{S} \right ) 
  \lgl S_j^\al \rgl \lgl S_j^\bt \rgl \;   .
\ee
This decoupling correctly interpolates between the quantum case of $S = 1/2$ and
$S \ra \infty$, when spins behave classically.

The other problem is that the mean-filed approximation is actually a semi-classical
approximation that can be employed for treating the coherent spin motion, although it
does not correctly describe the initial quantum stage of spin motion, when coherence has
not been imposed on the system. In the semi-classical approximation, there is no spin
motion at all, if the initial transverse projection is zero, that is, when $u(0) = 0$. Thus
the semi-classical approximation cannot characterize the development of coherence
and hence, to correctly define the delay time for the start of the coherent stage of motion.
This problem can be overcome by using the stochastic quantization
\cite{Yukalov_108,Yukalov_109,Yukalov_110,Yukalov_119,Yukalov_227}, when the local
fluctuating fields (\ref{4.91}) and (\ref{4.92}) are treated as stochastic variables. Note that
field (\ref{4.91}) is real, while (\ref{4.92}) is complex. These variables are assumed to be
zero-centered and representing white noise, so that their mutual correlations are given by
the stochastic averaging
$$
\lgl\lgl \xi_D(t) \rgl\rgl = \lgl\lgl \vp_D(t) \rgl\rgl = 0 \; , \qquad
\lgl\lgl \xi_D(t) \xi_D(t') \rgl\rgl = 2\gm_3\dlt(t-t') \; ,
$$
\be
\label{4.101}
 \lgl\lgl \xi_D(t) \vp_D(t') \rgl\rgl = \lgl\lgl \vp_D(t) \vp_D(t') \rgl\rgl = 0 \; ,
\qquad
\lgl\lgl \vp_D^*(t) \vp_D(t') \rgl\rgl = 2\gm_3\dlt(t-t') \; .
\ee
Here $\gamma_3$ is the attenuation caused by fluctuating dipolar fields, hence it is of
the order of $\rho \mu_F^2 S$.

To take into account spin correlations above the mean field, we introduce the transverse
attenuation $\gamma_2 = \rho \mu_F^2 \sqrt{S(S+1)}$ that is defined in the second-order
perturbation theory \cite{Haar_228}. Generally, there also exists the longitudinal attenuation
$\gamma_1$ that comes about because the Zeeman energy can be transferred to other
degrees of freedom.

In the equations of motion, the quadratic Zeeman parameter (\ref{4.90}) plays the role of
an effective anisotropy shifting the Zeeman frequency to the quantity
\be
\label{4.102}
\om_s ~ \equiv ~ \om_0 + \om_D s \; , \qquad \om_D  ~ \equiv ~ (2S - 1) D \;   .
\ee

Finally, we obtain the equation for the transition function,
$$
\frac{du}{dt} = - i ( \om_s + \xi_D - i\gm_2) u + fs \; +
$$
\be
\label{4.103}
  + \; \frac{i}{2}\; (2S-1)Q_ZH^2(u^* + u) s \; - \;
\frac{i}{2} \; (2S-1) Q_Z B_0 H ( w - 2s^2 + u^2 ) \;  ,
\ee
coherence intensity,
$$
\frac{dw}{dt} = - 2\gm_2 w + (u^* f + f^* u ) s +
$$
\be
\label{4.104}
 + \; \frac{i}{2}\; (2S-1)Q_ZH^2\left [ (u^*)^2 - u^2 \right ] s  +
i(2S-1)Q_Z B_0 H ( u^* - u ) s^2 \;,
\ee
and for the spin polarization,
$$
\frac{ds}{dt} = - \; \frac{1}{2}\; ( u^* f + f^* u ) \; + 
$$
\be
\label{4.105}
+\; \frac{i}{4}\; (2S-1) Q_ZH^2\left [ u^2 - (u^*)^2 \right ] \;
+ \; \frac{i}{2}\; (2S-1) Q_Z B_0 H ( u - u^* ) s  - \gm_1 ( s - s_\infty ) \;  ,
\ee
where $s_\infty$ is the equilibrium spin polarization.

These equations are to be complemented by the definition of the transverse magnetic
field $H$. Here we keep in mind that the field $H$ is a resonator feedback field created
by a resonant magnetic coil surrounding the sample. Then the equation for $H$ follows
from the Kirhhoff equation giving
\cite{Yukalov_108,Yukalov_109,Yukalov_110,Yukalov_119,Yukalov_122,Yukalov_227}
\be
\label{4.106}
 \frac{dH}{dt} + 2\gm H + \om^2 \int_0^t H(t') \; dt' = - 4\pi\eta_c\;
\frac{dm_x}{dt} \;  ,
\ee
where $\gamma$ is the resonator attenuation, $\omega$ is the resonator natural
frequency, $\eta_c = V/V_c$ is the coil filling factor and
\be
\label{4.107}
 m_x = \frac{\mu_F}{V} \sum_j \lgl S_j^x \rgl \;  .
\ee

The equation for $H$ can be represented in the integral form
\be
\label{4.108}
 H = - 4\pi\eta_c \int_0^t G(t-t') \dot{m}_x(t') \; dt' \;  ,
\ee
with the source
\be
\label{4.109}
\dot{m}_x = \frac{1}{2} \; \rho \mu_F S \; \frac{d}{dt} \; ( u^* + u )
\ee
and the transfer function
$$
 G(t) = \left [ \cos(\overline\om t) \; - \; 
\frac{\gm}{\overline\om} \; \sin(\overline\om t) \right ]
\exp(-\gm t) \;  ,
$$
where
$$
 \overline\om ~ \equiv ~ \sqrt{\om^2 - \gm^2} \;  .
$$

The effective interaction energy of the feedback field $H$ with the sample is proportional
to the parameter
\be
\label{4.110}
 \gm_0 ~ \equiv ~ \pi \eta_c \rho \mu_F^2 S \;  .
\ee

\subsection{Scale separation}

The efficient coupling of an electric circuit with the sample occurs only in the resonant case,
when the natural frequency of the circuit is tuned close to the Zeeman frequency of spins,
\be
\label{4.111}
 \left | \frac{\Dlt}{\om} \right | \ll 1 \qquad (\Dlt \equiv \om-\om_0 ) \;  .
\ee
The quadratic Zeeman effect shifts the effective frequency to the value (\ref{4.102}). In order
not to spoil the above quasi-resonance condition, this quadratic shift should be small, such
that
\be
\label{4.112}
 | A | \ll 1 \qquad \left ( A \equiv \frac{\om_D}{\om_0} \right ) \;  .
\ee
All attenuations are to be small as compared to the Zeeman frequency,
\be
\label{4.113}
  \frac{\gm}{\om_0} \ll 1 \; , \qquad   \frac{\gm_0}{\om_0} \ll 1 \; , \qquad 
 \frac{\gm_1}{\om_0} \ll 1 \; , \qquad  \frac{\gm_2}{\om_0} \ll 1 \; , \qquad 
 \frac{\gm_3}{\om_0} \ll 1 \;  .
\ee
And the quadratic Zeeman effect is assumed not to destroy the resonance situation, so that
the inequalities
\be
\label{4.114}
\left |  \frac{Q_Z H^2}{\om_0} \right | \ll 1 \; , \qquad 
\left | \frac{Q_Z B_0 H}{\om_0} \right | \ll 1
\ee
are valid.

Under these conditions, equation (\ref{4.108}) can be solved by the iterative procedure
starting with $u \propto \exp(- \omega_s t)$. After one iteration, we find
\be
\label{4.115}
 \mu_F H = i \left ( u X - X^* u^* \right ) \;  ,
\ee
with the coupling function
\be
\label{4.116}
 X = \gm_0 \om_s \left [ \frac{1-\exp\{-i(\om-\om_s)t-\gm t\}}{\gm+i(\om-\om_s)} +
 \frac{1-\exp\{-i(\om+\om_s)t-\gm t\}}{\gm-i(\om+\om_s)} \right ] \; .
\ee

The first, resonant, term here prevails over the second, nonresonant one, because of which
\be
\label{4.117}
 X ~ \cong ~ \gm_0 \om_s \; \frac{1-\exp(-i\Dlt_s t-\gm t)}{\gm+i\Dlt_s} \;  ,
\ee
where the dynamic detuning
\be
\label{4.118}
\Dlt_s ~ \equiv ~ \om-\om_s = \om-\om_0(1 + As)
\ee
is defined.

Introducing the dimensionless coupling parameter
\be
\label{4.119}
  g  ~ \equiv ~ \frac{\gm_0\om_0}{\gm\gm_2} \;  ,
\ee
we can represent the real part of the coupling function as
\be
\label{4.120}
\al ~ \equiv ~ {\rm Re} X = g \; \frac{\gm^2\gm_2}{\gm^2+\Dlt_s^2} \; ( 1 + As)
\left\{ 1 - \left [ \cos(\Dlt_s t) - \; 
\frac{\Dlt_s}{\gm}\; \sin(\Dlt_s t) \right ] e^{-\gm t} \right \}
\ee
and the imaginary part as
\be
\label{4.121}
 \bt ~ \equiv ~ {\rm Im} X = - g \; \frac{\gm\gm_2\Dlt_s}{\gm^2+\Dlt_s^2} \; ( 1 + As)
\left\{ 1 - \left [ \cos(\Dlt_s t) +
\frac{\gm}{\Dlt_s}\; \sin(\Dlt_s t) \right ] e^{-\gm t} \right \} \; .
\ee
The ratio of the imaginary part to the real part is $\beta/ \alpha \sim \Delta_s/ \gamma$.

Substituting the feedback field (\ref{4.115}) into equation (\ref{4.103}), with neglecting
counter-rotating and higher-harmonic terms, yields the form
\be
\label{4.122}
 \frac{du}{dt} = - i\Om u - i\vp_D s - i \xi_D u \;  ,
\ee
in which
$$
\Om = \om_s - i (\gm_2 - X s)\; - \; 
\frac{1}{2} \; (2S-1) \; \frac{Q_Z}{\mu_F^2} ( 2|X|^2 - X^2) ws \; -
$$
$$
- \;
\frac{i}{2}\;(2S-1) \; \frac{Q_Z}{\mu_F}\; B_0 \left ( X w - X^* w - 2X s^2 \right ) \; .
$$

In view of the small parameters, defined in inequalities (\ref{4.111}) to (\ref{4.114}), the
functional variable $u$ is to be treated as fast while the variables $w$ and $s$ as slow.
Then we can employ the scale separation approach
\cite{Yukalov_109,Yukalov_110,Yukalov_227,Yukalov_229,Yukalov_230,Yukalov_231,Yukalov_232}.
Following this approach, we solve equation (\ref{4.122}) for the fast variable, keeping
there the slow variables as integrals of motion, which gives
$$
u = u_0 \exp \left\{ - i\Om t - i \int_0^t \xi_D(t')\; dt' \right\} \; -
$$
\be
\label{4.123}
- \; is 
\int_0^t \vp_D(t') \exp\left\{ -i\Om(t-t') - i\int_{t'}^t \xi_D(t'')\; dt''\right\}\; dt' \; .
\ee
Then the solution for the fast variable, together with the expression for the feedback field,
is substituted into the equations for the slow variables, whose right-hand sides are averaged
over time and over the stochastic variables $\xi_D$ and $\vp_D$, which yields the
equations for the guiding centers. Thus we come to the equation for the coherence intensity
$$
\frac{dw}{dt} = - 2\gm_2 w + 2\al w s + 2\gm_3 s^2 \; +
$$
\be
\label{4.124}
  + \;
2(2S-1)\; \frac{Q_Z}{\mu_F^2} \; \al\bt w^2 s -
2(2S-1)\; \frac{Q_ZB_0}{\mu_F} \; \al w s^2
\ee
and for the spin polarization
$$
 \frac{ds}{dt} = - \al w - \gm_3 s \; -
$$
\be
\label{4.125}
- \;
(2S-1)\; \frac{Q_Z}{\mu_F^2} \; \al\bt w^2  +
(2S-1)\; \frac{Q_ZB_0}{\mu_F} \; \al w s - \gm_1(s - s_\infty ) \; .
\ee

Equations (\ref{4.124}) and (\ref{4.125}) characterize spin dynamics that can be regulated
by choosing the appropriate system parameters. The motion of spins also produces
electromagnetic radiation \cite{Yukalov_108,Yukalov_110,Yukalov_124,Yukalov_125,
Yukalov_233,Yukalov_234,Li_235,Yukalov_236} that, however, is rather small. The main
advantage in the feasibility of regulating the spin dynamics is that this can be used in
spintronics and in quantum information processing.

\subsection{Spin waves}

Dipolar interactions, in the presence of nonzero magnetization and an external magnetic
field can support spin waves \cite{Yukalov_110,Zhang_222,Yukalov_227}. Spin waves
correspond to small fluctuations of spins, which can be characterized by presenting the
spin operators as
\be
\label{4.126}
 S_j^\al = \lgl S_j^\al \rgl + \dlt S_j^\al \;  .
\ee

Let us consider the initial stage of spin dynamics, when the feedback field has not yet
being formed, $H = 0$, and spins, being polarized, have not yet started rotation, so that
\be
\label{4.127}
\lgl S_j^\pm \rgl = 0 \; , \qquad  \lgl S_j^z \rgl  \neq 0 \; .
\ee

Assuming that the averages $\langle S_j^\alpha \rangle$ do not depend on the index $j$,
we have
$$
 \xi_D = \sum_j \left ( a_{ij} \dlt S_j^z + c_{ij} \dlt S_j^+ + 
c_{ij}^* \dlt S_j^- \right ) \; , \qquad
 \vp_D = \sum_j \left ( \frac{a_{ij}}{2}\; \dlt S_j^- - 2b_{ij} \dlt S_j^+ -
2c_{ij} \dlt S_j^z \right ) \; .
$$

Substituting representation (\ref{4.126}) into the spin equations of motion (\ref{4.94})
and (\ref{4.95}), and linearizing the latter with respect to small deviations, we get the
equations for the spin deviations
\be
\label{4.128}
 \frac{d}{dt}\; S_j^- = - i\om_s S_j^- - i\vp_D \lgl S_j^z \rgl \; , \qquad
  \frac{d}{dt}\; \dlt S_j^z = 0 \; ,
\ee
where we take into account that $\delta S_j^- = S_j^-$. With the initial condition
$\delta S_j^z (0) = 0$, it follows that $\delta S_j^z(t) = 0$. Then
$$
  \xi_D = \sum_j \left ( c_{ij} S_j^+ + c_{ij}^* S_j^- \right ) \; , \qquad 
 \vp_D = \sum_j \left ( \frac{a_{ij}}{2}\; S_j^- -2b_{ij} S_j^+ \right ) \; .
$$

Introduce the Fourier transforms for the dipolar interaction parameters
\be
\label{4.129}
a_{ij} = \frac{1}{N_L} \sum_k a_k e^{i\bk\cdot\br_{ij} } \; , \qquad
b_{ij} = \frac{1}{N_L} \sum_k b_k e^{i\bk\cdot\br_{ij} } \; ,
\ee
and for the spin operators
\be
\label{4.130}
S_j^\pm = \sum_k S_k^\pm e^{i\mp\bk\cdot\br_j } \;   ,
\ee
where we use the notation ${\bf r}_j$ for the lattice vectors instead of ${\bf a}_j$ in 
order not to confuse with the dipolar parameter $a_{ij}$.

Then equation (\ref{4.128}) yields
\be
\label{4.131}
 \frac{d}{dt} \; S_k^- = -i A_k S_k^- + i B_k S_k^+ \;  ,
\ee
where
\be
\label{4.132}
 A_k ~ \equiv ~ \om_s + \frac{a_k}{2} \lgl S_j^z \rgl \; , \qquad 
B_k   ~ \equiv ~ 2b_k \lgl S_j^z \rgl \;  .
\ee
Looking for the solution in the form
$$
 S_k^- = u_k e^{-i\om_k t} + v_k^* e^{i\om_k t} \;  ,
$$
we obtain the spectrum of spin waves
\be
\label{4.133}
 \om_k = \sqrt{A_k^2 - | B_k|^2 } \;  .
\ee
In the long-wave limit $k \ra 0$, the spectrum enjoys the typical of spin waves quadratic
dependence on momentum,
\be
\label{4.134}
 \om_k ~ \simeq ~ | \om_s | \left [ 1 - \lgl S_j^z \rgl \sum_{\lgl j\rgl}
\frac{a_{ij}}{4\om_s} \; (\bk\cdot\br_{ij})^2 \right ] \;  .
\ee
Here the summation is over the nearest neighbours.

The spectrum is stable, provided that $|A_k| > |B_k|$, which implies the inequality
\be
\label{4.135}
\left | \om_0 + \left ( \frac{\om_D}{S} + \frac{a_k}{2} \lgl S_j^z \rgl \right ) \right |
> 2 \left | b_k \lgl S_j^z \rgl \right | \;   .
\ee
Hence, under sufficiently strong external magnetic field and nonzero spin polarization,
the spectrum of spin waves is well defined.

\subsection{Influence of tunneling}

In the previous sections, we have considered well defined isolating states in an
optical lattice, when the tunneling term can be neglected. Now our aim to is to 
take into account the influence of this term, considering an almost insulating state 
slightly perturbed by weak tunneling of atoms. If the tunneling is not negligible, 
then boson tunneling processes induce effective interactions between spins 
\cite{Imambekov_237,Yip_238,Chung_239}.

Let us consider the case, when the lower energy hyperfine manifold has $3$
magnetic sublevels, so that the total moment corresponds to spin $S = 1$. The
Hamiltonian part, including tunneling and the interaction term $H_F$ defined in
equation (\ref{4.77}) reads as
\be
\label{4.136}
 \hat H = - J \sum_{\lgl ij\rgl} \sum_m \hat c_{im}^\dgr \hat c_{jm} \; + \;
\hat H_F \; ,
\ee
where the summation is over the nearest neighbours. The interaction parameters
are given in equation (\ref{4.62}) and the tunneling parameter $J$ can be found from
equation (\ref{4.55}).

The usual expression for the optical lattice potential in $d$ dimensions is
\be
\label{4.137}
 U(\br) = \sum_{\al=1}^d V_\al \sin^2\left (k_0^\al r_\al \right ) \qquad
\left ( k_0^\al \equiv \frac{\pi}{a_\al} \right ) \;  .
\ee
To form an insulating state, the lattice potential depth has to be essentially larger
than the recoil energy
\be
\label{4.138}
 E_R ~ \equiv ~ \frac{1}{2m} 
\left ( \frac{1}{d} \sum_{\al=1}^d \frac{\pi^2}{a_\al^2} \right ) \;  .
\ee
For a cubic $d$-dimensional lattice, for which $V_\alpha  = V_0$ and the recoil
energy is
$$
 E_R = \frac{\pi^2}{2m a^2} \qquad (a_\al = a ) \;  ,
$$
there should be $E_R/V_0\ll 1$.

For an insulating state, one can resort to the tight-binding approximation treating the well
localized Wannier functions as Gaussians. Then from equation (\ref{4.55}) for a cubic
lattice, we find \cite{Yukalov_17,Yukalov_116} the tunneling parameter
\be
\label{4.139}
 J = \frac{d}{4} \left ( \pi^2 - 4 \right ) V_0 \exp\left ( -\; \frac{d\pi^2}{4}\;
\sqrt{ \frac{V_0}{E_R} } \right ) \;  .
\ee
And for the integral entering the interaction parameters, we have
$$
 \int | w(\br) |^4 \; d\br = \left ( \frac{\pi}{2} \right )^{d/2} \; \frac{1}{a^d}\;
\left ( \frac{V_0}{E_R} \right )^{d/4} \;  .
$$
Then these parameters in three dimensions take the form
\be
\label{4.140}
 \overline c_0 = \frac{\sqrt{8\pi}}{3} \; 
\left ( \frac{a_0+2a_2}{a} \right ) E_R \left ( \frac{V_0}{E_R} \right )^{3/4} \; , 
\qquad
\overline c_2 = \frac{\sqrt{8\pi}}{3} \; 
\left ( \frac{a_2-a_0}{a} \right ) E_R \left ( \frac{V_0}{E_R} \right )^{3/4} \; .
\ee
Respectively, for $d = 3$, the tunneling parameter is
\be
\label{4.141}
 J = \frac{3}{4} \left ( \pi^2 - 4\right ) V_0 \exp\left ( - \; \frac{3\pi^2}{4}\;
\sqrt{ \frac{V_0}{E_R} } \right ) \;  .
\ee

Let us consider the Mott state with odd number of atoms per lattice site, when the
filling factor can be written as
\be
\label{4.142}
 \nu ~ \equiv ~ \frac{N}{N_L} =  2n+1 \qquad ( n=0,1,2,\ldots) \;  .
\ee
The tunneling is assumed to be weak, such that
\be
\label{4.143}
 J\nu \ll \sqrt{\overline c_0 \overline c_2} \;  .
\ee

The effective spin Hamiltonian, valid for any odd number of atoms per site $\nu$,
in the limit of small tunneling between sites, can be found in the second order
perturbation theory with respect to the tunneling parameter \cite{Imambekov_237},
resulting in the Hamiltonian
\be
\label{4.144}
 \hat H = - J_0 - J_1 \sum_{\lgl ij\rgl} \bS_i\cdot \bS_j \; - \; 
J_2 \sum_{\lgl ij\rgl} (\bS_i\cdot \bS_j )^2 \; ,
\ee
in which
$$
\frac{J_0}{J^2} = \frac{4(15+20n+8n^2)}{45(\overline c_0 +\overline c_2)} \; - \;
\frac{4(1+n)(3+2n)}{9(\overline c_0 +2\overline c_2)} \; + \;
\frac{128(5+2n)}{225(\overline c_0 +4\overline c_2)}\; ,
$$
$$
\frac{J_1}{J^2} = \frac{2(15+20n+8n^2)}{15(\overline c_0 +\overline c_2)} \; - \;
\frac{16(5+2n)n}{75(\overline c_0 +2\overline c_2)} \; ,
$$
$$
\frac{J_2}{J^2} = \frac{2(15+20n+8n^2)}{45(\overline c_0 +\overline c_2)} \; + \;
\frac{4(1+n)(3+2n)}{9(\overline c_0 -2\overline c_2)} \; + \;
\frac{4(5+2n)n}{225(\overline c_0 +4\overline c_2)}\; .
$$
In the case of one atom per lattice site, when $\nu = 1$ and $n = 0$, one has
$$
\frac{J_0}{J^2} = \frac{4}{3(\overline c_0 +\overline c_2)}\; - \;
 \frac{4}{3(\overline c_0 +2\overline c_2)}\; + \; 
\frac{128}{45(\overline c_0 +4\overline c_2)}\; ,
$$
$$
\frac{J_1}{J^2} = \frac{2}{\overline c_0 +\overline c_2}\; ,
\qquad
\frac{J_2}{J^2} = \frac{2}{3(\overline c_0 +\overline c_2)}\; + \;
 \frac{4}{3(\overline c_0 - 2\overline c_2)}\; .
$$

In the general case, the total effective spin Hamiltonian for $S = 1$ and odd filling 
factor, omitting the constant terms but taking account of the Zeeman and dipolar terms
becomes
\be
\label{4.145}
  \hat H = - J_1 \sum_{\lgl ij\rgl} \bS_i\cdot \bS_j \; - \; 
J_2 \sum_{\lgl ij\rgl} (\bS_i\cdot \bS_j)^2 + \hat H_Z + \hat H_D \; .
\ee
The first two terms here describe effective spin interactions induced by atomic tunneling.
These terms can noticeably influence spin dynamics. Thermodynamics of the system with
Hamiltonian (\ref{4.145}) can also strongly differ from that characterized by the usual
Heisenberg interaction \cite{Gu_240}.

\section{Conclusion}

The paper reviews the main properties of bosonic atoms and molecules interacting through
nonlocal interaction potentials, such as the dipolar potential, and also through spinor
potentials. The specific points of the present review, distinguishing it from the previous
review articles are as follows.

The emphasis is not on briefly listing the numerous properties of the considered systems,
but rather on concentrating on the main of these properties, with thoroughly explaining the
methods of their description.

The necessity of regularizing the dipolar interaction potential is emphasized. Without this
regularization, one often comes to unphysical conclusions, when thermodynamic quantities,
such as chemical potential, free energy, and internal energy become not scalars, which
is certainly meaningless.

A special attention is paid to the derivation of effective spin Hamiltonians for dipolar and
spinor atomic systems. Methods of regulating spin motion are discussed. The possibility
for such an efficient manipulation of spin dynamics is important for applications in spintronics
and quantum information processing.

\vskip 5mm
{\bf Acknowledgements}

\vskip 2mm
The author is grateful for discussions and friendly support to V.S. Bagnato 
and E.P. Yukalova.

\end{document}